\documentclass{article}
\usepackage{amsmath,amssymb}
\usepackage{graphicx}%
\usepackage{verbatim}

\newcommand{\beq}[2]{\begin{equation}\label{#1}
#2 \end{equation}}

\newcommand{\dsp}{\displaystyle}
\newcommand{\Req}[1]{(\ref{#1})}

\newcommand{\kvadrskob}[1]{\left[#1\right]}
\newcommand{\figurskob}[1]{\left\{#1\right\}}

\newcommand{\Eps}{{\cal E}}

\newcommand{\anti}[1]{\overline{#1} \,}

\begin{document}

\begin{center}
{\bf \Large Statistical Cosmological Fermion Systems With Interparticle Fantom Scalar Interaction}. \\[12pt]
Yurii Ignat'ev, Alexander Agathonov and Dmitry Ignatyev \\
N.I. Lobachevsky Institute of Mathematics and Mechanics, Kazan Federal University, \\ Kremleovskaya str., 35, Kazan, 420008, Russia
\end{center}

\begin{abstract}
The article represents a research of the cosmological evolution of fermion statistical systems with fantom scalar interaction where ``kinetic'' term's contribution to the total energy of a scalar field is negative. As a result of analytical and numerical simulation of such systems it has been revealed a existence of four possible scenarios depending on parameters of the system and initial conditions. Among these scenarios there are scenarios with an early, intermediate and late non-relativistic stages of the cosmological evolution, all of which also have necessary inflation stage.

{\bf keywords}{\it physics of the early universe, particle physics - cosmology connection, inflation, phantom scalar interaction}

\end{abstract}

\title{Statistical Cosmological Fermion Systems With Interparticle Fantom Scalar Interaction}

\tableofcontents

\section{Introduction}
Fundamental scalar fields play an important role in understanding the dynamics of the early Universe (see e.g, \cite{Weinberg}, \cite{Gorb_Rubak}); with a help of various modifications of the Gravitation Theory suggested both by Einstein \cite{Einstein} (cosmological $\Lambda$ - term) and thereafter by  R. Utiyama and T. Fukuyama \cite{Utiyama}, A. Minkevich \cite{Minkevich}, \cite{Minkevich2} (Poincar\`{e} ga\-uge theory of gravity), A. Starobinsky \cite{Starobinsky} ($f(R)$ - gravitation), fundamental scalar fields, apparently can explain certain basic observational facts of the cosmology. Nevertheless, some important facts of observational cosmology are still not explained fairly convincingly within the framework of the standard cosmological scenario. For instance, the existence of a non-relativistic stage of the Universe extension, being a requirement for formation of its structure, is reckoned among such obvious facts. Particularly in this context at present time they consider a wide range of \emph{phenomenological theories} of fundamental nonminimum-coupled scalar field where various couplings between scalar and gravitational fields are introduced (potential, kinetic, combined). Corresponding theoretically - field constructions usually pursuit one goal: select such a phenomenological Lagrangian of interaction and its parameters, which will provide a cosmological scenario with all necessary stages namely inflation stage $\to$ ultrarelativistic stage $\to$ nonrelativistic stage $\to$ Secondary acceleration. Herewith, preceding standard cosmological scenario which was generally accepted in 60's-80's (Gamov's hot model) fits between early and late cosmological acceleration stages.

In this article we consider cosmological models which are based on fundamental scalar interaction. In contrast to phenomenological nonminimum models of scalar interaction we consider dynamic models of statistical systems of scalar charged particles where certain particle sorts can directly interact with a scalar field through a certain fundamental \textit{scalar charge}. A statistical system, possessing a scalar charge and being itself the source of the scalar field, can effectively influence on the scalar field managing its behavior. Such scalar interaction was introduced into general relativistic kinetic theory in 1982 by one of the article's authors \cite{Ignatev1}, \cite{Ignatev2}, \cite{Ignatev3}, \cite{Ignatev4} and later on by G.G. Ivanov \cite{Ivanov}. In particular, in articles  \cite{Ignatev2}, \cite{Ignatev3} on the basis of the kinetic theory it was obtained a self-consistent system of equations describing the statistical system of particles with scalar interaction. In \cite{kuza} there were investigated group properties of equilibrium statistical configurations with scalar interaction. In \cite{YuMif}, \cite{YuMif11} there were formulated the equations of cosmological model on basis of statistical Fermi systems with scalar interaction and attempts of numerical simulation of such systems were made. The macroscopic theory of statistical systems with scalar interaction was significantly improved and extended to the case of fantom scalar fields in papers \cite{YuNewScalar1}, \cite{YuNewScalar2}, \cite{YuNewScalar3} \cite{Ignatev14_1}, \cite{Ignatev14_2}\footnote{see also monographes \cite{Yubook1}, \cite{Yubook2} and review \cite{Yu_stfi14}.}. It was also extended to sector of negative effective masses of scalar interacting particles in \cite{Ignatev_stfi15}, \cite{Ignatev_15}; in the same works there were investigated the transformation properties of mathematical models of the statistical systems with scalar interaction relative to transformations of charge, chemical potential and other parameters of the model. The mathematical model of cosmological systems with scalar interaction was extended to the case of conformally invariant scalar field in \cite{Ignatev_16_1}, \cite{Ignatev_16_2}. In these papers the asymptotic transformation properties of this model  as well as collisionless kinetic Vlasov model were investigated. Article \cite{Ignatev_stfi16} provides an investigation of the cosmological model's asymptotic properties based on the statistical system of almost degenerated fermions with interparticle scalar interaction.

Let us now proceed to the questions of numerical simulation of cosmological evolution of the statistical systems with interparticle \emph{fantom scalar interaction},  possessing negative <<kinetic>> energy. From mathematical standpoint the system of differential equations describing the cosmological evolution of such dynamic systems falls into class of stiff systems and numerical simulation of such systems presents severe calculation difficulties. Stiff character of the system is due to the Einstein equations, in left side of which is a square of Hubble constant and in the right side of which contains an alternating expression which sign is determined by selection of two factors - negative kinetic energy and positive potential energy. Certain results of numerical simulation of such systems were obtained in \cite{Ignatev_Agathonov_2015} (cosmological evolution of degenerated Fermi gas), \cite{Ignatev_Mikhailov_2015} (cosmological evolution of single-component Boltzmann gas), \cite{Ignatev_AAS} (comprehensive analysis).

Results of numerical simulation enabled detection the series of unique properties of cosmological statistical models with interparticle fantom interaction, in particular (see e.g. \cite{Ignatev_AAS}):
\begin{enumerate}
\item \emph{burst of acceleration} -- presence of sharp bursts of invariant cosmological acceleration in sufficiently broad range of model parameters;
\item \emph{plateau phase} -- appearance of cosmological stages with constant acceleration;
\item \emph{thermal flash} -- rapid heating of the statistical systems at certain evolution stages.
\end{enumerate}
These and others properties of fantom scalar interaction differ it enormously from classic scalar interaction which in fact reduces to model of damped oscillations of the oscillator in a potential well. Mentioned properties of the cosmological statistical models with interparticle fantom scalar interaction can potentially explain a series of observational facts that are not quite clear at present time and which are related to both early and late stages of cosmological evolution. They can also probably reveal a nature of dark energy and dark matter. Given article is exactly devoted to detection of stable properties of the cosmological statistical models with scalar interaction which can be used in construction of dark sectors of energy and matter. Since, as it was shown earlier (see papers cited above), the cosmological evolution of the statistical systems with scalar interaction weakly depends on a type of statistics, in given work we carry out numerical simulation for degenerated or almost degenerated systems of fermions acting as scalar charge carrier.

\section{Mathematical Model Of the Statistical Systems With Interparticle Fantom Scalar Interaction}
\subsection{Microscopic Dynamics of Scalar Charged Particles}
The canonical equations of relativistic particle motion in the phase space $\Gamma$ have the following form (see e.g., \cite{Ignatev2}):
\begin{equation} \label{Eq1}
\frac{dx^{i} }{ds} =\frac{\partial H}{\partial P_{i} } ;\quad \quad \frac{dP_{i} }{ds} =-\frac{\partial H}{\partial x^{i} } ,
\end{equation}
 where $H(x,P)$ is a relativistically invariant\\ Hamilton function and $u^i=dx^i/ds$ is a particle velocity vector.
Calculating the full derivative of dynamic vari\-ables function $\Psi (x^{i} ,P_{k} )$ and taking into account (\ref{Eq1}) we obtain:
\begin{equation} \label{Eq2}
\frac{d\Psi }{ds} =[H,\Psi ],
\end{equation}
where the invariant Poisson brackets are introduced:
\begin{equation} \label{Eq3}
[H,\Psi ]=\frac{\partial H}{\partial P_{i}} \frac{\partial \Psi }{\partial x^{i} } -\frac{\partial H}{\partial x^{i} } \frac{\partial \Psi }{\partial P_{i} } \; .
\end{equation}
Let us note that Poisson bracket \eqref{Eq3} can be rewritten in the explicitly covariant form using
 {\it the operator of covariant Cartan differentiation}, %
$\widetilde{\nabla}_i$, \footnote{Covariant derivative in a bundle $\Gamma$ \cite{Cartan}.}, %
 (see e.g., \cite{Bogolyub})%
\footnote{Cartan covariant derivatives were first introduced into
relativistic statistics by A.A. Vlasov \cite{Vlasov}.}:
\begin{equation}\label{Cartan}%
\widetilde{\nabla}_i = \nabla_i +
\Gamma_{ij}^k P_k\frac{\partial}{\partial P_j},
\end{equation}
where $\nabla_i$ is an operator of covariant Ricci dif\-fe\-ren\-tiation and $\Gamma^k_{ij}$ are Christoffel symbols of the second kind with
respect to metrics $g_{ij}$ of base $X$. %
Operator $\widetilde{\nabla}$ is defined in such a way that
\begin{equation}\label{9.11}%
\widetilde{\nabla}_iP_k \equiv 0
\end{equation}
and the following {\it symbolic} rule of functions dif\-fe\-ren\-tiation is fulfilled:

\begin{equation}\label{9.13}%
\widetilde{\nabla}_i\Psi(x,P) = \nabla_i[\Psi(x],P).
\end{equation}
This rule means that in order to calculate Cartan derivative of function $\Psi(x,P)$ it would be enough to calculate its ordinary covariant
 derivative as if the momentum vector was covariant constant.
 Because of this equality the introduced operator is quite convenient
 for execution of differential and integral operations in the phase space %
$\Gamma$. Thus let us write the Poisson bracket \eqref{Eq3} in the explicitly covariant form:
\begin{equation}\label{H_Cart}
[H,\Psi ]\equiv \frac{\partial H}{tial P_{i}} \widetilde{\nabla}_i\Psi-\frac{\partial \Psi}{\partial P_{i}} \widetilde{\nabla}_i H,
\end{equation}

The Hamilton function and the normalization ratio for the generalized momentum take form\footnote{Throughout the paper, select the universal system of units $\hbar=c=G=1$.}:
\begin{equation}\label{H,m}
H(x,P)=\frac{1}{2} \left[m_*^{-1}(x)(P,P)-m_*\right]=0,
\end{equation}
\begin{equation}\label{norm}
H(x,P)=0\Rightarrow (P,P)=m^2_*,
\end{equation}
where $m_*=m_*(\Phi)$ is the \emph{effective mass of particle}.

Let us notice the identities that are valid for the Hamilton function \eqref{H,m} and which could be useful in future:
\begin{equation}\label{nabla_H}
\widetilde{\nabla}_iH=-\nabla_i m_*,
\end{equation}
\begin{equation}\label{HPsi}
[H,\Psi]=\frac{1}{m_*}P^i\widetilde{\nabla}_i\Psi+\partial_i m_*\frac{\partial \Psi}{\partial P_i},
\end{equation}
where $\Psi(x,P)$ is an arbitrary function.

Invariant action functional of a classic particle in scalar fields $\{\Phi_1,\ldots,\Phi_n\}$ has the following form \cite{Ignatev_15}
\begin{equation} \label{Sp}
S=\int  m_{*} ds,
\end{equation}
where $m_*(\Phi_1,\ldots,\Phi_n)$ is an \emph{invariant effective mass of a particle} in scalar fields. Due to additivity of the Lagrangian functino, the effective mass of a particle should have the form \cite{Ignatev_15}:
\begin{equation} \label{m_*}
m_*=m_0+\sum\limits_r  q^{(r)}\Phi_r,
\end{equation}
where $m_0$ is a certain initial rest mass and $q^{(r)}$ is
a charge of particle relative to the scalar field $\Phi_r$ which we presume to be functional independent. In particular, for a scalar singlet
$\{\Phi_1,\ldots,\Phi_n\}\equiv \Phi$
 it is:
\begin{equation}\label{m_*1}
m_*=m_0+q\Phi.
\end{equation}
As it was shown in \cite{Ignatev14_1}, the negativeness of the particle's effective mass function
does not lead to any contradictions at the level of microscopic dynamics since the observable momentum of particle
(as well as the 3-dimensional velocity $v^\alpha=u^\alpha/u^4$) conserves its orientation as opposed to the unobservable kinematic 4-velocity of a particle $u^i$:
\begin{equation}\label{Pp}
p^i=m_*\frac{dx^i}{ds}\equiv P^i.
\end{equation}

Let us notice that the effective particle mass (\ref{m_*}) could be a negative value, however this would not affect neither the equations of motion nor thr definitions of macroscopic currents of the dynamic values \cite{Ignatev_15} since these values and corresponding equations conclude only absolute magnitude of the particle effective mass. For instance, the Euler-Lagrange equations for a particle in scalar and gravitational fields take form:
\begin{equation} \label{EqLagrange}
\frac{d^{2} x^{i} }{ds^{2} } +\Gamma _{jk}^{i} \frac{dx^{j} }{ds} \frac{dx^{k} }{ds} =\partial _{,k} \ln |m_*|{\rm {\mathcal P}}^{ik} ,
\end{equation}
 where:
\begin{equation} \label{Proect}
{\rm {\mathcal P}}^{ik} ={\rm {\mathcal P}}^{ki} =g^{ik} -u^{i} u^{k};\quad u^i=\frac{dx^i}{ds}
\end{equation}
is a tensor of orthogonal projection on the direction $u$ so that:
\begin{equation} \label{Ortho}
{\rm {\mathcal P}}^{ik} u_{k} \equiv 0;\quad {\rm {\mathcal P}}^{ik} g_{ik} \equiv 3.
\end{equation}

\subsection{Macroscopic Averages and Equations of Dynamic Averages Transport}
Let $f_a(x,p)$ is an invariant function of $a$-sort particles distribution in a 7-dimensional phase space $\Gamma=X\times P$,
where
\begin{equation}\label{dP}
dP=\sqrt{-g}\frac{dp^1dp^2dp^3}{p_4}
\end{equation}
is an invariant differential of volume of a 3-dimensional momentum space where $p_4$ is a positive root of the equation (\ref{norm}). Macroscopic averages of a certain dynamic scalar function $\psi(x,p)$ are defined by means of the invariant distribution function:
\begin{eqnarray}\label{Psi}
\Psi(\tau)=\frac{2S+1}{(2\pi)^3}\int\limits_V  U_i dV\!\!\!
 \int\limits_{P(X)}\!\! p^i dP \psi(x,p)f(x,p),
\end{eqnarray}
where $U^i$ is a velocity vector of a macroscopic field of observers and $\tau$ is a time measured by clocks of these observers.
Inner integral over momentum space in (\ref{Psi}) represents itself a vector of the dynamic value's flux $\psi$.
In particular, at $\psi=1$ we obtain from (\ref{Psi}) a vector of particles' flux density\footnote{According to J. Synge \cite{Sing} numerical vector.}:
\begin{equation}\label{ni}
n^i(x)=\frac{2S+1}{(2\pi)^3}\int\limits_{P(X)} p^i f(x,p)dP,
\end{equation}
so that:
\begin{equation}\label{V}
N(\tau)=\int\limits_V n^iU_idV
\end{equation}
is a total number of particles in volume $V$ at $\tau$.

Let the following reactions run in the statistical system:
\begin{equation} \label{reactions}
\sum _{A=1}^{m} \nu _{A} a_{A} {\rm \rightleftarrows }\sum _{B=1}^{m'} \nu '_{B} a'_{B} ,
\end{equation}
where $a_{A} $ are particle symbols and $\nu _{A} $ are particle numbers in each channel of reactions. Thus the  momen\-tums of the initial (I) and final (F) states are equal:
\begin{equation}\label{p=p}
p_I=\sum\limits_{A=1}^m\sum\limits_\alpha^{\nu_A} p^\alpha_A,
\quad p_F=\sum\limits_{B=1}^{m'}\sum\limits_{\alpha'}^{\nu'_B} p'\
\!\!^{\alpha'}_B\rightarrow p_I=p_F.
\end{equation}
Then the strict consequences of general relativistic kinetic equations are the transport equation of dynamic values \cite{Ignatev_15}:
\begin{eqnarray}
\nabla _{i} \sum _{a} \int\limits _{P_0} \psi _{a} f_{a}
p^i dP_{a} -\sum _{a} \int\limits _{P_0} f_{a} m_*[H_{a} ,\psi _{a} ]dP_{a} =\nonumber\\
\label{transport_eqs}
-\sum _{by\; chanels} \int  \biggl(\sum _{A=1}^{m} \nu _{A} \psi _{A} -
\sum _{B=1}^{m'} \nu '_{B} \psi '_{B} \biggr)\times\nonumber\\
\delta ^{4} (p_{F} -p_{I} )(Z_{IF} W_{IF} -Z_{FI} W_{FI} )\prod _{I,F} dP,
\end{eqnarray}
where
$$W_{FI} =(2\pi )^{4} |M_{IF} |^{2} 2^{-\sum  \nu _{A} +\sum  \nu '_{b} } $$
is a scattering matrix of the reaction channel
 \eqref{reactions}, ($|M_{IF}|$ are invariant scattering amplitudes);
\begin{eqnarray}
Z_{IF} =\prod _{I} f(p_{A}^{\alpha } )\prod _{F} [1\pm f(p_{B}^{\alpha '} )];\nonumber\\
\quad Z_{FI} =\prod _{I} [1\pm f(p_{A}^{\alpha } )]\prod _{F} f(p_{B}^{\alpha '} ),\nonumber
\end{eqnarray}
the sign ``+'' corresponds to bosons and ``-'' corresponds to fermions (please see details in \cite{Ignatev3,Ignatev4}).

Putting $\Psi _{a}=g_{a} $ in  \eqref{transport_eqs}, where $g_{a} $ are certain fundamental charges being conserved in reactions \eqref{reactions}, with account of \eqref{HPsi} we obtain the transport equations of the statistical system's particle number flux densities:
\begin{equation} \label{GrindEQ__55_}
\nabla _{i} J_{G}^{i} =0,
\end{equation}
where:
\begin{equation} \label{GrindEQ__56_}
J_{G}^{i} =\sum _{a} \frac{2S+1}{(2\pi )^{3} } \; g_{a}
\int\limits_{P} f_a(x,p)p^{i} dP.
\end{equation}
is a density vector of the fundamental current corres\-ponding to charges $g_{a} $. Particularly, the con\-ser\-vation law
\eqref{GrindEQ__55_} always is held for each particle sort
$b$ ($g_{a} =\delta _{a}^{b} $) given their collisions are elastic.

Let us put  $\Psi _{a} =P^{k} $ in \eqref{transport_eqs}. Then as a result of the conservation law of the momentum at collisions \eqref{reactions}, the integrand in big parentheses \eqref{transport_eqs} is equal to:
\begin{equation}
\sum _{A=1}^{m} \nu _{A} \Psi _{A} -
\sum _{B=1}^{m'} \nu '_{B} \Psi '_{B}\equiv P_I-P_F=0.\nonumber
\end{equation}
Thus we obtain the transport equations of statistical system energy-momentum:
\begin{equation} \label{GrindEQ__57_}
\nabla _{k} T_{p}^{ik} -\sum\limits_r\sigma_{(r)}\nabla ^{i} \Phi_{r} =0,
\end{equation}
where there are introduced the \textit{statistical system energy - mo\-mentum tensor}
\begin{equation}\label{Tpl}
T^{ik}_p=\sum\limits_{a} \frac{2S+1}{(2\pi )^{3}}
\int\limits_{P} f_a(x,p)p^ip^kdP
\end{equation}
and the \textit{scalar densities of statistical system charge relative to scalar field $\Phi_r$}, $\sigma^{(r)}$ :
\begin{equation}\label{sr}
\sigma^{(r)}=\sum\limits_a \sigma^{(r)}_a,
\end{equation}
where $\sigma^{(r)}_a$ are the \textit{scalar charge densities of $a$ - component of statistical system relative to scalar field $\Phi_r$}:
\begin{equation} \label{GrindEQ__58_}
\sigma^{(r)}_a =\frac{2S+1}{(2\pi )^{3} } m^*_a q^{(r)}_a
\int\limits_{P} f_a(x,p)dP_0,
\end{equation}

Particularly, for charge singlet $(q,\Phi)$
the con\-ser\-vation law \eqref{GrindEQ__57_} takes form:
\begin{equation} \label{T_pikk}
\nabla _{k} T_{p}^{ik} -\sigma \nabla ^{i} \Phi =0,
\end{equation}
where it is (see \cite{Ignatev3,Yu_stfi14}):
\begin{equation} \label{GrindEQ__59_}
\sigma =\Phi \frac{2S+1}{(2\pi )^{3} } q^{2}
\int\limits_{P)} f(x,p)dP.
\end{equation}
It should be noted that the form of the energy-momentum tensor \eqref{Tpl} and charge scalar density
\eqref{GrindEQ__58_} which was found for scalar charged particles at given Hamilton function, is a direct consequence of the canonical equations and the assumption
about conservation of total momentum in local collisions of particles.

\subsection{Fantom Scalar Fields with Attraction}
The Lagrangian function of a massive fantom scalar field with attraction of like-charged particles has the form: \cite{Ignatev14_2}\footnote{In \cite{YuNewScalar3} it was shown that classic scalar fields with attraction of like-charged scalar charged particles are incompatible with the Einstein equations as well as fantom scalar fields with repulsion of like-charged particles.}:
\begin{equation}\label{Ls}
L_s=-\frac{1}{8\pi}\bigl(g^{ik}\Phi_{,i}\Phi_{,k}+m^2_s\Phi^2\bigr),
\end{equation}
and the corresponding tensor of energy-momentum is equal to:
\begin{equation}\label{Tik_s}
T^{ik}_s=\frac{1}{8\pi}\bigl(-2\Phi^{,i}\Phi^{,k}+g^{ik}\Phi_{,j}\Phi^{,j}+g^{ik}m^2_s\Phi^2\bigr).
\end{equation}
In this paper we consider a scalar singlet. In this case the conservation law of total energy-momentum tensor of the system of ``scalar charged particles + scalar field''  $T^{ik}=T^{ik}_p+T^{ik}_s$ has the form:
\begin{equation}\label{Tik,k=0}
\nabla_k T^{ik}= -\frac{1}{4\pi}\bigl(\Box\Phi-m^2_s\Phi-4\pi\sigma\bigr)\nabla^i\Phi=0,
\end{equation}
wherefrom we obtain the equation of the scalar field with a source (\ref{EqPhi}.
\begin{equation}\label{EqPhi}
\Box\Phi-m^2_s\Phi=4\pi\sigma.
\end{equation}
The solution of the field equation for point scalar source rests in the origin of coordinates of the Euclidean space
\[f(x,p)=\delta^{(3)}(\mathbf{r})\delta^{(3)}(\mathbf{p}) ,\]
and has the form \cite{YuNewScalar3}:
\begin{equation}\label{single_solve}
\Phi=q\frac{\sin m_sr}{r},
\end{equation}
and actually coincides with the solution of the Einstein equations for small spherical gravitational disturbances against the background of Friedmann metrics \cite{Ignat_Popov}.
From (\ref{single_solve}) it follows that the eigen mass of scalar particle at $m_0=0$ is defined by its potential energy in scalar eigenfield \footnote{Let us notice that he scalar charge of particle is a dimensionless value ($[q]=0$).}:
\begin{equation}\label{m_self}
m_*^0=\lim\limits_{r\to0}q\Phi(r)=qm_s.
\end{equation}
Let us notice that the solution of the field equation for the fantom field in the case of single charge differs from the corresponding solution of the field equation for the classical field in substitution $\sin m_sr\to \mathrm{e}^{-m_sr}$.

Further, the force of two scalar charged particles interaction with charges $q_1$ and $q_2$, located in a distance $r$ from each other, according to (\ref{EqLagrange}) and (\ref{Proect}) is equal to:
\begin{equation}\label{F12}
\mathbf{F}_{12}=-q_1 q_2\frac{\mathbf{r}}{r^3}(m_sr\cos m_sr-\sin m_sr).
\end{equation}
At small distances
\begin{equation}\label{F120}
\mathbf{F}_{12}=-q_1 q_2\frac{\mathbf{r}}{3}m_s^3, \quad (rm_s\ll 1)
\end{equation}
this force is a force of attraction of like-charged scalar particles and acts likewise elastic force with elasticity modulus  $k=q_1 q_2 m^3_s/3$. Hence we can state that at small distances pairs of like-charged scalar particles will likewise oscillator perform harmonic oscillations with the frequency $\sqrt{k/m_*}=m_s\sqrt{q/3}$. At large distances
\begin{equation}\label{F128}
\mathbf{F}_{12}=-q_1 q_2\frac{\mathbf{r}}{r^2}\cos m_s r, \quad (rm_s\gg 1)
\end{equation}
the force is alternating and falls in proportion to $1/r$. For opposite charged particles the force $F_{12}$ at small distances corresponds to repulsion.
The vacuum solutions of the equation (\ref{EqPhi}) correspond to retarded and advanced waves at wave numbers $\mathbf{k}$ which are greater than Compton ones,
\begin{equation}\label{k>ms}
\Phi=\sum\limits_{\pm}C_\pm+ \mathrm{e}^{\pm i\sqrt{\mathbf{k}^2-m^2_s}t+i\mathbf{kr}};\quad \mathbf{k}^2>m^2_s,
\end{equation}
spreading with a phase velocity which is less than velocity of light. For wave numbers $\mathbf{k}$, which are less than Compton ones, the vacuum solutions correspond to increasing and decaying standing waves \footnote{These properties of the solutions again emphasize the unique ''exotic'' character of the fantom scalar field.}:
\begin{equation}\label{k<ms}
\Phi=\sum\limits_\pm C_\pm \mathrm{e}^{\pm\sqrt{m^2_s-\mathbf{k}^2}t+i\mathbf{kr}}; \quad \mathbf{k}^2<m^2_s.
\end{equation}
\subsection{The Locally Equilibrium Statistical Systems of Scalar Interacting Particles}
In the case of locally thermodynamical equilibrium (LTE) which is the only case being considered in this article \footnote{The collisionless kinetic model was considered in \cite{Ignatev_16_1}, \cite{Ignatev_16_2}}, the distribution function has the locally-equilibrium form:
\begin{equation}\label{Func_distr_equil}
f^0_{(a)}(x,p_a) = \figurskob{\exp
\kvadrskob{\frac{\dsp{- \mu_a + (v,P_a)}}{\theta}} \mp 1}^{ -
1},
\end{equation}
where signs $-$ and $+$ correspond to particles with integer and half-integer spin, $v^i$ is a timelike vector of the macroscopic velocity of the statistical system, so that:
\begin{equation}\label{(v,v)=1}
(v,v)=1,
\end{equation}
$\theta$ is a local temperature of the statistical system.
Next, $\mu_a$ is a chemical potential of $a$ -sort particles wherein chemical potentials must satisfy the conditions of the chemical equilibrium:
\begin{eqnarray}
\label{chem_eq}
\sum _{A=1}^{m} \nu _{A} \mu_{A} =\sum _{B=1}^{m'} \nu '_{B} \mu'_{B} .
\end{eqnarray}

\subsection{The Moments of The Equilibrium Distribution}
The moments of the distribution \Req{Func_distr_equil} are equal to \cite{Ignatev_15}:
\begin{eqnarray}\label{ni} n^i_{(a)}(x) = n_{(a)}(x) v^i\,;\\
\label{T_pik}
T^{ik}_p(x) = (\Eps_p + \mathcal{P}_p) v^i v^k - \mathcal{P}_p g^{ik}\,,
\end{eqnarray}
where $n_{(a)}$ is a ``$a$''-sort particle number density, $\Eps_p=\sum\Eps_{(a)}$ and $\mathcal{P}_p=\sum\mathcal{P}_{(a)}$ are total energy density and pressure of the statistical system:
\begin{eqnarray} \label{na_LTE}
n_{(a)} =\frac{2S+1}{2\pi ^{2} } m_{*}^{3} \int _{0}^{\infty } \frac{{\rm sh}^{2} x{\rm ch}
xdx}{e^{-\gamma _{a} +\lambda _{*} {\rm ch} x} \pm 1} ;\\
\label{Ep_LTE}
{\mathcal E}_{p} = \sum _{a} \frac{2S+1}{2\pi ^{2} } m_{*}^{4}
\int _{0}^{\infty } \frac{{\rm sh}^{2} x{\rm ch}^{2} xdx}{e^{-\gamma _{a} +
\lambda _{*} {\rm ch} x} \pm 1}; \\
\label{Pp_LTE}
{\mathcal P}_{p}=\sum _{a} \frac{2S+1}{6\pi ^{2} } m_{*}^{4}
\int_{0}^{\infty }\frac{{\rm sh}^{4} xdx}{e^{-\gamma _{a} +\lambda _{*} {\rm ch} x} \pm 1}; \\
\label{Tp_LTE}
T_{p} =\sum _{a} \frac{2S+1}{2\pi ^{2} } m_{*}^{2}
\int _{0}^{\infty }
\frac{{\rm sh}^{2} xdx}{e^{-\gamma _{a} +\lambda _{*} {\rm ch} x} \pm 1};\\
\label{s_LTE}
\sigma =\sum _{a} \frac{2S+1}{2 \pi ^{2} } q(m+q_{(a)}\Phi)^{3}
\int _{0}^{\infty } \frac{{\rm sh}^{2} xdx}{e^{-\gamma _{a} +\lambda _{*} {\rm ch} z} \pm 1} ,
\end{eqnarray}
where two dimensionless scalar functions are introduced:
\begin{equation}\label{lm}
\lambda _{*} =\frac{m_{*}}{\theta};\quad \gamma_{(a)}=\frac{\mu_{(a)}}{\theta},
\end{equation}
and macroscopic scalars  ${\mathcal E}_{p}$, ${\mathcal P}_{p}$, $T_{p}$ and $\sigma$ are obtained through summation of the corresponding values over components of the systems.

Let us notice that the chemical potential of massless particles possessing zero fundamental charges at LTE conditions is equal to zero. This conclusion follows from the fact that numbers $\nu^a_A$ of such particles participating in reactions \Req{reactions} can be completely arbitrary. Then from the fact of existence of the reaction of particles and antiparticles annihilation it follows the well-known relation \cite{Landau_Stat}:
\beq{3.1.24}{\anti{\mu}_{(a)} = - \mu_{(a)}\,. }
Let us also notice that the relativistic chemical potential is connected to the Fermi momentum $p_f$ by means of standard relativistic relation:
\begin{equation}\label{mu0}
\mu=\sqrt{m^2_*+p^2_f}.
\end{equation}

\section{The Mathematical Model of the Locally Equilibrium Self - Gra\-vitating Statistical System of Scalar Charged Particles}
\subsection{The Complete System of Equations}
The complete system of self-consistent macroscopic equations describing the self-gravitating statistical system of scalar charged particles, comprises of:

\begin{itemize}
\item First of all, the Einstein equations:
\begin{equation}\label{Einst_Scalar}
R^{ik}-\frac{1}{2}Rg^{ik}=8\pi (T^{ik}_p+T^{ik}_s),
\end{equation}
where $T^{ik}_p$ is a determined above tensor of energy-momentum of the statistical system (\ref{T_pik}), (\ref{Ep_LTE}), (\ref{Pp_LTE}), and $T^{ik}_s$
is a tensor of energy-momentum of the scalar field (\ref{Tik_s});
\item Second, transport equations of energy-momentum of particles (\ref{T_pikk}).
\item Third, equations of the scalar field with a source (\ref{EqPhi}):
\item Fourth, the conservation law of particles (\ref{GrindEQ__55_}).
\begin{equation}\label{ni,i}
\nabla_i\sum\limits_{(a)} {\rm e}_{(a)} n^i_{(a)}=0,
\end{equation}
\end{itemize}
Let us find out what are the consequence of the conservation laws (\ref{T_pikk}) and (\ref{ni,i}) at LTE conditions (see \cite{Ignatev_AAS}). %
Using (\ref{ni}) we reduce the conservation law of particles number \ref{ni,i}) to the form:
\begin{equation}\label{III.7b}
\nabla_k (\Delta n v^k)=0,\quad \Delta n\equiv \sum\limits_{(a)} {\rm e}_{(a)} n_{(a)}.
\end{equation}
From the relation ration of the velocity vector (\ref{(v,v)=1}) it follows the well-known identity law:
\begin{equation}\label{III.6}
v^k_{~,i}v_k\equiv 0.
\end{equation}
Next, with account of definition (\ref{T_pik}) the conservation laws of the energy-momentum tensor of the statistical system(\ref{T_pikk}) can be reduced to:
\begin{eqnarray}\label{III.7}
(\mathcal{E}_{p}+\mathcal{P}_{p})v^i_{~,k}v^k=(g^{ik}\!\!\!-v^iv^k)(\mathcal{P}_{p,k}+\sigma\Phi_{~,k});\\
\label{III.7a}
\nabla_k(\mathcal{E}_{p}+\mathcal{P}_{p})v^k=(\mathcal{P}_{p,k}+\sigma\Phi_{~,k})v^k,
\end{eqnarray}

Thus, formally on 3 macroscopic scalar functions $\Eps_p, \mathcal{P}_p, n_e$ and 3 %
independent components of the velocity vector  $v^i$ the macroscopic conservation laws provide 5 independent equations (\ref{III.7b}) -- (\ref{III.7a}),since one of the equations (\ref{III.7}) is dependent on others in consequence of the identity (\ref{III.6}).
However, not all cited above macroscopic scalars are functionally independent since all they are determined through locally equilibrium distribution functions (\ref{Func_distr_equil}).%
At resolved series of chemical equilibrium conditions, when the only one chemical potential remains independent as well as at resolved equation of mass surface and given scalar potential and scale factor, four macroscopic scalars, %
$\mathcal{E}_p, \mathcal{P}_p, n_e, \sigma$, are determined through two thermodynamic scalars --- a certain chemical potential $\mu$ and local temperature $\theta$.
Thus, the system of equations(\ref{III.7b}) -- (\ref{III.7a}) proves to be completely defined.

\subsection{The Cosmological Model}

Let us consider the formulated above self-consistent mathematical model with regard to the cosmological situation for the space-flat Friedmann model:
\[ds^2=dt^2-a^2(t)(dx^2+dy^2+dz^2),\]
In such a case all thermodynamic functions depend only on time. One can easily make sure that $v^i=\delta^i_4$
turns equations (\ref{III.7}) into identities, the system of equations (\ref{III.7b}) --  (\ref{III.7a})%
is reduced to two differential equations relative to two thermodynamic functions $\mu$ è $\theta$ :
\begin{equation}\label{III.7a1}
\dot{\mathcal{E}}_{p}+3\frac{\dot{a}}{a}(\mathcal{E}_{p}+\mathcal{P}_{p})=\sigma\dot{\Phi};
\end{equation}
\begin{equation}\label{III.7b1}
\dot{\Delta n} +3\frac{\dot{a}}{a}\Delta n=0\Rightarrow \Delta n a^3=\mathrm{Const}.
\end{equation}
At $\Phi=\Phi(t)$ the tensor of scalar field's energy-momentum also takes form of energy-momentum tensor of the ideal isotropic homogenous flux:
\begin{equation} \label{MET_s}
T_{s}^{ik} =({\rm {\mathcal E}}_s +{\rm {\mathcal P}}_{s} )v^{i} v^{k} -{\rm {\mathcal P}}_s g^{ik} ,
\end{equation}
where
\begin{eqnarray}\label{Es}
{\mathcal E}_s=\frac{1}{8\pi}(-\dot\Phi^2+m_s^2\Phi^2);\\
\label{Ps} {\mathcal P}_{s}=-\frac{1}{8\pi}(\dot\Phi^2+m_s^2\Phi^2),
\end{eqnarray}
so that:
\begin{equation}\label{e+p}
{\mathcal E}_s+{\mathcal P}_{s}=-\frac{\dot{\Phi}^2}{4\pi}.
\end{equation}
The equation of the scalar field in the Friedmann metrics takes form:
\begin{equation}\label{Eq_S_t}
\ddot{\Phi}+3\frac{\dot{a}}{a}\dot{\Phi}-m^2_s\Phi= 4\pi\sigma.
\end{equation}
The non-trivial Einstein equation should be added to these equations:
\begin{equation}\label{Einstein_a}
3\frac{\dot{a}^2}{a^2}=8\pi{\mathcal E},
\end{equation}
where ${\mathcal E}$ is a total energy density of fermions and the scalar field. This system of equations (\ref{III.7a1}), (\ref{III.7b1}), (\ref{Eq_S_t}) and (\ref{Einstein_a}) relative to  $\theta(t), \mu(t), \Phi(t), a(t)$ describes a closed mathematical model of the cosmological evolution of the statistical system with interparticle fantom scalar interaction (see \\ 
\cite{Ignatev14_2} ). In this equations one, (\ref{III.7b1}), is an algebraic equation, two, (\ref{III.7a1}) and  (\ref{Einstein_a}), are ordinary differential equations of the first order and one equation, (\ref{Eq_S_t}), is an ordinary differential equation of the second order. Thus, reducing this system to normal form we obtain the system of four ordinary differential equations of the first order with algebraic coupling relative to five functions:
\begin{equation}\label{funcs}
\mu(t);\; \theta(t);\; \Phi(t);\; Z(t)=\dot{\Phi};\; a(t).
\end{equation}
\subsection{Simplification of Equations of the Cosmological Model}
Let us notice the following important circumstances:
\begin{enumerate}
\item The system of equations (\ref{III.7a1}), (\ref{Eq_S_t}) and (\ref{Einstein_a}) with an account of definitions (\ref{na_LTE}) -- (\ref{s_LTE}), as well as
(\ref{Es}) and (\ref{Ps}) is an autonomous system of ordinary differential equations relative to functions (\ref{funcs}) since time variable is not explicitly included in these equations. Hence, this variable can be excluded while processing to a new system of variables:
\begin{equation}\label{funcs_new}
\mu(a);\; \theta(a);\; \Phi(a);\; \Phi'_a; \; \dot{a}(a),
\end{equation}
supposing
\begin{eqnarray}\label{t->a}
\frac{d\phi(t)}{dt}=\frac{d\phi }{da}\dot{a}; \quad \ddot{a}=\frac{d\dot{a}}{da}\dot{a}\equiv \frac{1}{2}\frac{d\dot{a}^2}{da}; \nonumber\\ \frac{d^2\phi}{dt^2}=\frac{d^2}{da^2}+\frac{1}{2}\frac{d\phi}{da}\frac{d\dot{a}^2}{da}.
\end{eqnarray}
\item The statistical system's energy-momentum conservation law in these variables (\ref{III.7a1}) takes form:
\begin{equation}\label{Cons_Ep}
\frac{d\Eps_p}{da}+\frac{3}{a}(\Eps_p+\mathcal{P}_p)=\sigma \Phi'_a.
\end{equation}
\item The scalar field equation in this variables (\ref{Eq_S_t}) takes form:
\begin{equation}\label{EqPhi(a)}
\Phi''_{aa}\dot{a}^2+\frac{1}{2}\Phi'_a\frac{d\dot{a}^2}{da}+3\frac{\dot{a}^2}{a}\Phi'_a-m^2_s\Phi=4\pi\sigma.
\end{equation}

\item In conclusion, the Einstein equation (\ref{Einstein_a}) becomes an algebraic equation relative to variables (\ref{funcs_new}):
\begin{equation}\label{Einstein(a)}
\left(\frac{3}{a^2}+\Phi'_a\ \!^2\right)\dot{a}^2=m^2_s\Phi^2+8\pi\Eps_p.
\end{equation}
\end{enumerate}
This equation can easily be resolved relative to $\dot{a}$ and result can be substituted into the field equation (\ref{EqPhi(a)}).
As a result, there will remain the system of three ordinary differential equations relative to variables $\Phi(a)$, $\Phi'_a(a)$ and
of the thermodynamic functions. Having found the solution of these equations, we should resolve the equation
\begin{equation}\label{dota(a)}
\frac{da}{dt}=\dot{a}\Rightarrow a=a(t).
\end{equation}

\section{The Mathematical Model of the Relativistic Almost-Degenerated Fermi System}

\subsection{The Macroscopic Scalars for ALmost-Degenerated Single-Component Fermi System}
In this chapter we will investigate analytic properties of almost-degenerated locally equilibrium Fermi system with interparticle scalar interaction in detail. Let us notice that these properties, being related directly to particles, do not depend on a character of scalar interaction: they coincide with system with classical and fantom scalar interaction. Since in this article we consider the cosmological models with ultrarelativistic start, it is necessary to take into account both particles and antiparticles under thermodynamic equilibrium conditions. Let us consider a problem of necessity of accounting antiparticles in the symmetric model for the simplicity, where $m^+_*=m^-_*=|q\Phi|$ (see e.g., \cite{Ignatev_AAS}). According to (\ref{3.1.24}) under LTE conditions the condition of antisymmetry of chemical potentials of particles and antiparticles should be fulfilled. Thus, for macroscopic scalars (\ref{na_LTE}) -- (\ref{Pp_LTE}) we have:
\begin{eqnarray}
\label{Eqdeg2}
n_\pm =& \displaystyle\frac{2S+1}{2\pi^2}\int\limits_0^\infty \frac{1}{{\rm e}^{\mp \gamma+\frac{\sqrt{m^2_*+p^2}}{\theta}}+1}p^2dp;\\
\label{Eqdeg3}
\Eps_\pm =& \displaystyle\frac{2S+1}{2\pi^2}\int\limits_0^\infty \frac{p^2\sqrt{m^2_*+p^2}dp}{{\rm e}^{\mp \gamma+\frac{\sqrt{m^2_*+p^2}}{\theta}}+1};\\
\label{Eqdeg4}
\mathcal{P}_\pm =& \displaystyle\frac{2S+1}{6\pi^2}\int\limits_0^\infty \frac{1}{{\rm e}^{\mp \gamma+\frac{\sqrt{m^2_*+p^2}}{\theta}}+1}\frac{p^4dp}{\sqrt{m^2_*+p^2}}.
\end{eqnarray}

Supposing the fermion charge conservation law (\ref{III.7b1}), let us write down the following:
\begin{equation}\label{Eqdeg6}
\Delta n=\frac{2S+1}{2\pi^2}\int\limits_0^\infty\left(\frac{p^2}{\mathrm{e}^{ -\gamma+\frac{\varepsilon(p)}{\theta}}+1}-\frac{p^2}{ \mathrm{e}^{\gamma+\frac{\varepsilon(p)}{\theta}}+1}\right) dp,
\end{equation}
where  $\Delta n$ is a constant in the right side of (\ref{III.7b1}), which is proportional to excess of fermions over anti-fermions; $\varepsilon(p)=\sqrt{m^2_*+p^2}$ is a kinetic energy of particles. Further we suppose the condition of strong degeneracy to be fulfilled:
\begin{equation}\label{Eqdeg8}
\gamma\gg 1 \Rightarrow \mathrm{e}^{-\gamma}\ll 1.
\end{equation}
Let us expand the corresponding integrals by this small exponent. In consequence of (\ref{Eqdeg8}) in integrands for antiparticles the member with exponent is larger one as compared to figure one which can be neglected. As follows from (\ref{Eqdeg8}) under conditions of strong degeneracy the antiparticle number is exponentially small in the statistical system even if it is ultrarelativistic.\\[12pt]
\emph{Antiparticles}:\\
Thus, the macroscopic densities for antipaticles under conditions of strong degeneracy are easily found \cite{Ignatev_stfi16} and reduced to classical expressions for corresponding densities with an account of factor ${\rm e}^{-\gamma}$:
\begin{eqnarray}
\label{na_B}
n_{-} &=& \frac{2S+1}{2\pi ^{2} } m_{*}^{3} {\rm e}^{-\gamma} \frac{\mathrm{K}_{2} (\lambda )}{\lambda } ; \\                                                                                                   \label{E_B}
\mathcal{E}_{-} &=& \frac{(2S+1)}{2\pi ^{2} } m_{*}^{4} {\rm e}^{-\gamma}%
\left(\frac{\mathrm{K}_{3} (\lambda )}{\lambda } -\frac{\mathrm{K}_{2} (\lambda )}{\lambda^{2} } \right);  \\                                                                   \label{P_B}
\mathcal{P}_{-} &=& \frac{(2S+1)}{2\pi ^{2} }m_{*}^{4} {\rm e}^{-\gamma} \frac{\mathrm{K}_{2} (\lambda )}{\lambda^{2} } ;\\                                                                                           \label{s_B}
\sigma_{-} &=& \frac{2S+1}{2\pi ^{2} } q m_*^{3} {\rm e}^{-\gamma} \frac{\mathrm{K}_{1}(\lambda )}{\lambda},
\end{eqnarray}
where
\begin{equation} \label{Kn}
\mathrm{K}_{n} (z)=\frac{\sqrt{\pi } z^{n} }{2^{n} \Gamma \left(n+\frac{1}{2} \right)} \int _{0}^{\infty }e^{-z\cosh t} \sinh^{2n} tdt
\end{equation}
are Bessel functions of imaginary argument (see e.g., \cite{Lebed}), $\Gamma (z)$ is a gamma-function.\\[12pt]
\emph{Particles}:\\
To find macroscopic scalars for scalar charged fermions we will use the Sommerfeld method (see e.g. \cite{Landau_Stat}) of approximate calculation of the integral of the following form:
\begin{eqnarray}\label{Land1}
\int\limits_0^\infty \frac{f(\varepsilon)d\varepsilon}{{\rm e}^{-\gamma+\varepsilon/\theta}+1}\approx
\displaystyle\int\limits_0^\mu f(\varepsilon)d\varepsilon+ \nonumber\\
\hspace{5mm}\displaystyle\frac{\pi^2}{6}\theta^2 f'_\varepsilon(\mu)+\frac{7\pi^4}{360}\theta^4 f'''_{\varepsilon\varepsilon\varepsilon}(\mu)+\ldots.
\end{eqnarray}
for small $\theta$. Herewith we should take into account the relativistic coupling (\ref{mu0}) in integrals (\ref{Land1}), that will lead to change of limits of integration. In the case of particle number density $n_+$ function $f(\varepsilon)$ takes form:
\begin{equation}
f(\varepsilon)=\frac{2S+1}{2\pi^2}\varepsilon\sqrt{\varepsilon^2-m^2_*},
\end{equation}
therefore integral (\ref{Eqdeg2}) accurate within $\theta^2$ can be approximated by the following expression:
\begin{equation}\label{n+}
n_+=\frac{m_*^3\psi^3}{3\pi^2}+\theta^2\frac{m_*}{6}\frac{(1+2\psi^2)}{\psi},
\end{equation}
where the following dimensionless function is introduced
\begin{equation}\label{psi}
\psi = \frac{p_f}{m_*}.
\end{equation}

Let us calculate now the Fermi particles energy density (\ref{Eqdeg3}). In this case the function $f(\varepsilon)$ for the integral (\ref{Land1}) takes form:
\begin{equation}
f(\varepsilon)=\frac{2S+1}{2\pi^2}\varepsilon^2\sqrt{\varepsilon^2-m^2_*}.
\end{equation}
Then within  the accuracy of summand, which is quadratic  over temperature, the energy density of fermions is equal to:
\begin{eqnarray}\label{E+}
{\rm {\mathcal E}}_{+}= &\!\!\!{\displaystyle\frac{m_*^4}{8\pi^2}}\left[\psi\sqrt{1+\psi^2}(1+2\psi^2)-\ln(\psi+\sqrt{1+\psi^2}) \right] \nonumber\\
&+\displaystyle\theta^2\frac{m_*^2 }{6} \frac{\sqrt{1 + \psi^2} (1 + 3\psi^2)}{\psi}.
\end{eqnarray}

Let us calculate, finally, the Fermi - particles' pressure (\ref{Eqdeg4}). In this case the function $f(\varepsilon)$ for the integral (\ref{Land1}) is equal to:
\begin{equation}
f(\varepsilon)=\frac{2S+1}{6\pi^2}(\varepsilon^2-m^2_*)^{3/2},
\end{equation}
and within the accuracy of summand, which is quad\-ratic over temperature, the pressure of fermions is equal to:
\begin{eqnarray}\label{P+}
{\rm {\mathcal P}}_{+}={\displaystyle\frac{m_*^4}{24\pi^2}}\left[\psi\sqrt{1+\psi^2}(2\psi^2-3)\right.\nonumber\\
+\displaystyle\left.3\ln(\psi+\sqrt{1+\psi^2}) \right]
+\theta^2\frac{m_*^2}{6} \psi \sqrt{1 + \psi^2}.
\end{eqnarray}
Summands in square brackets(\ref{E+}) and (\ref{P+}) coincide with corresponding expressions for completely degenerated single-component Fermi system.
Is follows from (\ref{n+}) that the condition of temperature corrections' smallness is equialent to the condition (\ref{Eqdeg8}), i.e.,
\begin{equation}\label{theta_to_0}
\gamma\to\infty\Rightarrow \frac{p_f}{\theta}\to \infty \Rightarrow \lambda\psi\gg 1.
\end{equation}
\emph{Photons}:\\

For the establishment of the thermodynamic equilibrium between particles and antiparticles at high energies, it is necessity to take into account photons and other massless particles which can be products of fermion annihilation: the condition (\ref{3.1.24}) is valid when accounting reactions of particles and antiparticles annihilation. Herewith photons, that should be accounted in the model of two-component Fermi-system are created in the annihilation reactions.

Under LTE conditions the energy density and photon pressure are set using the expressions:
\begin{equation}\label{E_g}
\mathcal{E}_{\gamma}= \frac{\pi^2}{15}\theta^4,\quad \mathcal{P}_{\gamma}= \frac{\pi^2}{45}\theta^4.
\end{equation}
\subsection{The Fermion Charge Conservation Law}

In the considered approximation (\ref{Eqdeg8}) the fermion number conservation law (\ref{III.7b1}) takes form:
\begin{eqnarray}\label{Eqdeg30}
a^3 \Delta n = a^3 (n_+ - n_-)=
a^3 \left[\frac{m_*^3\psi^3}{3\pi^2}+\right. \nonumber\\ \left.\frac{m_*^3}{6\lambda^2}\frac{(1+2\psi^2)}{\psi} - \frac{m_*^3 {\rm e}^{-\lambda\sqrt{1+\psi^2}}}{\pi^{2}} \frac{\mathrm{K}_{2}(\lambda )}{\lambda}\right]  = {\rm Const}.
\end{eqnarray}
Let us expand $\Delta n$ over temperature correction smallness:
\begin{equation}\label{dn}
\Delta n = n_0 + \delta n(\theta),
\end{equation}
where
\begin{equation}\label{n0}
n_0 = \frac{m_*^3\psi^3}{3\pi^2}
\end{equation}
is a particle number density of the completely degenerated Fermi system. Calculating the correction $\delta n$, we find:
\begin{equation}
\delta n = \frac{m_*^3}{6\lambda^2}\frac{(1+2\psi^2)}{\psi} - \frac{m_*^3 {\rm e}^{-\lambda\sqrt{1+\psi^2}}}{\pi^{2}} \frac{\mathrm{K}_{2}(\lambda )}{\lambda}.
\end{equation}
In the approximation, which is a zero-order one over $1/\gamma$, due to the conservation law of particle number (\ref{Eqdeg30}) and (\ref{n0}) we find the integral of motion:
\begin{equation}\label{integr}
m_* \psi a  = {\rm Const}\Rightarrow p_fa=\mathrm{Const}.
\end{equation}

\subsection{The Energy-Momentum Conservation Law}

Let us find the total energy density and the pressure of the Fermi system consisting of scalar charged fermions, antifermions and photons. Using the relations (\ref{E_B}), (\ref{P_B}), (\ref{E+}), (\ref{P+}) and (\ref{E_g}), we find the expression for the complete energy density of the statistical system:
\begin{eqnarray}
{\rm {\mathcal E}}_{p}=&\!\!{\displaystyle\frac{m_*^4}{8\pi^2}}\left[\psi\sqrt{1+\psi^2}(1+2\psi^2)\!-\!\ln(\psi+\sqrt{1+\psi^2}) \right]\nonumber\\
&  + \displaystyle\theta^2\frac{m_*^2}{6}\frac{ \sqrt{1 + \psi^2} (1 + 3\psi^2)}{\psi}\nonumber\\
&  + \displaystyle {\rm e}^{-\gamma}\frac{m_{*}^{4}}{\pi ^{2} } %
\left(\frac{\mathrm{K}_{3} (\lambda )}{\lambda } -\frac{\mathrm{K}_{2} (\lambda )}{\lambda^{2} } \right) + \frac{\pi^2}{15}\theta^4
\end{eqnarray}
and its pressure:
\begin{eqnarray}
{\mathcal P}_{p}=&\displaystyle\frac{m_*^4}{24\pi^2}\bigl[\psi\sqrt{1+\psi^2}(2\psi^2-3)+\nonumber\\
&\displaystyle3\ln(\psi+\sqrt{1+\psi^2})\bigr]+\theta^2\frac{m_*^2}{6} \psi \sqrt{1 + \psi^2} \nonumber\\
&+\displaystyle\mathrm{e}^{-\gamma}\frac{m_{*}^{4}}{\pi ^{2}}  \frac{\mathrm{K}_{2} (\lambda )}{\lambda^{2}} + \frac{\pi^2}{45}\theta^4.
\end{eqnarray}
Proceeding to dimensionless functions $\lambda$, $\psi$
\begin{eqnarray}\label{gamma}
\theta=\frac{m_*}{\lambda}, \quad \gamma = \lambda\sqrt{1+\psi^2},
\end{eqnarray}
with an account of the recurrent relations with the Bessel function
\begin{equation}
\mathrm{K}_{3}(\lambda) = \mathrm{K}_{1}(\lambda)+\frac{4\mathrm{K}_{2}(\lambda)}{\lambda}
\end{equation}
we obtain the final expressions for macroscopic charges ${\mathcal E}_{p}, \ {\mathcal P}_{p}$:

\begin{eqnarray}\label{E_p_theta}
{\mathcal E}_{p} =\displaystyle\frac{m_*^4}{8\pi^2}\bigl[\psi\sqrt{1+\psi^2}(1+2\psi^2)-\ln(\psi+\sqrt{1+\psi^2}) \bigr]\nonumber\\
 +\displaystyle \frac{m_*^4}{6\lambda^2}\frac{ \sqrt{1 + \psi^2} (1 + 3\psi^2)}{\psi} + \nonumber\\
 \displaystyle\frac{m_{*}^{4} {\rm e}^{-\lambda\sqrt{1+\psi^2}}}{\pi^{2}} %
\frac{\lambda\mathrm{K}_{1}(\lambda)+3\mathrm{K}_{2}(\lambda)}{\lambda^2} + \frac{\pi^2}{15}\frac{m_*^4}{\lambda^4};\quad\\
\label{P_p_theta}
{\mathcal P}_{p}=\displaystyle \frac{m_*^4}{24\pi^2}\bigl[\psi\sqrt{1+\psi^2}(2\psi^2-3)+\nonumber\\
3\ln(\psi+\sqrt{1+\psi^2}) \bigr] + \frac{m_*^4}{6\lambda^2} \psi \sqrt{1 + \psi^2} + \nonumber\\
\frac{m_{*}^{4} {\rm e}^{-\lambda\sqrt{1+\psi^2}}}{\pi^{2}} \frac{\mathrm{K}_{2}(\lambda)}{\lambda^{2}} + \frac{\pi^2}{45}\frac{m_*^4}{\lambda^4}.\quad
\end{eqnarray}
Similarly, let us find the scalar charge density:
\begin{eqnarray}\label{sigma_theta}
\sigma = \frac{q m_*^3}{2\pi^2}\bigl[\psi\sqrt{1+\psi^2}-
\ln(\psi+\sqrt{1+\psi^2})\bigr] + \nonumber\\
\frac{q m_*^3}{6\lambda^2}\frac{\sqrt{1 + \psi^2}}{\psi} + \frac{q m_*^{3} {\rm e}^{-\lambda\sqrt{1+\psi^2}}}{\pi^{2}}\frac{\mathrm{K}_{1}(\lambda)}{\lambda}.
\end{eqnarray}
Let us expand the obtained macroscopic charges over temperature corrections smallness:
\begin{eqnarray}
{\mathcal E}_{p}={\mathcal E}_0 + \delta{\mathcal E}(\theta);\;
{\mathcal P}_{p}={\mathcal P}_0 + \delta{\mathcal P}(\theta);\;
\sigma=\sigma_0 + \delta\sigma(\theta), \nonumber
\end{eqnarray}
where ${\mathcal E}_0$, ${\mathcal P}_0$ and $\sigma_0$ are first terms in right sides of the relations (\ref{E_p_theta}), (\ref{P_p_theta}) and (\ref{sigma_theta}), corresponding to completely degenerated Fermi system. Thus, for the completely degenerated Fermi system the conservation law of the particles' energy-momentum(\ref{III.7a1}) takes the following form:
\begin{eqnarray}
\dot{\mathcal{E}}_{0}+3\frac{\dot{a}}{a}(\mathcal{E}_{0}+\mathcal{P}_{0})-\sigma_0\dot{\Phi}  = 0 \Rightarrow \nonumber\\
\frac{m_*^4\psi^3\sqrt{1 + \psi^2}}{\pi^{2}}\frac{d}{dt} \ln (m_* \psi a ) = 0
\end{eqnarray}
and is identically fulfilled as a result of equation of motion (\ref{integr}). Obtained integral of motion should be used in temperature corrections. It is important to highlight that the integral of motion (\ref{integr}) is obtained as the integral of the energy conservation law of the completely degenerated Fermi system; fermion number conservation law can be obtained from this law. Thus, the illusory contradiction between the number of independent equations and number of unknown functions under conditions of complete degeneracy, is removed\footnote{local temperature $\theta=0$ is gone}.

Thus, the autonomous system of equations (\ref{III.7a1}), (\ref{Eq_S_t}) and (\ref{Einstein_a}) with an account of definitions (\ref{E_p_theta}) -- (\ref{sigma_theta}) and integral (\ref{integr}) describes a self-consistent cosnological model on basis of relativistic statistical system, consisting of almost degenerated scalar charged fermions, photons and fantom scalar field.

\section{The Asymptotic Behavior of the System in the Ultrarelativistic Limit}
\subsection{The Ultrarelativistic Limit and the Condition of Strong Degeneracy}

Let us now investigate the asymptotic behavior of the system in the ultrarelativistic limit when $p_{F} \gg m$, $\theta \gg m$. Then:

\begin{eqnarray}\label{ultra}
\mu \to p_f, \quad \gamma \to \frac{p_f}{\theta}.
\end{eqnarray}
The condition of fermion strong degeneracy $\gamma \gg 1$ leads to the limitation on functions:
\begin{equation}
\gamma = \frac{p_f}{\theta} \gg 1 \quad \Rightarrow \quad p_f \gg \theta .
\end{equation}
Let us expand the macroscopic scalars of the statistical system (\ref{E_p_theta}) -- (\ref{sigma_theta})
by small parameter of deviation from degeneracy of the relativistic Fermi system:
\begin{equation}\label{xi}
\xi = \left( \frac{\theta}{ p_f } \right)^2\equiv \frac{1}{\psi^2\lambda^2}\ll 1.
\end{equation}
In the approach linear over $\xi$ the macroscopic scalars for the Fermi system (\ref{E_p_theta}) -- (\ref{sigma_theta}) take form:
\begin{eqnarray}\label{plasma_ultra_smpar}
\Delta n &=& \frac{p_f^3}{3 \pi^2} + \xi \frac{p_f^3}{3},
\nonumber\\
\mathcal{E}_{p} &=& \frac{p_f^4}{4 \pi^2} + \xi \frac{p_f^4}{2} + \frac{\pi^2}{15} (p_f^2 \xi)^2,
\nonumber\\
\mathcal{P}_{p} &=& \frac{p_f^4}{12 \pi^2} + \xi \frac{p_f^4}{6} + \frac{\pi^2}{45} (p_f^2 \xi)^2,
\nonumber\\
\sigma &=& 0.
\end{eqnarray}

\noindent
Thus, in this approximation $\sigma = 0$, hence ultrarelativistic almost-degenerated fermions interact with a scalar field non-minimally, i.e. scalar field becomes à free field.

\subsection{The Conservation Laws}
\begin{enumerate}
\item
From the fermion number conservation law (\ref{III.7b1}) it follows:
\begin{eqnarray}
 a^3 p_f^3 \left( 1 + \pi^2\xi \right) = {\rm Const} \Rightarrow\nonumber\\
 p_f = \frac{p_0}{a} \left(1 + \pi^2\xi \right)^{-1/3}.
\end{eqnarray}
Thus, we finally obtain the following relation for the Fermi momentum:
\begin{eqnarray}\label{conservI_ultra_small}
p_f = \frac{p_0}{a} \left( 1 - \frac{\pi^2}{3} \xi \right)
\end{eqnarray}
\item
The energy-momentum conservation law can be re-written in the next form:
\begin{eqnarray}
\frac{p_f^4}{\pi^2} \left[ \frac{\dot{a}}{a} + \frac{\dot{p}_F}{p_f} + 2 \pi^2 \xi \left( \frac{\dot{a}}{a} + \frac{\dot{p}_F}{p_f} + \frac{\dot{\xi}}{4\xi} \right) \right] \nonumber\\
+ \frac{4 \pi^2}{15} (p_f^2 \xi)^2 \left[ \frac{\dot{a}}{a} + \frac{\dot{p}_F}{p_f} + \frac{\dot{\xi}}{2\xi} \right] = 0.\nonumber
\end{eqnarray}
Expanding the obtained expression by smallness of the parameter $\xi$, let us finally get:
\begin{eqnarray}\label{conservII_ultra_small}
\frac{p_f^4}{\pi^2}\biggl[\frac{d}{dt} \ln \left( a p_f \right) + 2 \pi^2 \xi \frac{d}{dt} \ln \left( a p_f\xi^{1/4} \right) \biggr] \nonumber\\
+ \frac{4 \pi^2}{15} (p_f^2 \xi)^2 \frac{d}{dt} \ln \left( a p_f \xi^{1/2} \right) = 0
\end{eqnarray}
\end{enumerate}
In the case of complete degeneracy $\xi = 0$ the expression (\ref{conservI_ultra_small}) leads to the cited above integral (\ref{integr}).
In the approximation, linear by smallness of $\xi$, the expression (\ref{conservII_ultra_small}) can be re-written in the following way:
\begin{eqnarray}\label{tmp_conservII_ultra_small}
\frac{{\rm C_1}}{a^4} \left[ -\frac{\pi^2}{3} \dot{\xi} \left(1\! +\! \frac{\pi^2}{3} \xi \right)\! +\! 2\pi^2 \xi \dot{\xi} \left( \frac{1}{4 \xi} - \frac{\pi^4}{9} \xi - \frac{\pi^2}{3} \right) \right]\\
 + \frac{{\rm C_2}}{a^4} \left(\xi^2 - \frac{4\pi^2}{3} \xi^3 \right) \dot{\xi} \left( \frac{1}{2 \xi} - \frac{\pi^4}{9} \xi - \frac{\pi^2}{3} \right) = 0.\nonumber
\end{eqnarray}
Since the degeneracy parameter $\xi$ is a free parameter and does not depend on other functions included in the relation (\ref{tmp_conservII_ultra_small}), the last one can be fulfilled only in the following case:
\begin{eqnarray}\label{xi=const}
\dot{\xi} = 0 \quad \Rightarrow \quad \xi = {\rm Const},
\end{eqnarray}
where from we obtain the evolution law of the Fermi momentum and temperature in the ultrarelativistic limit:
\begin{eqnarray}\label{pf_T_ultra_small}
p_f = \frac{p_0}{a} \left( 1 - \frac{\pi^2}{3} \xi \right) = \frac{\bar{p}_0}{a}, \quad \theta = \frac{\theta_0}{a}.
\end{eqnarray}

\subsection{Accounting of the Non-Relativistic Corrections}
In the ultrarelativistic limit for almost degenerated fermions the following relations should be fulfilled
\begin{equation}
p_f \gg m_* \quad \Rightarrow\quad \frac{1}{\psi} = \frac{m_*}{p_f} \to 0.
\nonumber\\
\end{equation}
For almost degenerated fermions $\gamma \gg 1$ we have:
\begin{eqnarray} %
\gamma=\frac{\mu}{\theta}=\frac{\sqrt{m_*^2+p_f^2}}{\theta}\equiv \frac{p_f\sqrt{1+(1/\psi)^2}}{\theta}. \nonumber\\
\end{eqnarray}
In the ultrarelativistic limit($1/\psi \to 0$) it is:
\begin{equation}
\gamma \to \bar{\gamma}\equiv \frac{p_f}{\theta}.
\nonumber\\
\end{equation}
Let us expand the macroscopic scalars by smallness of non-relativistic corrections:
\begin{eqnarray}\label{plasma_ultra_rel}
\Delta n &=& \frac{m_*^3 \psi^3}{3 \pi^2} \left( 1 + \pi^2 \xi \right),
\nonumber\\
\mathcal{E}_{p} &=& \frac{m_*^4 \psi^4}{4 \pi^2} \left( 1 + \frac{1}{\psi^2} + 2 \pi^2 \xi \right) + \frac{\pi^2}{15} (m_*^2 \psi^2 \xi)^2,
\nonumber\\
\mathcal{P}_{p} &=& \frac{m_*^4 \psi^4}{12 \pi^2} \left( 1 - \frac{1}{\psi^2} + 2 \pi^2 \xi \right) + \frac{\pi^2}{45} (m_*^2 \psi^2 \xi)^2,
\nonumber\\
\sigma &=& \frac{q m_*^3 \psi^4}{2 \pi^2} \frac{1}{\psi^2}.
\end{eqnarray}
Fraction of antiparticles in the approximation $\psi \gg 1$, $\lambda \to 0$ is equal to:
\begin{equation}\label{dn/n}
\frac{n_{-}}{n_{+}} = 6 {\rm e}^{-1/\sqrt{\xi}} \xi^{3/2}\ll 1
\end{equation}
and depends only on $\xi$, therefore as a result of (\ref{xi=const}) the relative concentration of antiparticles remains constant.

Summarizing the results of this chapter, let us notice the following properties of the \emph{almost degenerated} statistical system's behavior:
\begin{enumerate}
\item In the course of the cosmological evolution, the Fermi momentum changes by the following law (\ref{integr}):
\begin{equation}\label{ap}
ap_f=\mathrm{Const}.
\end{equation}
\item In the course of the cosmological evolution, the temperature of almost degenerated Fermi system changes by the following law:
\begin{equation}\label{atheta}
a\theta=\mathrm{Const},
\end{equation}
while its degree of degeneracy remains constant.
\item In the course of the cosmological evolution, the degree of polarization of ultrarelativistic almost degenerated Fermi system (\ref{dn/n}) remains small and constant.
\end{enumerate}

Let us further notice the following circumstance. The degree of relativity of almost degenerated statistical system, is apparently defined by the condition:
\begin{equation}\label{ultra}
p^2_f+\theta^2\gg m_*^2\Rightarrow \psi^2(1+\xi^2)\gg 1,
\end{equation}
i.e. under conditions of strong degeneracy ($\xi\to 0$) is practically reduced to the single condition:
\begin{equation}\label{psi>>1}
\psi\gg1\Rightarrow p_f\gg m_*.
\end{equation}
Therefore the statistical system of almost degenerated fermions can easily be violated if only the scalar field's potential does not fall proportionally to $1/a$, as it does in the case of conformally invariant scalar field \cite{Ignatev_16_1}, \cite{Ignatev_16_2}. As can be seen below the relativity condition actually is violated sufficiently fast as a result of scalar field's potential growth. Along with that the statistical system becomes non-relativistic and its degree of degeneracy grows.

\section{The Mathematical Model of the Cosmological Evolution of the Completely Degenerated Statistical System with a Fantom Scalar Interaction}
\subsection{The Complete System of Equations}
Thus, under conditions of complete degeneracy, the macroscopic scalars of single-component statistical system are equal to:
\begin{equation}\label{3}
n=\displaystyle\frac{p_f^3}{3\pi^2}\equiv \frac{m^3_*\psi^3}{3\pi^2};
\end{equation}
\begin{eqnarray}
\label{3a}
{\mathcal E}_{p} =&\displaystyle\frac{m_*^4}{8\pi^2}
\bigl[\psi\sqrt{1+\psi^2}(1+2\psi^2)\hskip 1.7cm\\
&-\displaystyle\ln(\psi+\sqrt{1+\psi^2})\bigr];\\
\label{3b}
{\mathcal P}_{p} =&\displaystyle\frac{m_*^4}{24\pi^2}
\bigl[\psi\sqrt{1+\psi^2}(2\psi^2-3)\hskip 1.7cm\\
&+\displaystyle 3\ln(\psi+\sqrt{1+\psi^2})\bigr];\\
\label{3c}
\sigma=&\displaystyle q\frac{m_*^3}{2\pi^2}\bigl[\psi\sqrt{1+\psi^2}
-\ln(\psi+\sqrt{1+\psi^2})\bigr],
\end{eqnarray}
herewith it is:
\begin{eqnarray}\label{p_f(a)}
\psi=\frac{p_f}{m_*}\quad m_*=|m_0+q\Phi|;\nonumber\\
p_f=\frac{p_0}{a},\; p_0=\left.p_f\right|_{a=1}.
\end{eqnarray}
Thus, under conditions of complete degeneracy all the macroscopic scalars are explicitly defined through elementary scalar functions $\Phi(t)$ è $a(t)$.

In the ultrarelativistic limit the values of these scalars are defined by formulas (\ref{plasma_ultra_smpar}), where one should put $\xi=0$, in the ultrarelativstic limit
\begin{equation}\label{nonrel}
m_* \gg p_{F} \Rightarrow\quad \psi \ll 1
\end{equation}
formulas (\ref{3a}) -- (\ref{3c}) have the following asymptotics:
\begin{eqnarray}\label{plasma_nonrel}
\mathcal{E}_{p}=\frac{m_* p_f^3}{3 \pi^2}=\frac{|m_0+q\Phi|^3p^3_0}{3 \pi^2 a^3};\nonumber\\
\mathcal{P}_{p} =0;\quad \sigma= \frac{q p_f^3}{3 \pi^2}=\frac{q p_0^3}{\pi^2 a^3}.
\end{eqnarray}

In turn, scalar functions $\Phi(t)$ è $a(t)$ are defined by the system of two ordinary differential equations, (\ref{Eq_S_t}) and  (\ref{Einstein_a}):
\begin{eqnarray}\label{Eq_Phi(t)}
\ddot{\Phi}+3\frac{\dot{a}}{a}\dot{\Phi}-m^2_s\Phi= 4\pi\sigma;\\
\label{Eq_a(t)}
3\frac{\dot{a}^2}{a^2}=8\pi({\mathcal E}_p+\mathcal{E}_s),
\end{eqnarray}
where $\mathcal{E}_p$ is the Fermi system's energy density (\ref{3a}), and $\mathcal{E}_s$ is the fantom scalar field's energy density (\ref{Es}):
\begin{equation}\label{E_s}
{\mathcal E}_s=\frac{1}{8\pi}(-\dot\Phi^2+m_s^2\Phi^2).
\end{equation}
Introducing the following variables:
\begin{equation}\label{Z,lambda}
Z(t)=\dot{\Phi}; \quad \Lambda(t)=\ln a\Leftrightarrow a=\mathrm{e}^\Lambda,
\end{equation}
let us reduce equations (\ref{Eq_Phi(t)}) and (\ref{E_s}) to the following form:
\begin{eqnarray}
\label{dZ}
\dot{Z}&=&-3\dot{\Lambda}Z+m^2\Phi+4\pi\sigma(\Lambda,\Phi);\\
\label{dL}
\dot{\Lambda}&=&8\pi\Eps_p(\Lambda,\Phi)-Z^2+m_s^2\Phi^2.
\end{eqnarray}
In order to get from this system a normal system of differential equations convenient for numerical simulation, it is necessary to substitute the expression for $\dot\Lambda$ from equation (\ref{dL}) into the left part of equation (\ref{dZ}).

\subsection{Cauchy Problem}\label{remark_t0}
Let us state now Cauchy problem for the system (\ref{Z,lambda}), (\ref{dZ}) and (\ref{dL}). Using admissible transformations, let us choose a scale factor and time in such a way that:
\begin{eqnarray}\label{Coshe}
\Lambda(0)=0; \; p_f(0)=p_0;\;\Phi(0)=\Phi_0;\; Z(0)=Z_0.
\end{eqnarray}
Since $a(0)=1$, the chosen instant of time does not coincide with a time singularity. In order to reformulate the final results in terms of cosmological time  $t$, which is calculated starting from the cosmological singularity, two algorithms can be suggested. Let us describe the first one. Numerical integration needs to be extended to the negative interval of time until we get
$a(-t_0)=0$. Then all calculated functions are required to be transformed in the following way: $\Psi(t)\to \psi(t+t_0)$. Let us now explain the second algorithm. Since we consider here a class of cosmological models with ultrarelativistic start where on the early stages of the cosmological evolution the contribution of the scalar field in the energy-balance is vanishingly small, we can use the well-known relation for the ultrarelativstic Universe (see e.g. \cite{Land_Field}):
\begin{equation}\label{ultra_e}
\Eps=\frac{3}{32\pi \tau^2},
\end{equation}
where $\tau$ is a truly cosmological time. Then we can realize the next re-calibration of the time variable $t$. Let $\Eps(t_0)$
is a value of total energy density on the early stages, obtained as a result of numerical simulation. Then it should be:
\begin{equation}\label{t->tau}
\Eps(t_0)=\frac{3}{32\pi \tau_0^2}\Rightarrow \tau_0=\sqrt{\frac{3}{32\pi\Eps(t_0)}}.
\end{equation}
Let us notice that such a re-calibration of time to cosmological time was not carried out in the cited above papers \cite{Ignatev_Agathonov_2015} -- \cite{Ignatev_AAS}. The time variable $t$ on plots of the cosmological evolution in these papers is a relative value. To reduce these plots to cosmological time it is required to carry out the following shift of the time scale:
\begin{equation}\label{t+tau}
t\to t+(\tau_0-t_0).
\end{equation}
In this article we will carry out the mentioned re-calibration of the time scale on all plots in such a way that time $t$ is everywhere counted from the cosmological singularity. Further, for the sake of simplicity we will everywhere assume:
\begin{equation}\label{Z0}
Z_0=0.
\end{equation}
\subsection{The Analysis of the Mathematical Model at Small Times $m_s t\ll 1$}
As it is cited above we consider the cosmological models with ultrarelativistic start where inflationary mode of expansion can be realized at earlier stages. In the most extreme, ultrarelativistic limit $\psi\to\infty$ formulas (\ref{plasma_ultra_smpar}) for the macroscopic scalars take form:
\begin{eqnarray}\label{plasma_ultra}
\mathcal{E}_{p} = \frac{p_F^4}{4 \pi^2} ;\quad \mathcal{P}_{p}=\frac{p_F^4}{12 \pi^2};\quad \sigma= 0.
\end{eqnarray}
Next, the following relation corresponds to early stages:
\begin{equation}\label{t-to-0}
m_st\ll 1.
\end{equation}
In this case the field equation takes the following form:
\begin{equation}\label{Eq_Phi_t->0}
\ddot{\Phi}+3\frac{\dot{a}}{a}\dot{\Phi}=0
\end{equation}
and has its first integral:
\begin{equation}\label{dot{Phi}}
a^3\dot{\Phi}=\mathrm{C}_1,
\end{equation}
where $C_1$ is an arbitrary constant. In the same approximation the energy density of the scalar field (\ref{E_s}) is equal to:
\begin{equation}\label{E_s-t->0}
{\mathcal E}_s=-\frac{\mathrm{C}^2_1}{8\pi a^6}
\end{equation}
and the Einstein equation (\ref{Eq_a(t)}) with an account of (\ref{plasma_ultra}) and (\ref{p_f(a)}) takes the form:
\begin{equation}\label{Eq_Einst_ultra}
3\frac{\dot{a}^2}{a^2}=-\frac{\mathrm{C}^2_1}{a^6}+\frac{2p^4_0}{a^4}.
\end{equation}
In the case of scalar field the main contribution near singularity is made by the scalar field having in this case the ultimate stiff equation of state $\Eps_s=\mathcal{P}_s$ and leading to the evolution law of the scale factor $a\sim t^{1/3}$ \cite{Ignatev_stfi15a}.
For the fantom scalar field  $\mathrm{C}_1\not=0$ there is no such possibility exactly because of prevalence of negative scalar field's contribution in this case near singularity. Therefore for the fantom field as opposed to classical one, there remains the single possibility near singularity:
\begin{equation}\label{Phi=C}
\Phi=\Phi_0=\mathrm{Const}, \quad (m_st\ll1).
\end{equation}
In this case the Einstein equation takes the form:
\begin{equation}\label{Eq_Einst_Ultra_1}
3\frac{\dot{a}^2}{a^2}=m^2_s\Phi^2_0+\frac{2p^4_0}{a^4}.
\end{equation}
and has it solution
\begin{equation}\label{a(t)-t->0}
a=\sqrt{\frac{\sqrt{2}p^2_0}{m_s\Phi_0}}\sqrt{\sinh\left(\frac{2m_s\Phi_0}{\sqrt{3}}t\right)}.
\end{equation}
Equation (\ref{a(t)-t->0}) belongs to the class of solutions obtained in \cite{Ignatev_13b} and describing a smooth transition from the ultrarelativstic stage $a\sim t^{1/2}$ to the inflation stage $a\sim e^{v_0t}$. As is seen from the solution (\ref{Eq_Einst_Ultra_1}), transition to the inflation stage happens at times of the order:
\begin{equation}\label{t_inf}
t\gtrsim t_{inf}=\frac{\sqrt{3}}{4m_s\Phi_0}.
\end{equation}
From the other hand, (\ref{t-to-0}) is a condition of the considered approximation's validity. Hence we can make a conclusion:
at $t_{inf}m_s<1$ inflation can develop on the considered time interval, in the opposite case the entire early stage of the Universe is ultrarelativistic. Thus, the condition of early inflation appearance (at times less than $m^{-1}_s$) in the case of system with a fantom field is a smallness of the initial scalar potential\footnote{in spite of apparent paradoxicality of this conclusion at first glance}:
\begin{equation}\label{ultra_sol}
\Phi_0\ll\frac{\sqrt{3}}{4}.
\end{equation}
\subsection{The Massless Fantom Field}

It is obvious that the case of massless fantom field related to the case of small time $m_s t\ll 1$. It is remarkable that the scalar field's mass $m_s$ disappears from the condition (\ref{ultra_sol}). Particularly, for a massless scalar field, assuming in (\ref{a(t)-t->0}) $m_s\to0$ we find:
\begin{equation}\label{a(t)-t->0,ms=0}
a(t)=\left(\frac{8}{3}\right)^{1/4}\sqrt{t}
\end{equation}
the ultrarelativstic asymptotics ($\Omega=-1$). As is seen from the previous analysis, the ultrarelativstic asymptotics will be valid for the massless fantom field while fermions stay ultrarelativstic.
In the non-relativistic limit $\psi\to0$ formulas (\ref{plasma_ultra_smpar}) for the macroscopic scalar take form:
\begin{eqnarray}\label{E_nonrel}
\mathcal{E}_{p} = \frac{m_* p_f^3}{3 \pi^2} =\frac{|m_0+q\Phi|^3p^3_0}{3 \pi^2 a^3};\\
\label{sigma_nonrel}
\mathcal{P}_{p} = 0; \quad \sigma = \frac{q p_f^3}{3 \pi^2}=\frac{q p_0^3}{3 \pi^2 a^3}.
\end{eqnarray}
In this case we again obtain the equation of the massless scalar field without a source (\ref{Eq_Phi_t->0}) and hence for mentioned above reasons its unique feasible solution (\ref{Phi=C}) $\Phi=\Phi_0$. Thus, only non-relativistic fermions contribute to the energy. As a result of this, we find the solution
\begin{equation}\label{a=t2/3}
a(t)=\sqrt{\frac{2m^{(0)}_*\ \!\!^3 p^3_0 }{\pi}}t^{2/3}\quad \bigl(m^{(0)}_*=|m_0+q\Phi_0|\bigr),
\end{equation}
corresponding to the non-relativistic mode of expansion ($\Omega=-1/2$). Thus in the case of massless scalar field the cosmological evolution can start from the ultrartelativistic mode and end with a non-relativistic one. Numerical simulation is required for investigation of the behavior of this model with massless fantom field at intermediate stages.

\subsection{The Properties of the Cosmological Model with a Massive Fantom Field at Great Times $m_s t\gg 1$}
Let now $m_s\not\equiv 0$. Let us consider the behavior of the cosmological model at great times
\begin{equation}\label{mst>>1}
m_s t\gg 1,
\end{equation}
supposing that at these times the Fermi system becomes non-relativistic. However, since at these times scale factor becomes a great value then according to (\ref{sigma_nonrel}) scalar charge's density $\sigma$  again becomes a small value and can be casted out. However in the field equation (\ref{Eq_Phi(t)}) it is not possible to cast out the first and the second derivatives compared with a massive terms as in such a case we would get
$m^2\Phi=0$.
\begin{equation}\label{Eq_Phi(t)1}
\ddot{\Phi}+3\dot{\Lambda}\dot{\Phi}-m^2_s\Phi= 0;
\end{equation}
The contribution of particles in the energy density can be casted out In the Einstein equation (\ref{Eq_a(t)}) for the same reason:
\begin{equation}\label{Eq_Einst_t->8}
3\dot{\Lambda}^2=-\dot{\Phi}^2+m^2_s\Phi^2.
\end{equation}
This system has the following solution, asymptotic at $m_st\to\infty$:
\begin{eqnarray}\label{Phi0-L0}
\Phi^{(0)}(t)=\frac{m_s t}{\sqrt{3}}+\Phi_0,\; (m_s\not\equiv 0);\nonumber\\
\Lambda^{(0)}(t)=\frac{m^2_s t^2}{6}+\Phi_0\frac{m_s t}{\sqrt{3}},
\end{eqnarray}
where $\Phi_0=\mathrm{Const}$ is an arbitrary constant. Let us notice that the solution (\ref{Phi0-L0}) turns the field equation (\ref{Eq_Phi(t)1}) into the identity and leads to relative error of the order of $(m_s t)^{-2}$ at substitution into the Einstein equation (\ref{Eq_Einst_t->8}).

Let us calculate the invariant cosmological acceleration relative to the solution found:
\begin{eqnarray}\label{Omega_t->8}
\Omega = \frac{\ddot{a}^{(0)} a^{(0)}}{\dot{a}^{(0)\,2}} = 1 + \frac{\ddot{\Lambda}^{(0)}}{\dot{\Lambda}^{(0)\,2}} =
1 + \frac{1}{\bigl(\Phi^{(0)}(t)\bigr)^2}.
\end{eqnarray}
Thus, at great times ($m_s t \to \infty$) the systems comes to inflation expansion mode ($\Omega \to 1$). Let us notice that the stability of the obtained asymptotic solution at great times can be proved.

It is not too difficult to see that the effective mass of fermions grows as a linear function of time:
\begin{equation}\label{m_*->8}
m_* \equiv |m + q\Phi| \approx  q\frac{m_s t}{\sqrt{3}},
\end{equation}
so that fermions soon become non-relativistic ($\Psi\to0$). Let us make the following important notices in this context:
\begin{enumerate}
\item A cold completely degenerated Fermi system with very big effective masses of scalar charged fermions can be a good model of dark matter.
\item At the certain stage of the cosmological evolution gravitational instabilities in non-relativistic matter can lead to appearance of isolated ranges with dark matter.
\item The standard Cooper's mechanisms in Fermi systems with attraction of particles can lead to creation of bosons from fermion pairs and hence to 	superfluidity of dark matter ranges.
\item At growth of the effective masses of fermions above the Planck value, i.e. according to (\ref{m_*->8}) at
\begin{equation}
m_* \gg m_\mathrm{Pl}\Leftrightarrow q\frac{m_s t}{\sqrt{3}}\gg 1,
\end{equation}
massive fermions basically can generate stable primary black holes. With an account of Hawking's theorems these will most likely be the cases with superfluid quasi-bosons and zero spin.
\end{enumerate}

\section{Numerical Simulation of the Cosmological Evolution}
Now let us proceed to the results of numerical integration of the system of two differential equations (\ref{Eq_Phi(t)}), (\ref{Eq_a(t)} with initial conditions (\ref{Coshe}) and the following definitions of energy and scalar charge densities (\ref{3a}), (\ref{3c}) and (\ref{E_s}). We should keep in mind the notice from chapter  \ref{remark_t0} about re-calibration of the time instant $t_0$. On the plots below this re-calibration is done so time is counted from the point of cosmological singularity on the Planck time scale. All values on plots are measured also in the Planck units. Further we will cal a relation of total density of system to its total energy density a
\emph{barotropic coefficient}, $\kappa$:
\begin{equation}\label{kappa}
\kappa=\frac{\mathcal{P}}{\mathcal{E}}=\frac{\mathcal{P}_p+\mathcal{P}_s}{\mathcal{E}_p+\mathcal{E}_s}.
\end{equation}
As is known (see e.g. \cite{Ignatev_AAS}), the value of the invariant cosmological acceleration $\Omega$ is connected with a barotropic coefficient $\kappa$ by the following relation:
\begin{equation}\label{Omega(kappa)}
\Omega=-\frac{1}{2}(1+3\kappa),
\end{equation}
so that the ultrarelativistic equation of state $\kappa=1/3$ is corresponded by value $\Omega=-1$; non-relativistic equation of state $\kappa=0$ is corresponded by $\Omega=-1/2$; value $\kappa=-1/3$ is corresponded by $\Omega=0$; the inflation (vacuum) equation of state $\kappa=-1$ is corresponded by $\Omega=+1$, values $\kappa<-1$ are corresponded by hyperinflation $\Omega>1$.

Let us also introduce a dimensionless parameter which will be important further on
\begin{equation}\label{eta_s}
\eta_s=\frac{\mathcal{E}_s}{\mathcal{E}_p},
\end{equation}
and which is a relation of the scalar field's energy density to fermions' energy density.

Let us introduce the following characteristic time instants which are important for understanding the mechanism of the cosmological evolution of fermion system with a fantom scalar field:
\begin{enumerate}
\item The Compton time instant relative to the scalar field's quanta mass:
\begin{equation}\label{t_compt}
t_s=\frac{1}{m_s}, \quad m_s\not=0.
\end{equation}
\item For the massless scalar field with a source we can find an identical time instant relative to the scalar field's effective mass.
In works \cite{Ignatev_16_1}, \cite{Ignatev_16_2} (and also in\\ \cite{Ignatev_stfi16}) it has been shown using analytical methods that scalar charge's density $\sigma$ at certain conditions can play a role of a massive term in the equation of the scalar field. Herewith the effective mass of the scalar field can be introduced even in the case $m_s=0$:
\begin{equation}
\label{m_s_*}
m_s^*=\sqrt{\frac{4\pi\sigma}{\Phi}},
\end{equation}
Therefore even at zero mass of scalar field quanta at presence of source function $\sigma$ the scalar field behave itself in many cases as massive scalar field the only difference being the effective mass (\ref{m_s_*}) dependent on cosmological time $m_s^*(t)$.
Let us introduce the Compton time instant correspondingly to the effective mass:
\begin{equation}
t^*_s=\frac{1}{m^*_s}, \quad m_s=0.
\end{equation}
\item Time instant $t_\eta$, when energy densities of fermions and scalar field become equal:
\begin{equation}\label{t_eta}
\mathcal{E}_s(t_{\eta})=\mathcal{E}_p(t_{\eta})\Leftrightarrow \eta_s(t_{\eta})=1.
\end{equation}
Let us notice that there could be two such time instants.
\item Time instant $t_r$ of fermions' transition from the ultrarelativistic state to non-relativistic:
\begin{equation}\label{t_r}
\psi(t_r)=1.
\end{equation}
\end{enumerate}

Further, for plotting simulated results we will use author's function:
\begin{equation}\label{eqn:Lig}
\mathrm{Lig}(x)\equiv \mathrm{sgn}(x)\lg(1+|x|),
\end{equation}
which is required to represent results on a logarithmic scale when displayed function %
can change its sign. Introduced function $\mathrm{Lig}(x)$ is convenient due to its %
behavior, since for small argument values the function coincides with argument value, %
and for large argument values the function coincides with its decimal logarithm taken %
with the sign of the argument:
\[\mathrm{Lig}(x) \approx \left\{%
\begin{array}{ll}%
x, &  |x|\to 0;\\
\mathrm{sgn}(x)\ln|x|, & |x|\to\infty.
\end{array}
\right.
\]
It is easy to show that
\[\frac{d\mathrm{Lig}(x)}{dx}\geq 0,\]
and that provides a continuously differentiability of the function $\mathrm{Lig}(x)$, %
and this, in turn, provides bijective mapping of (\ref{eqn:Lig}). Inverse transformation %
can be done by expression:
\[%
x=\left\{\begin{array}{ll}
10^{\mathrm{Lig}(x)}-1, & x\geq 0;\\
1-10^{-\mathrm{Lig}(x)}, & x<0.
\end{array}
\right. \]
\subsection{The Case of Massive Fantom Scalar Field with Minimal Interaction ($\sigma = 0$)}\label{sigma=0}
Numerical integration in this case reveals three characteristic stages of the cosmological evolution:\footnote{Main regularities of the evolution at early and later stages confirm cited above results of the analytic research.}
\begin{enumerate}
\item $t \lesssim t_s$: \textit{fermions' prevalence}\\
\noindent
\emph{Characteristic properties}: Small values of scalar field potential and its derivative - influence of field on system evolution is minor.
\item $t_s \lesssim t \lesssim t_{\eta}$: \textit{concurrence of fermions and the scalar field}\\
\noindent
\emph{Characteristic properties}: Sharp growth of the scalar field's potential and its derivative in the time instant $t_{\eta}$  (\ref{t_eta}) (Fig.  \ref{Fig3}). In this time instant the next functions reach extremum: $\dot{\Phi}$(maximum); barotropic coefficient, $\kappa$,(minimum); invariant cosmological \\ acceleration, $\Omega$, (maximum); density of total \\energy, $\mathcal{E}_{pl}+\mathcal{E}_{s}$, (minimum). At peak time, the barotropic coefficient becomes less than -1 -- fantom equation of state.\\
\item $t_{\eta}\lesssim t < +\infty)$: \textit{prevalence of the scalar field}\\
\noindent
\emph{Characteristic properties}: The derivative of the scalar field's potential tends to the constant $m_s/\sqrt{3}$; the potential grows linearly; scalar field's energy density significantly exceeds fermions' energy and defines further evolution of the system; barotropic coefficient $\kappa \to -1$; invariant cosmological acceleration $\Omega \to 1$ -- a runout to inflation.
\end{enumerate}
Below the plots of numerical simulation of the system with the following parameters are shown:\\[12pt]
$p_0 = 100$, $m_0=0.001$, $\Phi(0)= 5\cdot 10^{-7}$. Everywhere on plots Fig. 1 -- Fig. 10 it is: heavy line is $m_s = 10^{-2}$, thin line is $m_s = 10^{-4}$, normal dotted line is $m_s = 10^{-6}$, fine dotted line is $m_s = 10^{-8}$. Also, everywhere on these and all other plots along the X-axis there are the values of decimal logarithm  starting from the point of cosmological singularity in the Planck scale, $\log_{10}t$.
\begin{center}\refstepcounter{figure}
\includegraphics[width=120mm]{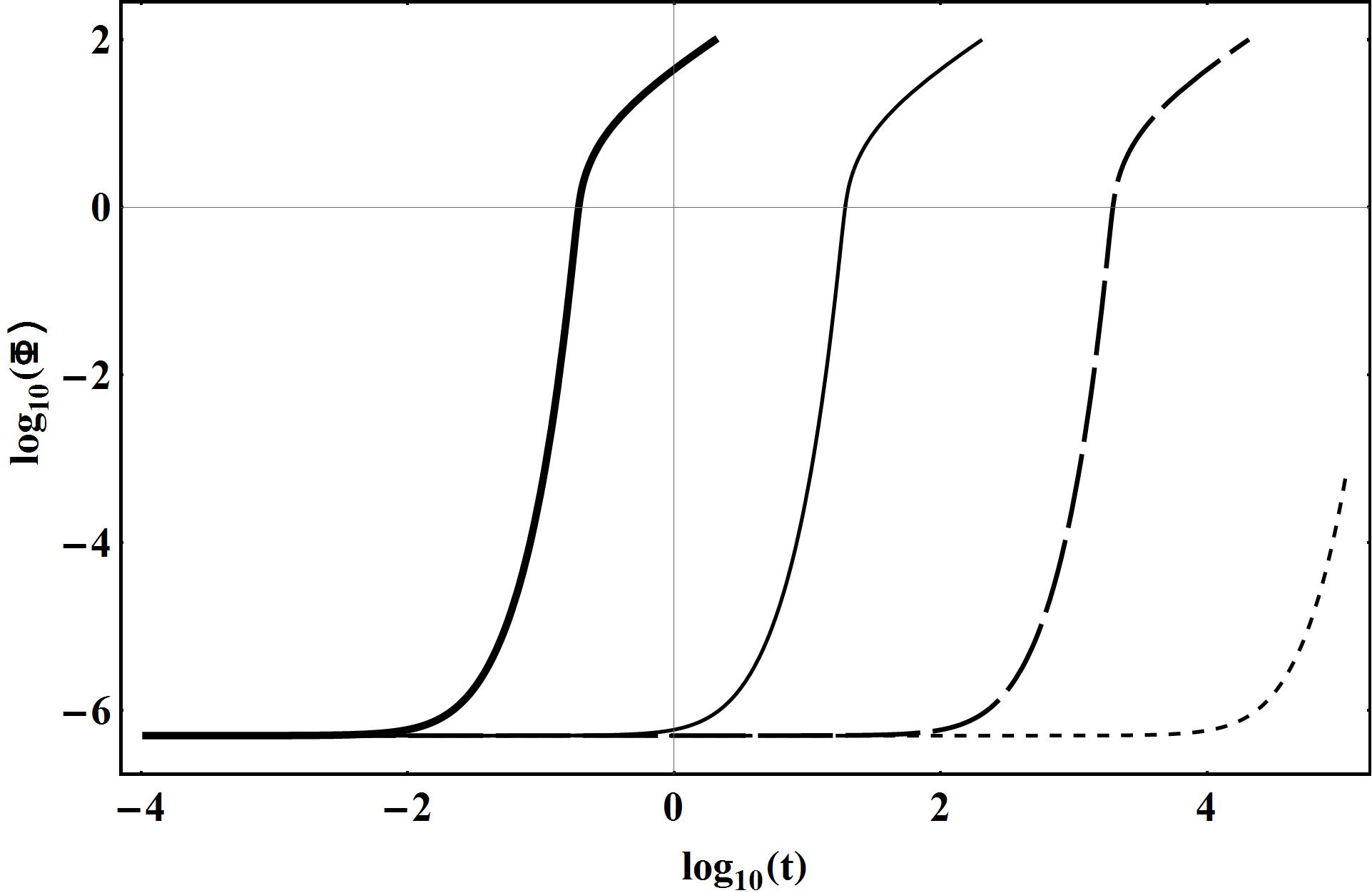}
\end{center}
{\small \textbf{Fig.  \thefigure}. \label{Fig1}The evolution of potential $\Phi$. Along the Y-axis there are plotted the values of common logarithm of the scalar potential, $\log_{10}\Phi$. }\vspace{12pt}

\begin{center}\refstepcounter{figure}
\includegraphics[width=120mm]{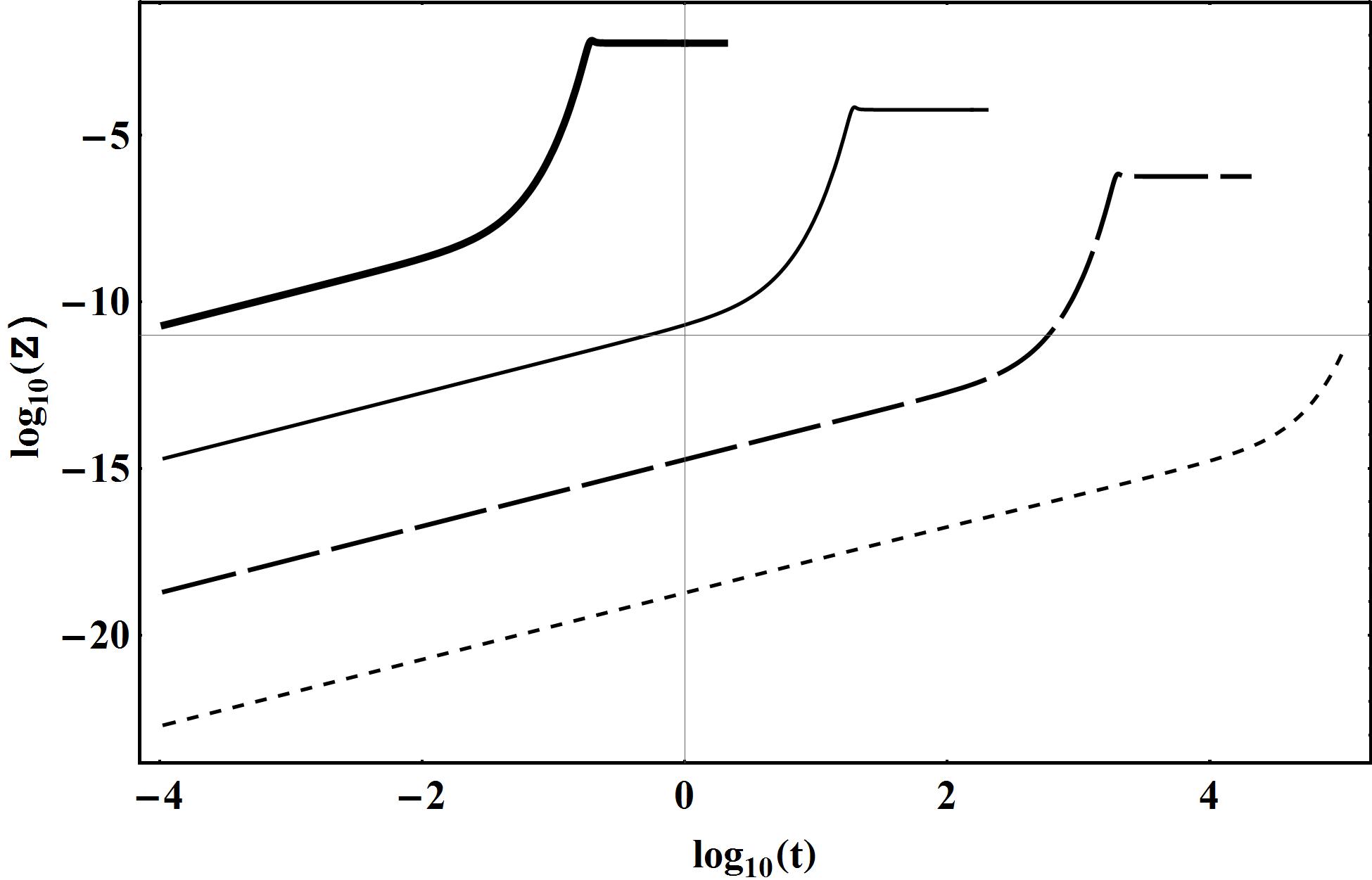}\label{Fig2}
\end{center}
{\small \textbf{Fig.  \thefigure}.\label{Fig2} The evolution of the potential's derivative $Z=\dot{\Phi}$. Along Y-axis there are plotted the values of coomon logarithm of the scalar potential, $\log_{10}\Phi$.\vspace{12pt} }

\begin{center}\refstepcounter{figure}
\includegraphics[width=120mm]{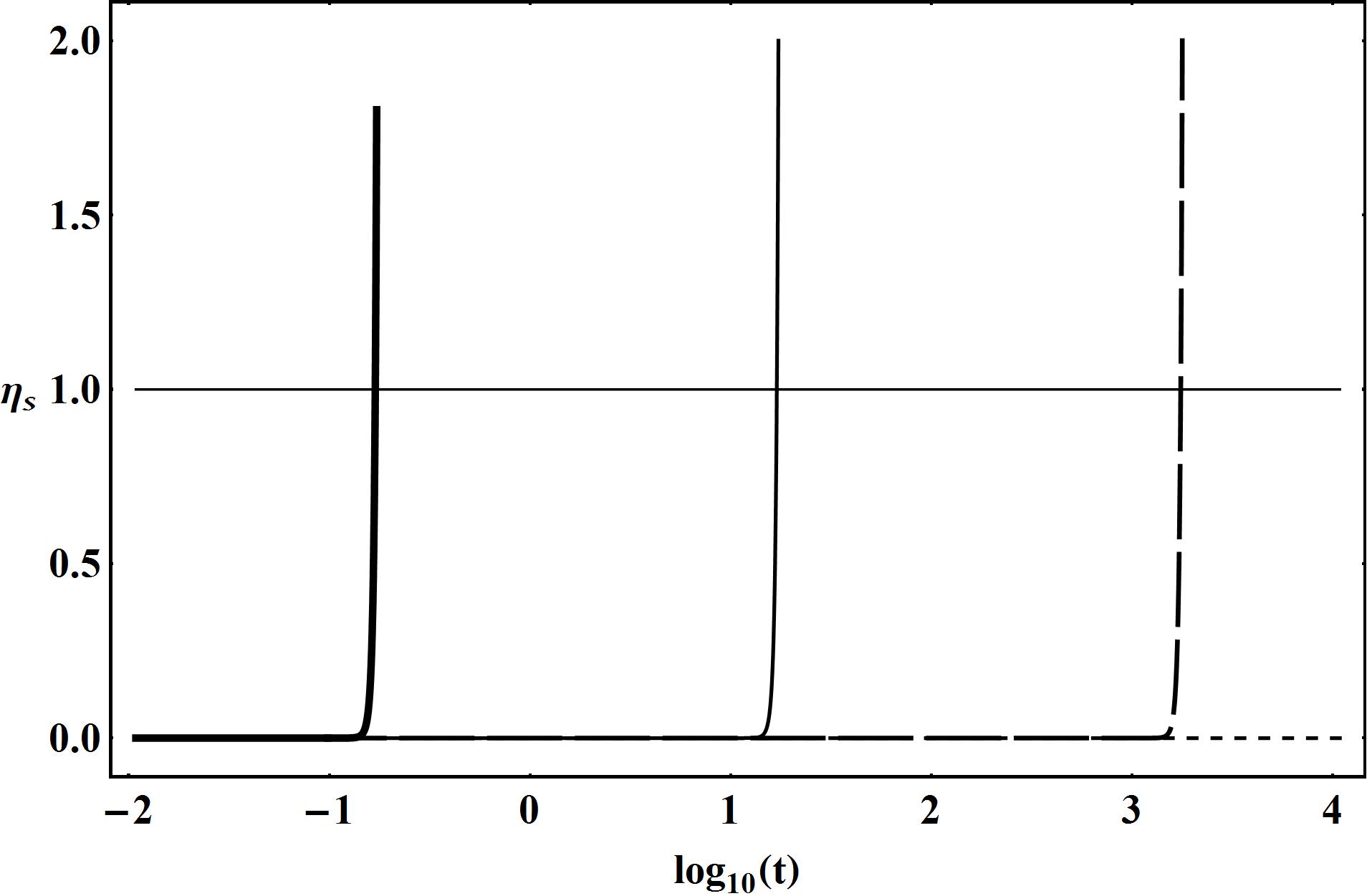}
\end{center}
{\small \textbf{Fig.  \thefigure}.\label{Fig3} The evolution of the relation of scalar field's and fermions' energy densities $\eta_s$. \vspace{12pt}}

\begin{center}\refstepcounter{figure}
\includegraphics[width=120mm]{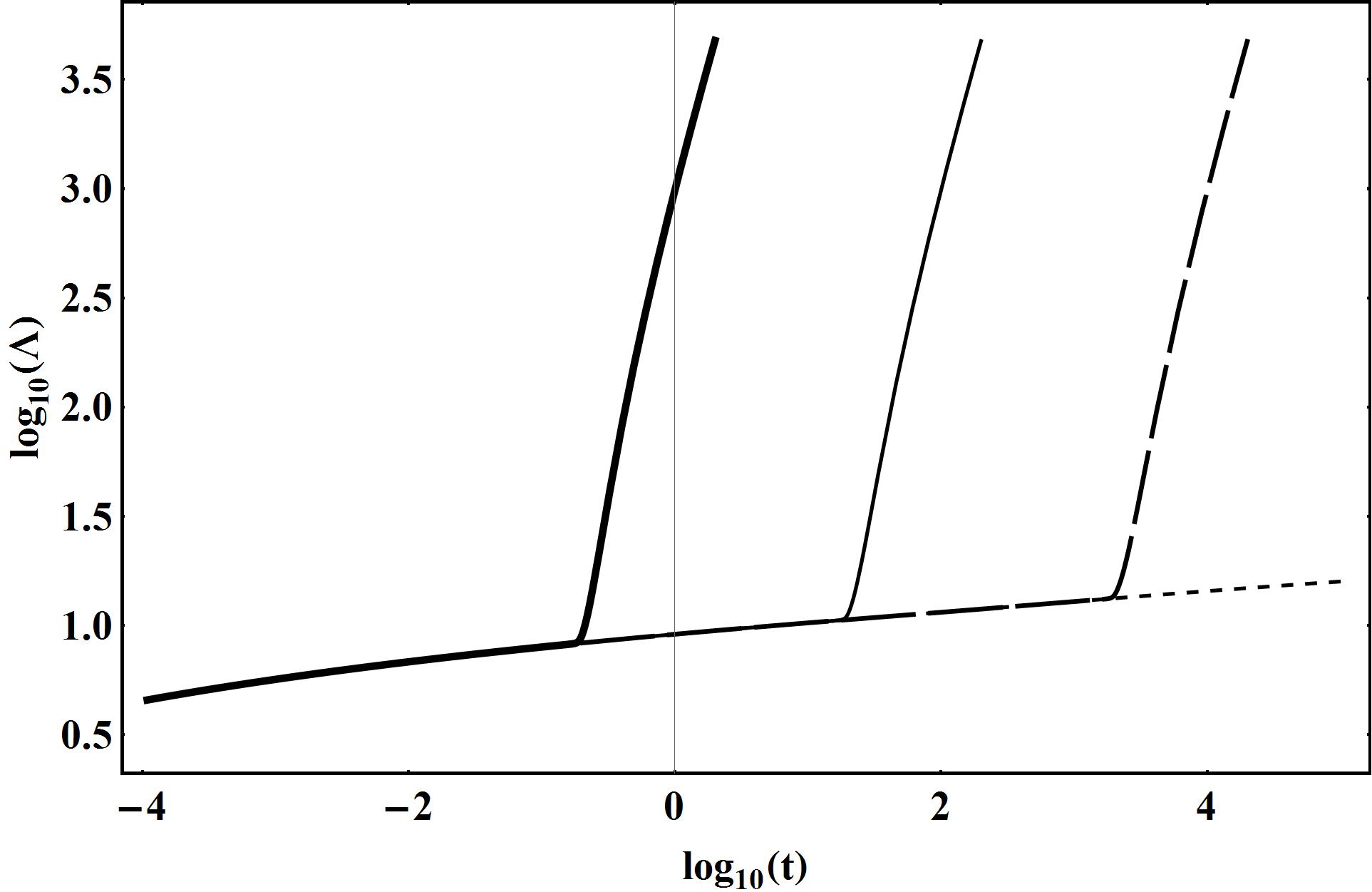}
\end{center}
{\small \textbf{Fig.  \thefigure}.\label{Fig4} The evolution of the scale function's logarithm $\Lambda(t)=\exp a(t)$. Along Y-axis there are the values of common logarithm, $\log_{10}\Lambda$, ò.å., $\log_{10}(\ln a(t))$.\vspace{12pt}}

Below, on the phase diagram Fig. 9, one can see plainly all three mentioned above stages of the cosmological evolution.
Next, on Fig.  3 we see instants of the cosmological time $t_{\eta}$, where $\eta_s(t_{\eta})=1$. A comparison with Fig.  2 shows that time instant $t_{\eta}$ corresponds to the extremum of the scalar field potential's derivative, $Z$.

Fig.  5 -- Fig. 7 are the plots showing the evolution of the barotropic coefficient $\kappa(t)$ and invariant cosmological acceleration $\Omega(t)$ related to it. These bursts appear exactly in the instants of time $t_{\eta}$.

As is seen from plots on Fig. 6, bursts are preceded by ultrarelatistic expansion which changes to inflation after burst. Fig. 7 shows a detailed structure of the cosmological acceleration's burst. This figure demonstrates that the burst is not a computer fantom.

\begin{center}\refstepcounter{figure}
\includegraphics[width=120mm]{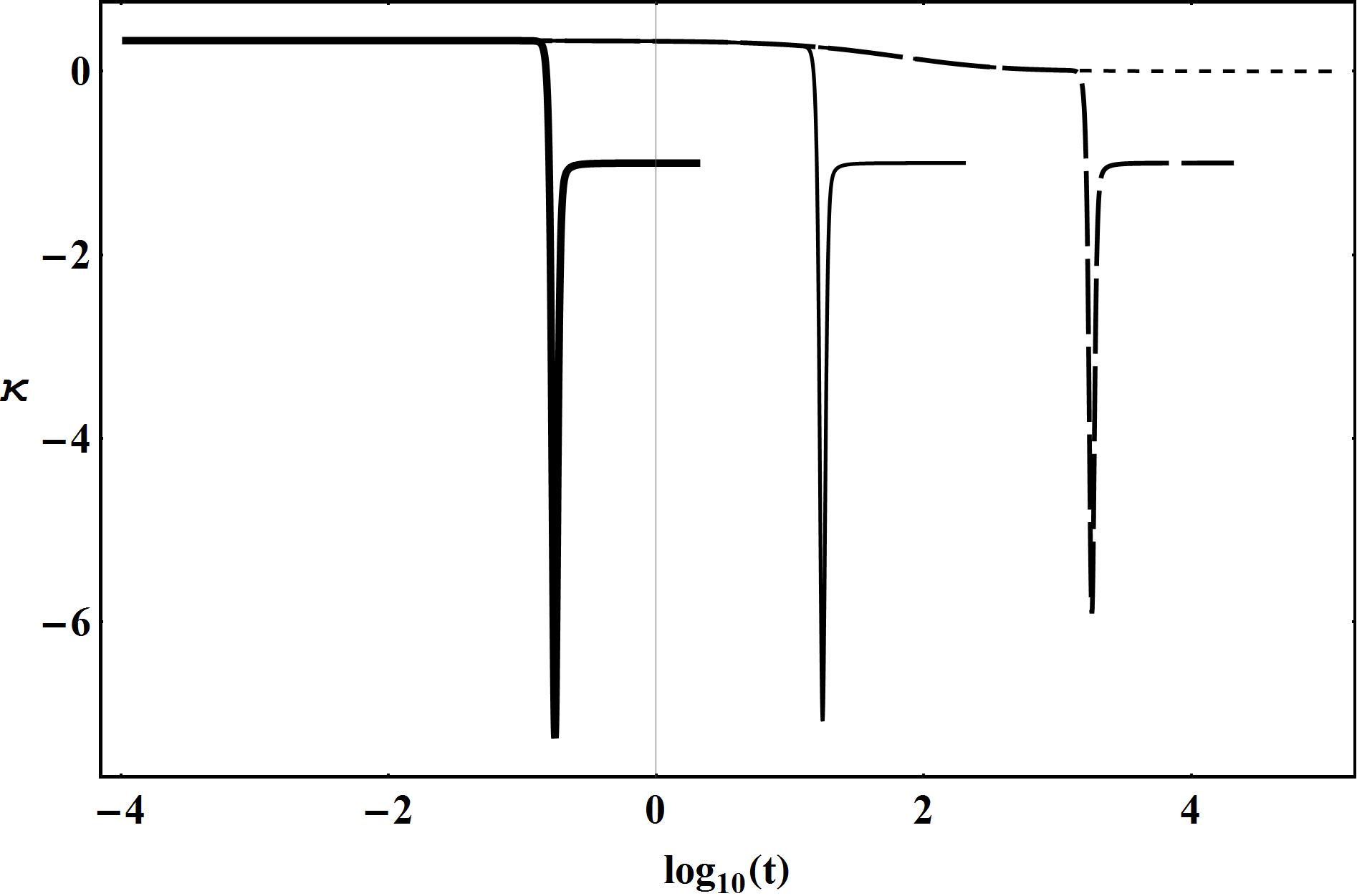}
\end{center}
{\small \textbf{Fig.  \thefigure}.\label{Fig5} The evolution of the barotropic coefficient $\kappa$.\vspace{12pt} }

\begin{center}\refstepcounter{figure}
\includegraphics[width=120mm]{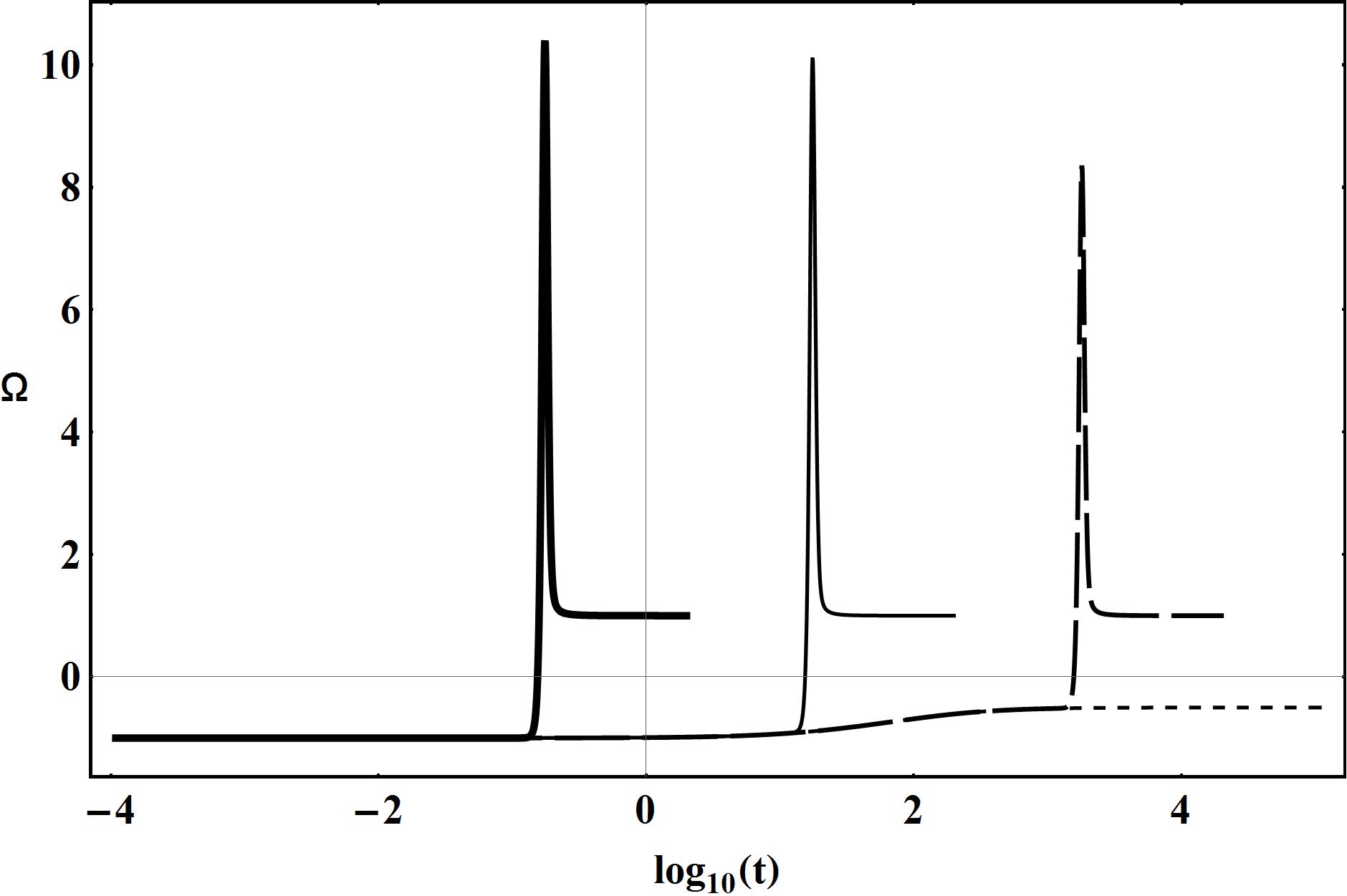}
\end{center}
{\small \textbf{Fig.  \thefigure}.\label{Fig6}The evolution of the invariant cosmological acceleration %
$\Omega$.\vspace{12pt} }

\begin{center}\refstepcounter{figure}
\includegraphics[width=120mm]{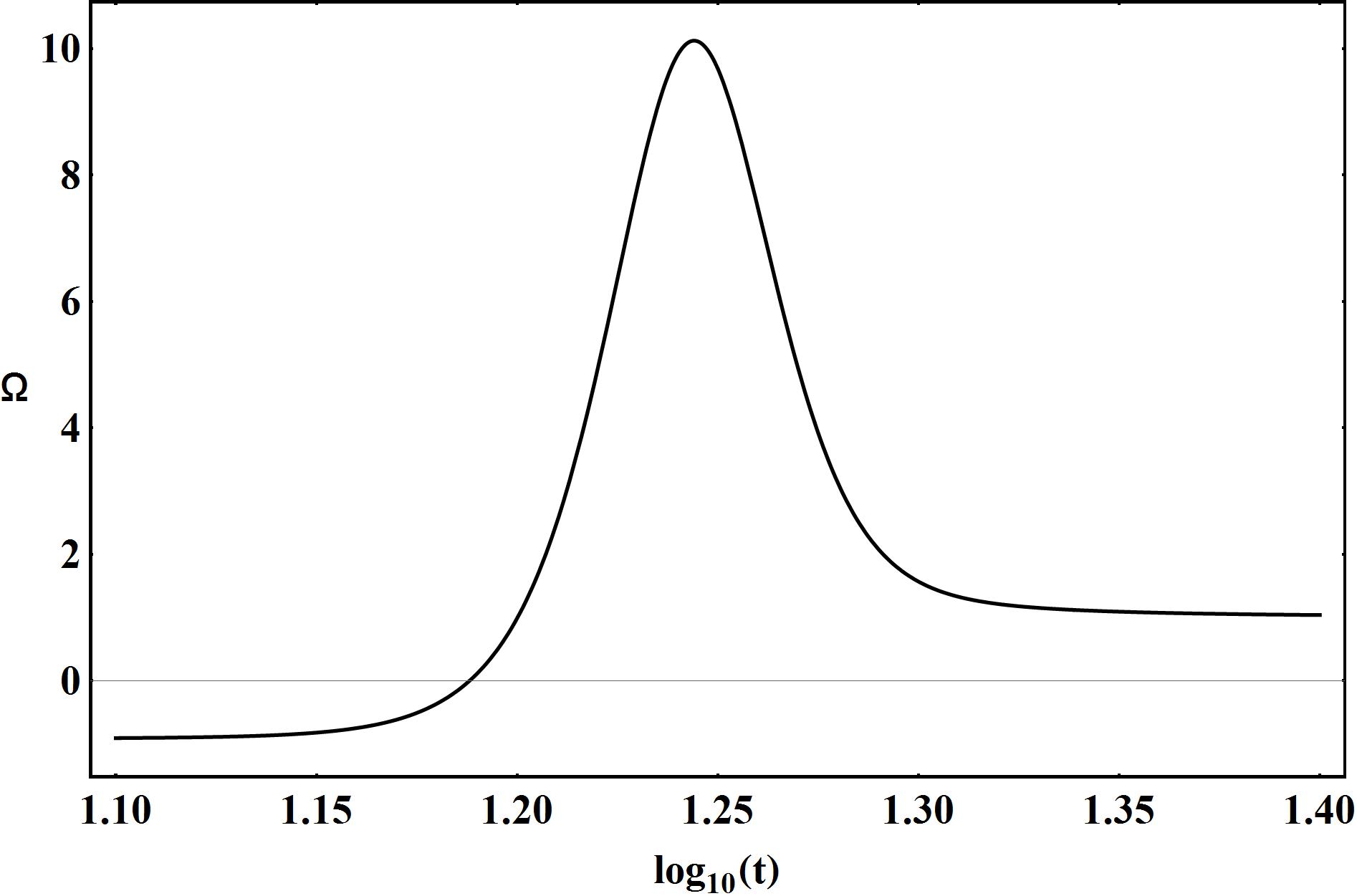}
\end{center}
{\small \textbf{Fig.  \thefigure}. \label{Fig7}\hfill Detailed structure of the second fantom burst of the invariant cosmological acceleration $\Omega$ shown on Fig. 6:
$p_0 = 100$, $m_0 = .001$, $\Phi(0) = 5\cdot 10^{-7}$, $m_s = 10^{-4}$.\vspace{12pt} }

\begin{center}\refstepcounter{figure}
\includegraphics[width=120mm]{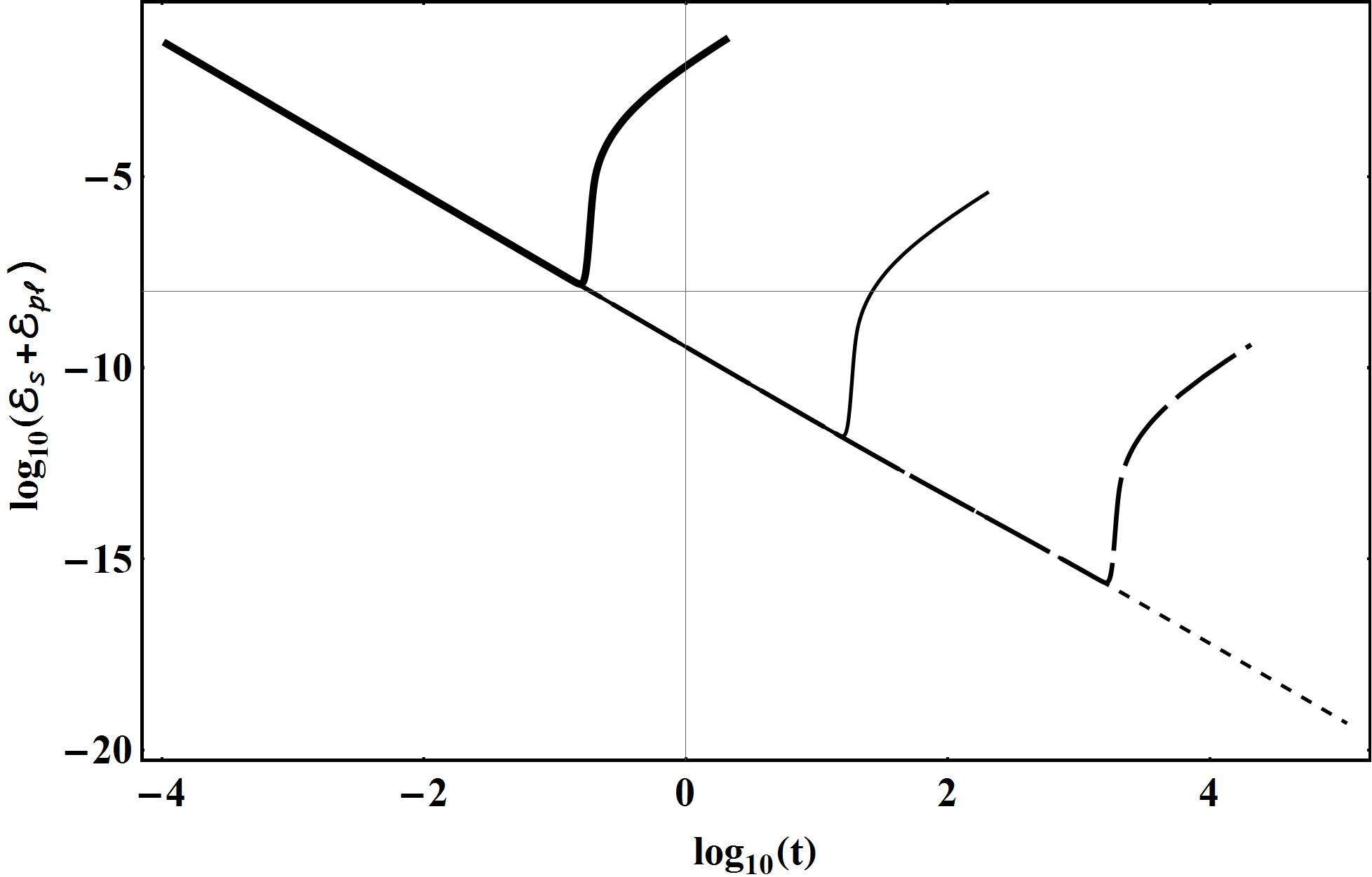}
\end{center}
{\small \textbf{Fig.  \thefigure}. \label{Fig8}The evolution of the total energy density's logarithm $\log_{10} \mathcal{E}$.\vspace{12pt} }

\begin{center}\refstepcounter{figure}
\includegraphics[width=120mm,height=8cm]{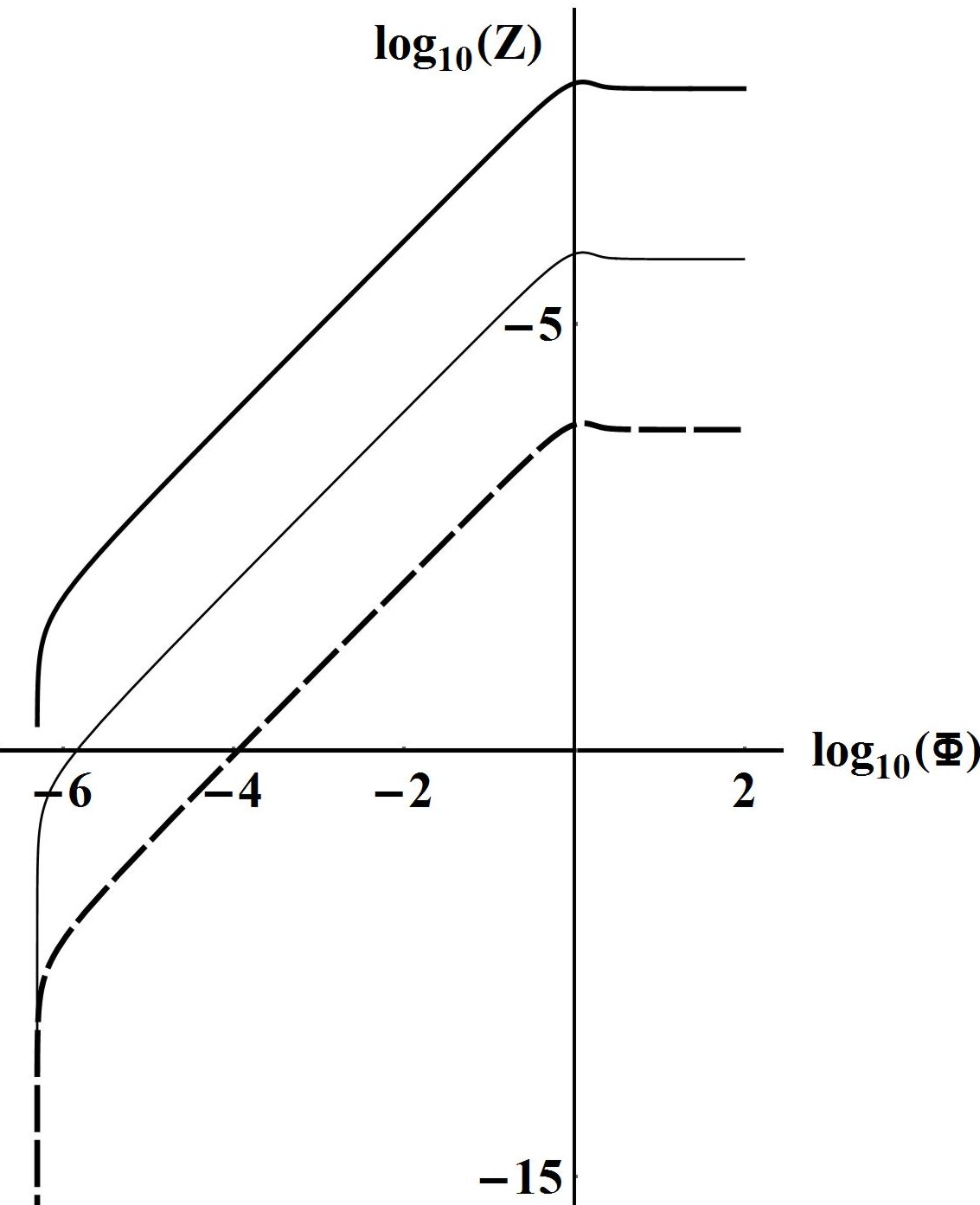}
\end{center}
{\small \textbf{Fig.  \thefigure}. \label{Fig9}Phase portrait. Left vertical branches of plots correspond to the first stage of the system's evolution, inclined parts - to the second stage and horizontal right branches - to the third stage.\vspace{12pt} }

\begin{center}\refstepcounter{figure}
\includegraphics[width=120mm]{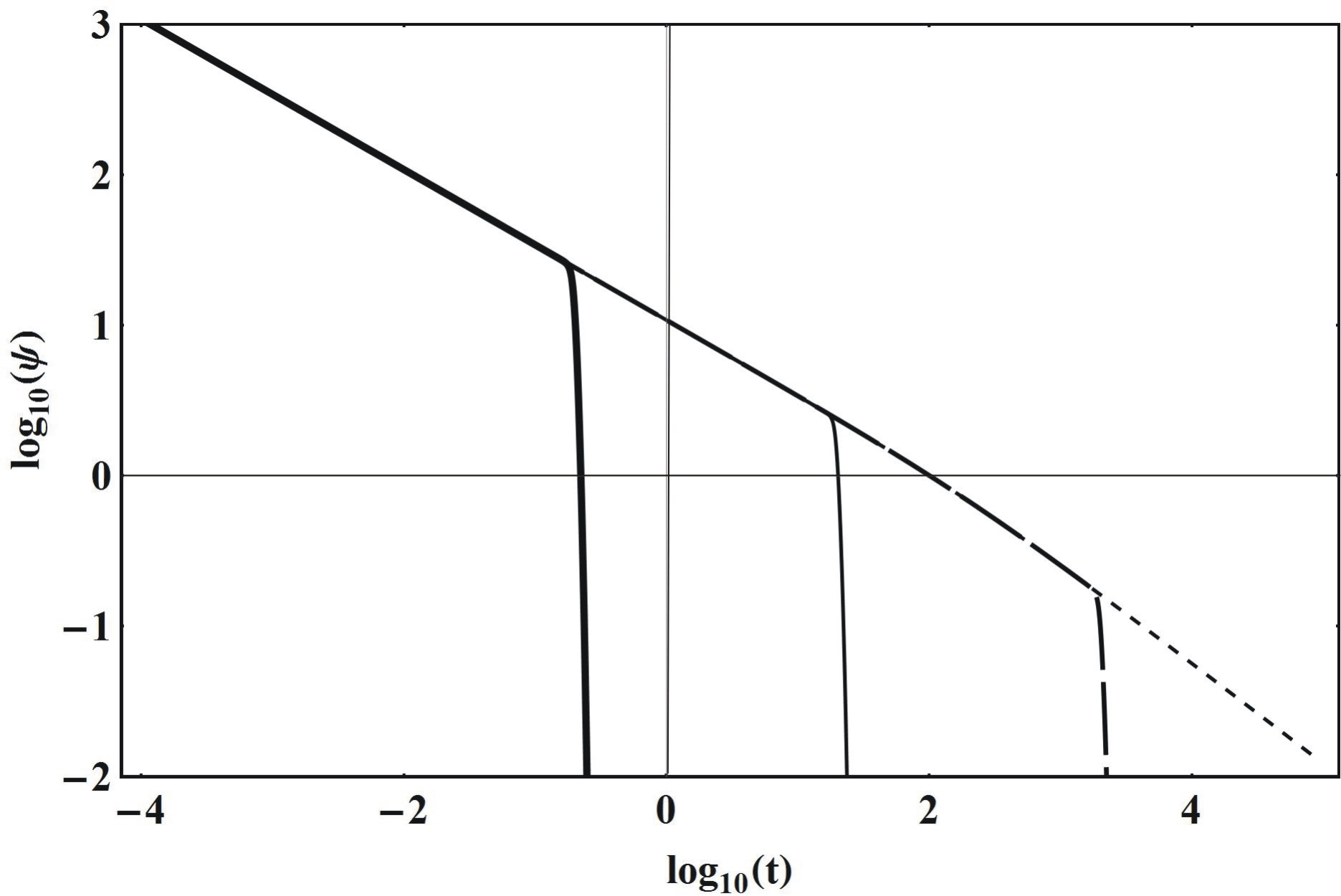}
\end{center}
{\small \textbf{Fig.  \thefigure}. \label{Fig10}The evolution of function $\psi = p_f / m_*$. Horizontal line corresponds to  $\log_{10}\psi=0\rightarrow\psi=1$, i.e. transition from ultrarelativistic equation of fermions' state to non-relativistic.\vspace{12pt} }

\subsection{The Case of Massless  ($m_s = 0$) Fantom Scalar Field with a Source}
In this case the system of differential equations takes the following form:
\begin{eqnarray}
\label{scalar_ms=0}
&&\ddot{\Phi}+3\dot{\Lambda}\dot{\Phi} - 4\pi \sigma= 0;\\
&&\dot{\Lambda}^2=\frac{8\pi}{3}(\mathcal{E}_{p}+\mathcal{E}_{s}),
\end{eqnarray}
where scalar field's energy density is equal to
\begin{equation}
{\rm {\mathcal E}}_s=-\frac{\dot\Phi^2}{8\pi}.
\end{equation}
and can be only a negative values; herewith in the initial time instant $t=0$ it equals zero. As a result, the cosmological scenario for the massless scalar field from chapter \ref{sigma=0} is changed. There are still three stages however on the last stage fermions prevail:
\begin{enumerate}
\item $t \lesssim t_s^*)$: \textit{prevalence of ultrarelativistic fermions}.\\
\emph{Characteristic properties}: small values of the scalar field's potential -- minor contribution of the field into system evolution.\\
\item $t_s^* < t\lesssim  t_r $: \textit{prevalence of the scalar field}.\\
\emph{Characteristic properties}: Sharp growth of the scalar field's potential and its derivative, maximum influence on evolution of the system. In the instant of time when fermions transit from relativistic stage to non-relativistic, $t_r$, scalar charge's density reaches maximum. In this instant of time the following functions reach maximum: $Z=\dot{\Phi}$ (maximum); barotropic coefficient, $\kappa$,(minimum); invariant cosmological acceleration, $\Omega$, (maximum); total energy's density, $\mathcal{E}_{p}+\mathcal{E}_{s}$, (maximum). In peak instant of time the barotropic coefficient depending on system parameters can become less than -1 -- fantom equation of state and inflation. It should be noted the important fact: contribution of fermions in the energy density at this stage is minor however the scalar field is managed exactly by fermions with a help of scalar density of charges $\sigma$. In this sense, we can draw an analogy between fermions and chemical reactions' catalysts.
\item $t_r<t< +\infty)$: \textit{prevalence of non-relativistic fermions}\\
\noindent
\emph{Characteristic properties}: Scalar charge's density falls, influence of the scalar field on the cosmological evolution becomes vanishingly small. The derivative of the scalar field's potential tends to zero ($m_s^*\to0$), potential tends to constant value; non-relativistic fermions defined future evolution of the system, barotropic coefficient $\kappa \to 0$\footnote{Thus, the considered case shows that given model can withdraw a problem of early inflation's stop and therefore ensure a necessary for structure generation non-relativistic stage of the Universe expansion.}.

\end{enumerate}

On the below plots (Fig. 11 -- Fig. 18) the results of numerical modelling of the cosmological evolution of the system of degenerated fermions with massless fantom scalar field are shown. Everywhere on the plots the following parameter values are accepted:\\
$p_0 = 0.01$, $m_0 = 0$, $\Phi(0) = 5\cdot 10^{-7}$. Heavy line is $q = 0.01$, thin line is $q = 0.1$, normal dotted line is $q = 1$, fine dotted line is $q = 5$.

\begin{center}\refstepcounter{figure}
\includegraphics[width=120mm]{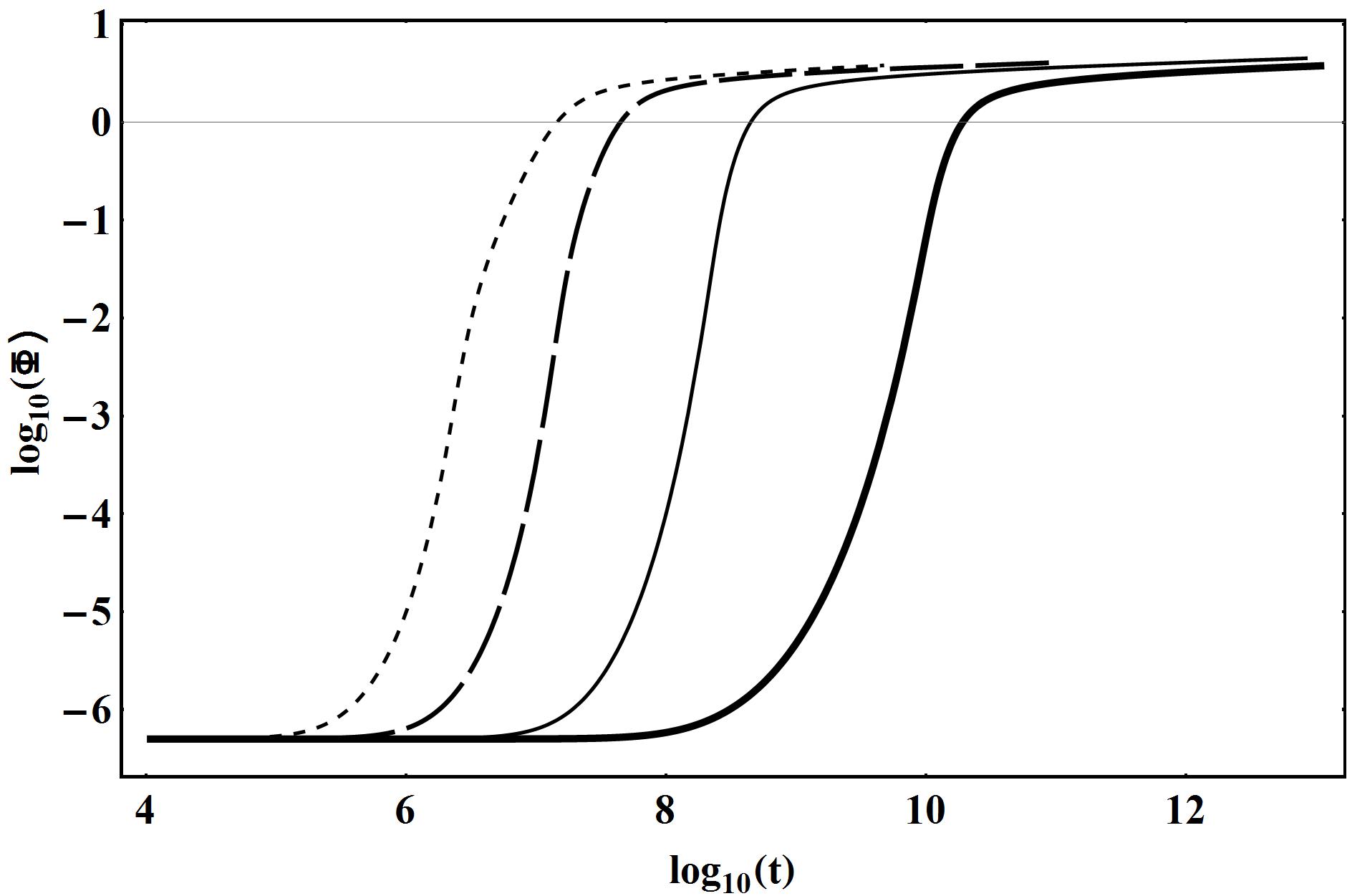}
\end{center}
{\small \textbf{Fig.  \thefigure}. \label{Fig11}The evolution of the potential's logarithm $\log_{10}\Phi$.\vspace{12pt} }

\begin{center}\refstepcounter{figure}
\includegraphics[width=120mm]{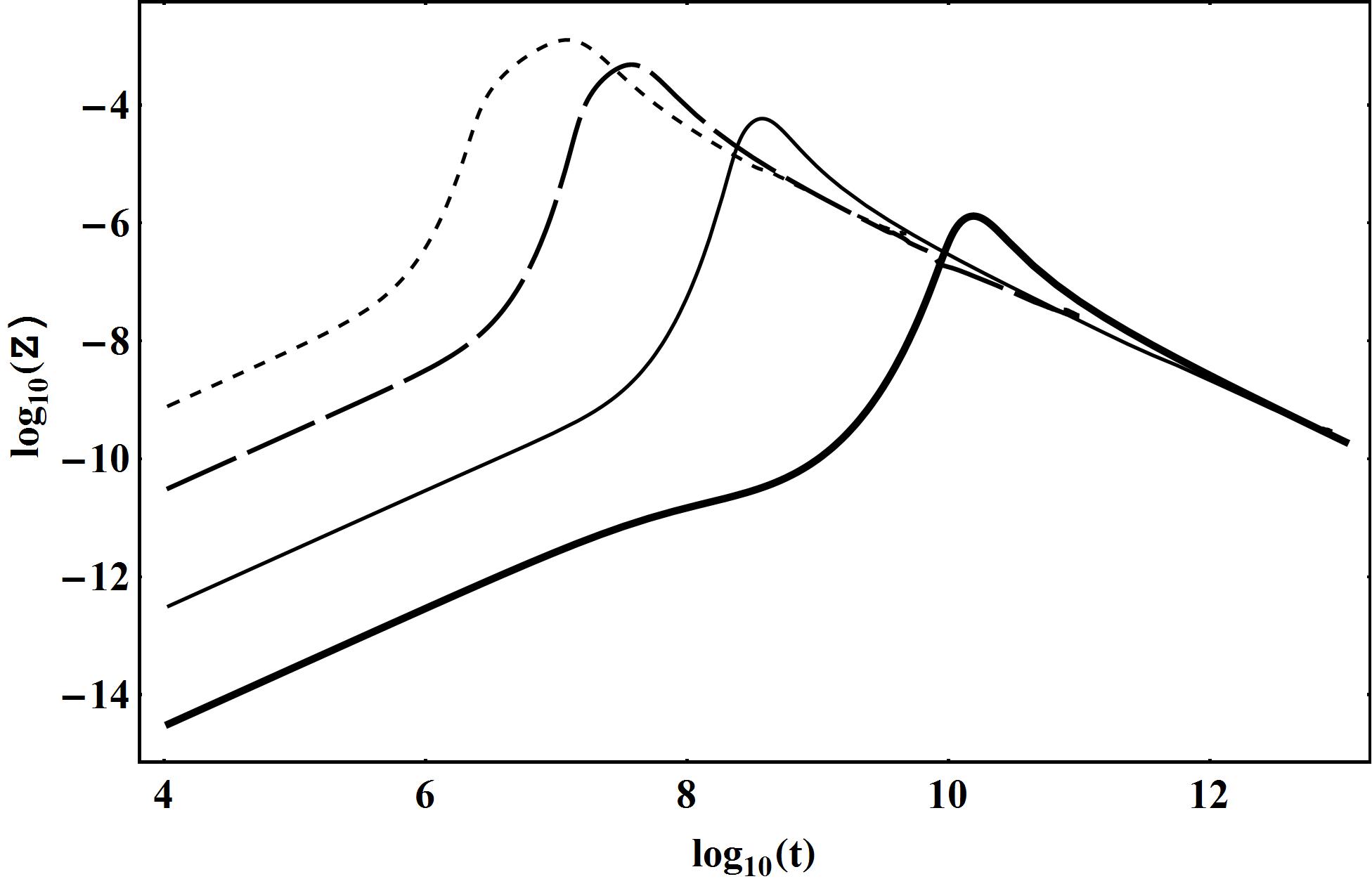}
\end{center}
{\small \textbf{Fig.  \thefigure}. \label{Fig12}The evolution of logarithm of the potential's derivative $\log_{10} Z=\log_{10} \dot{\Phi}$.\vspace{12pt} }

\begin{center}\refstepcounter{figure}
\includegraphics[width=120mm]{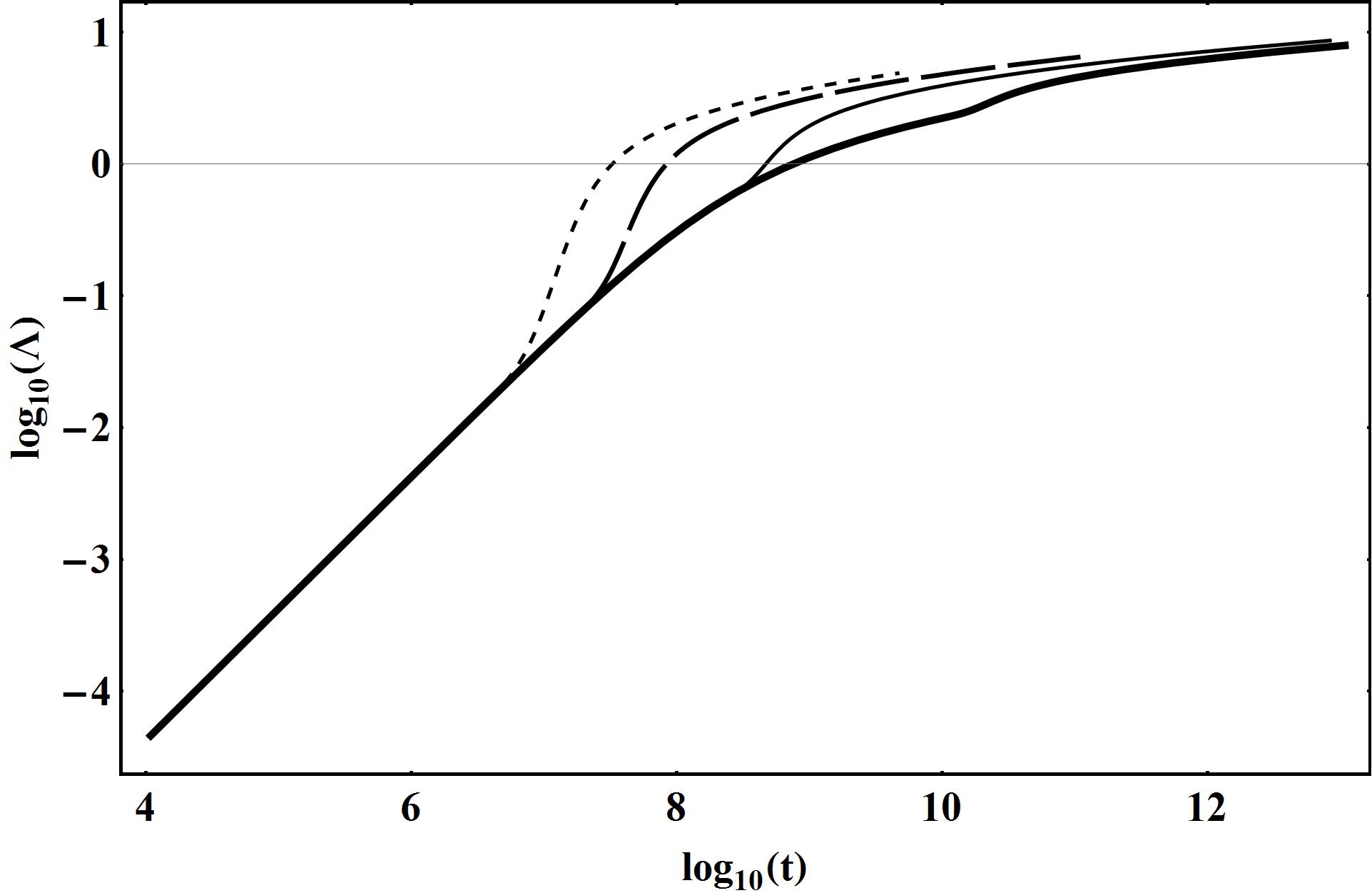}
\end{center}
{\small \textbf{Fig.  \thefigure}. \label{Fig13}The evolution of the scale function's logaritm $\log_{10}\Lambda(t)$.\vspace{12pt}}

\begin{center}\refstepcounter{figure}
\includegraphics[width=120mm]{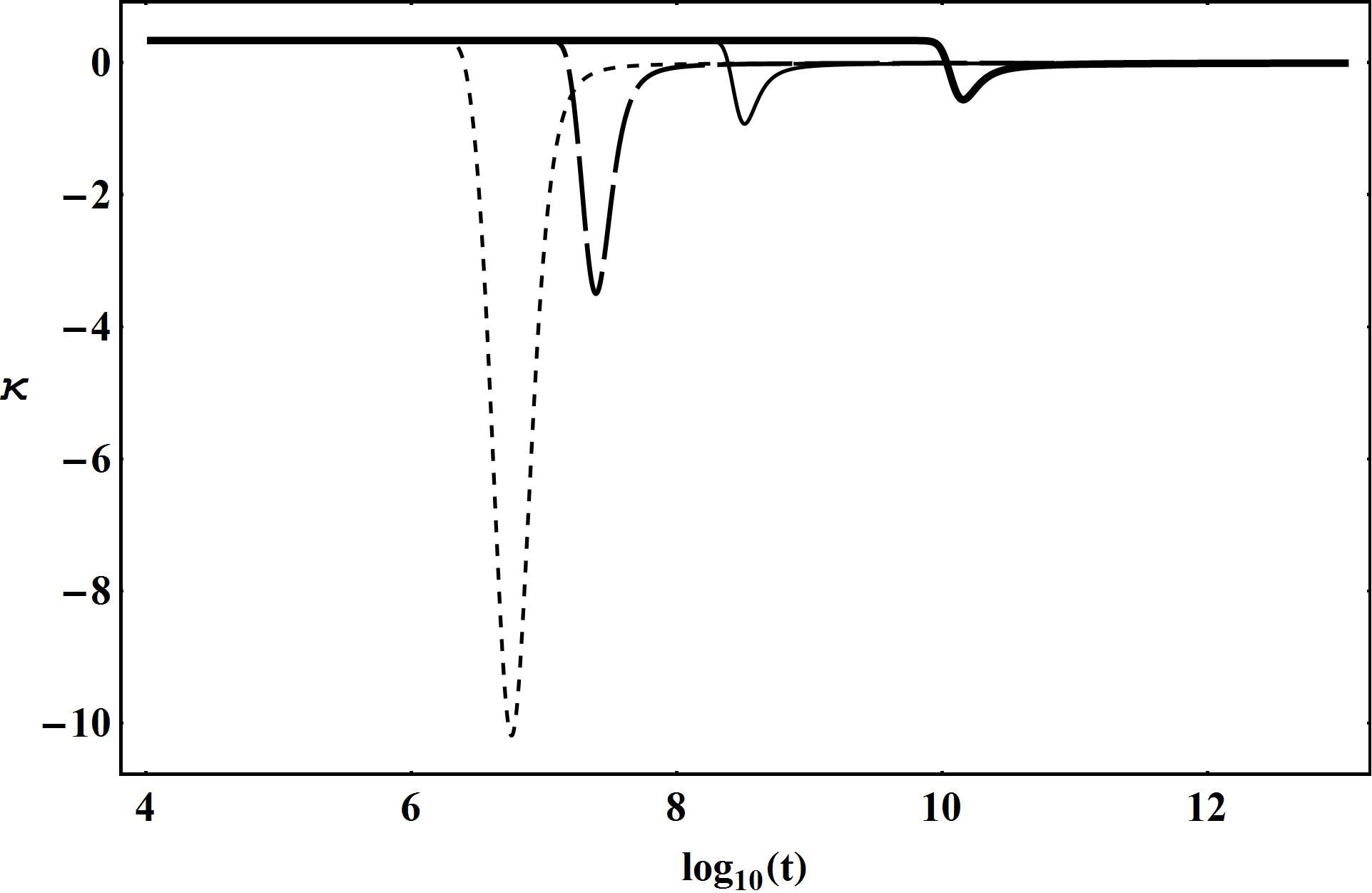}
\end{center}
{\small \textbf{Fig.  \thefigure}. \label{Fig14}The evolution of barotropic coefficient $\kappa$.\vspace{12pt}}

\begin{center}\refstepcounter{figure}
\includegraphics[width=120mm]{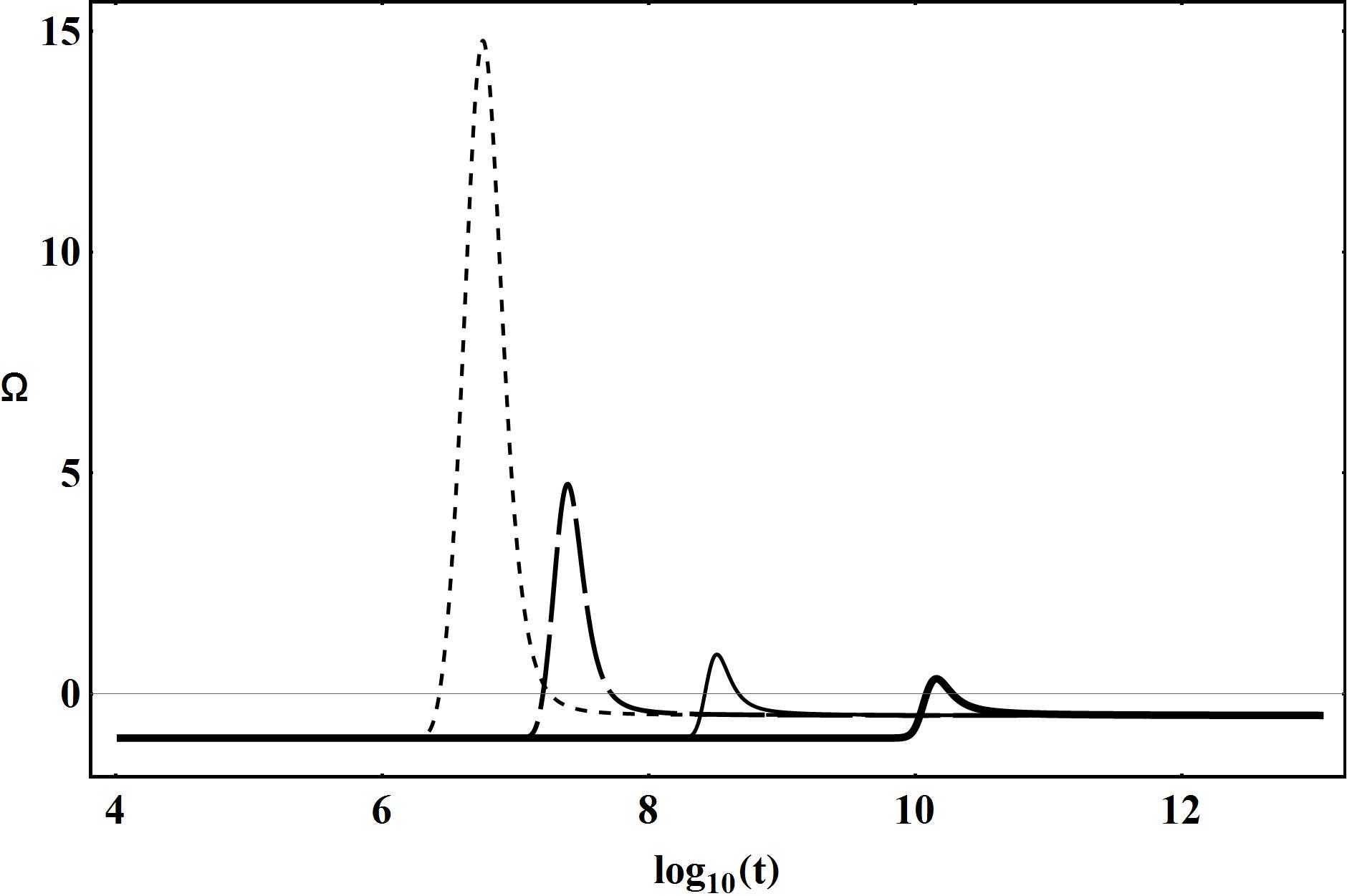}
\end{center}
{\small \textbf{Fig.  \thefigure}. \label{Fig15}The evolution of the invariant cosmological acceleration %
$\Omega$.\vspace{12pt}}

\begin{center}\refstepcounter{figure}
\includegraphics[width=120mm]{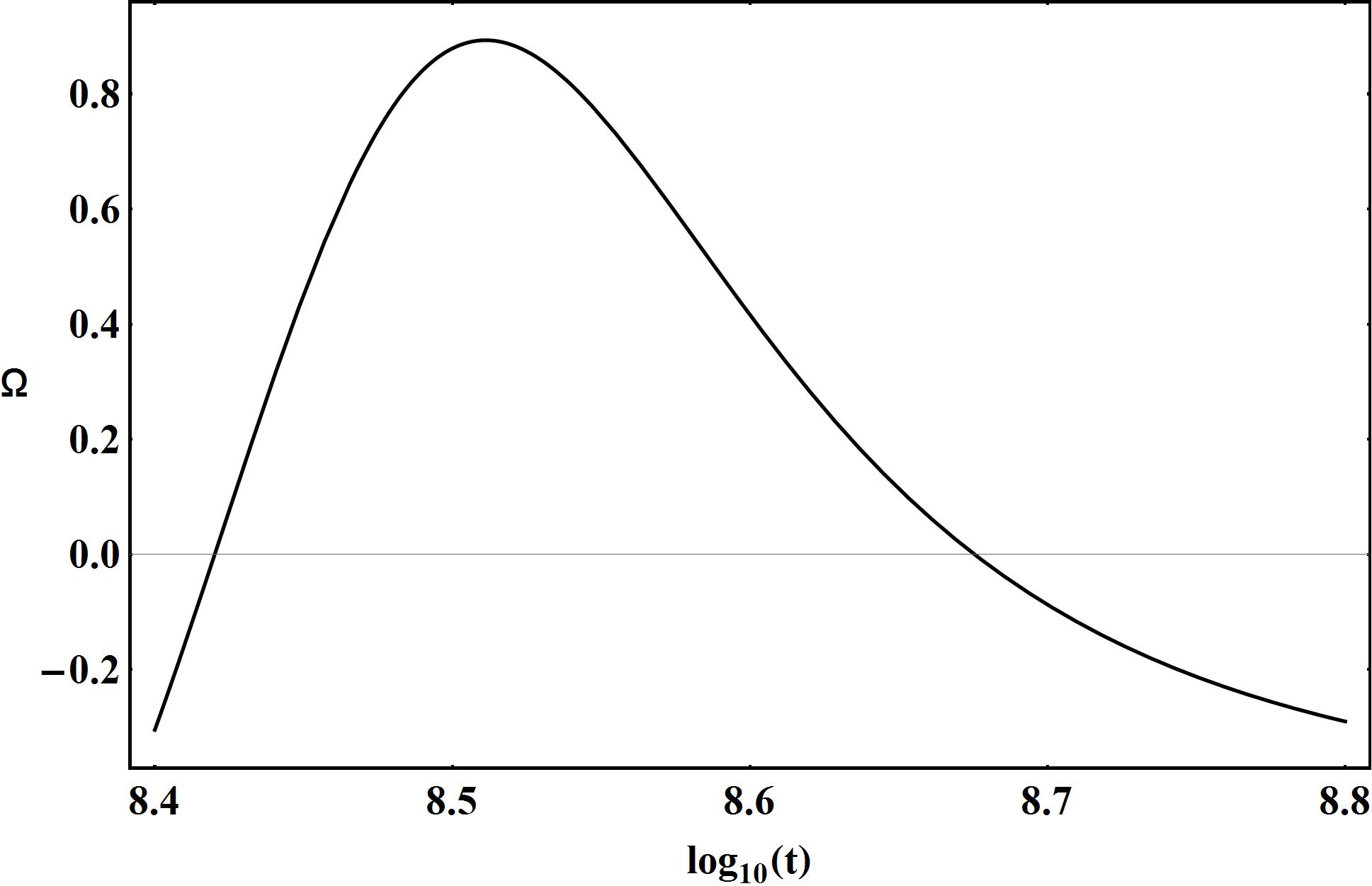}
\end{center}
{\small \textbf{Fig.  \thefigure}. \label{Fig16}Detailed structure of the third dantom burst of the invariant cosmological acceleration shown on Fig. 15: %
$\Omega$;  $q = 1$.\vspace{12pt}}

\begin{center}\refstepcounter{figure}
\includegraphics[width=120mm]{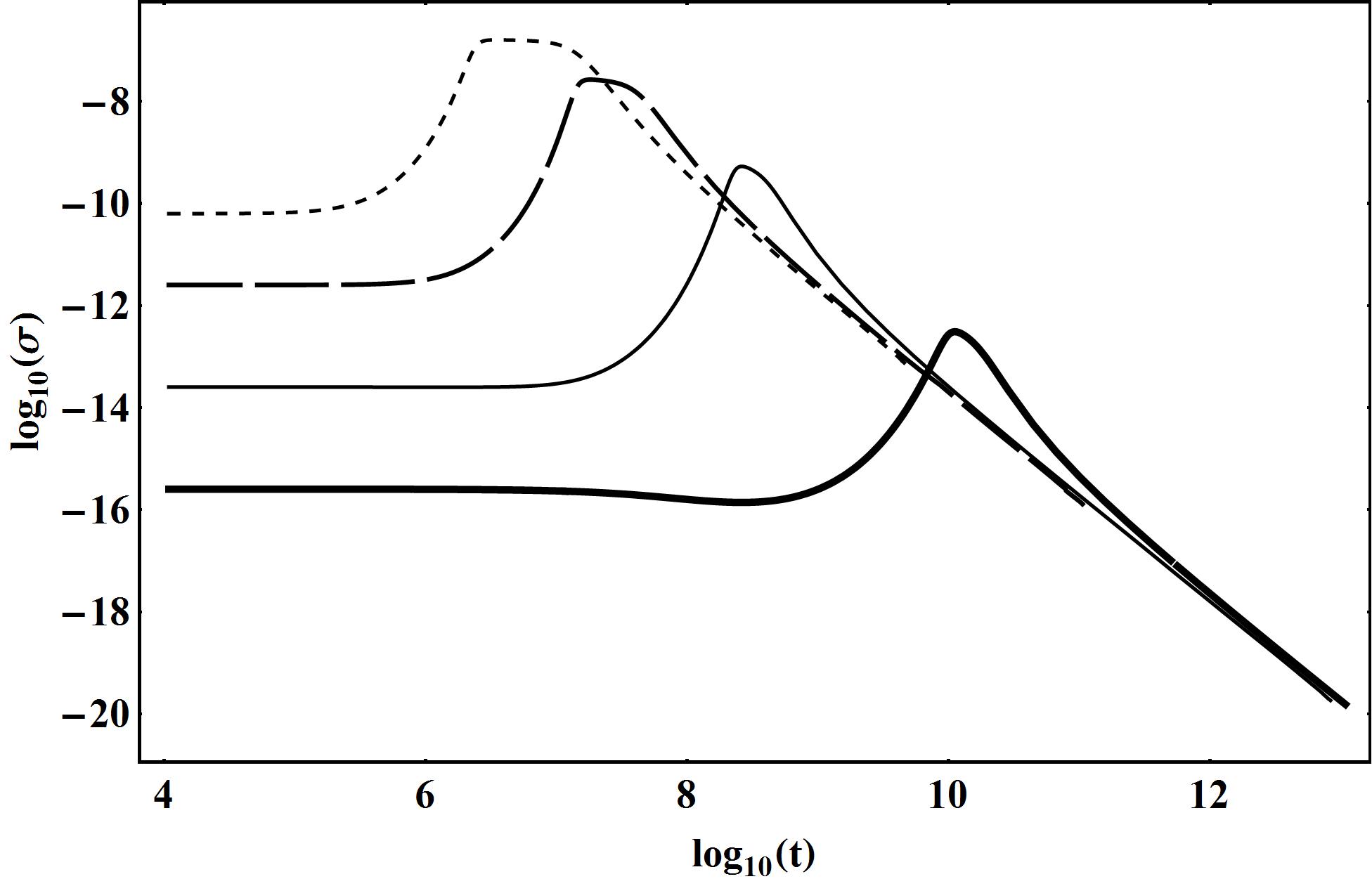}
\end{center}
{\small \textbf{Fig.  \thefigure}. \label{Fig17}The evolution of logarithm of the scalar charge's density $\log_{10}\sigma$.\vspace{12pt}}

\begin{center}\refstepcounter{figure}
\includegraphics[width=120mm]{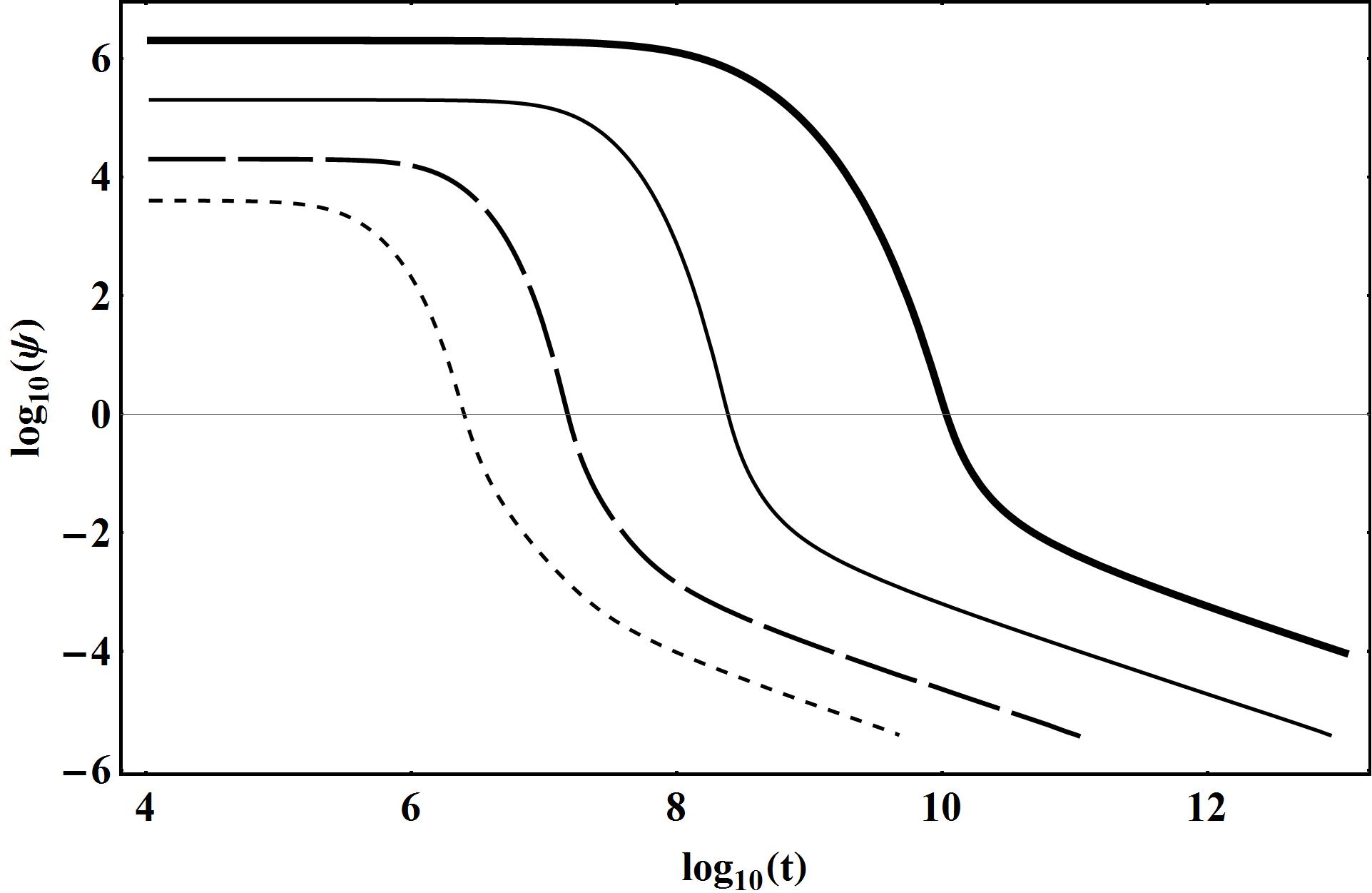}
\end{center}
{\small \textbf{Fig.  \thefigure}. \label{Fig18}The evolution of logarithm of the function $\psi = p_0 / m_*a$.\vspace{12pt}}

\subsection{The Case of Massless Fantom Scalar Field ($m_s\not=0$) with the Source ($\sigma\not=0$)}
In this, the most common case, we need to resolve the system of ordinary differential equations (\ref{Z,lambda}), (\ref{dZ}) and (\ref{dL}) with initial conditions (\ref{Coshe}) with an account of definitions (\ref{3a}) -- (\ref{3c}) and (\ref{E_s}).
The results of numerical integration of the system with the following parameters are shown below (Fig. 19 -- Fig. 26):
$p_0 = 0.01$, $m_0 = .001$, $\Phi(0) = 5\cdot 10^{-7}$. Heavy line is $m_s = 10^{-4}$, $q = 0$; thin line is $m_s = 10^{-4}$, $q = 0.1$; fine dotted line is $m_s = 0$, $q = 0.01$.\\

\begin{center}\refstepcounter{figure}
\includegraphics[width=120mm]{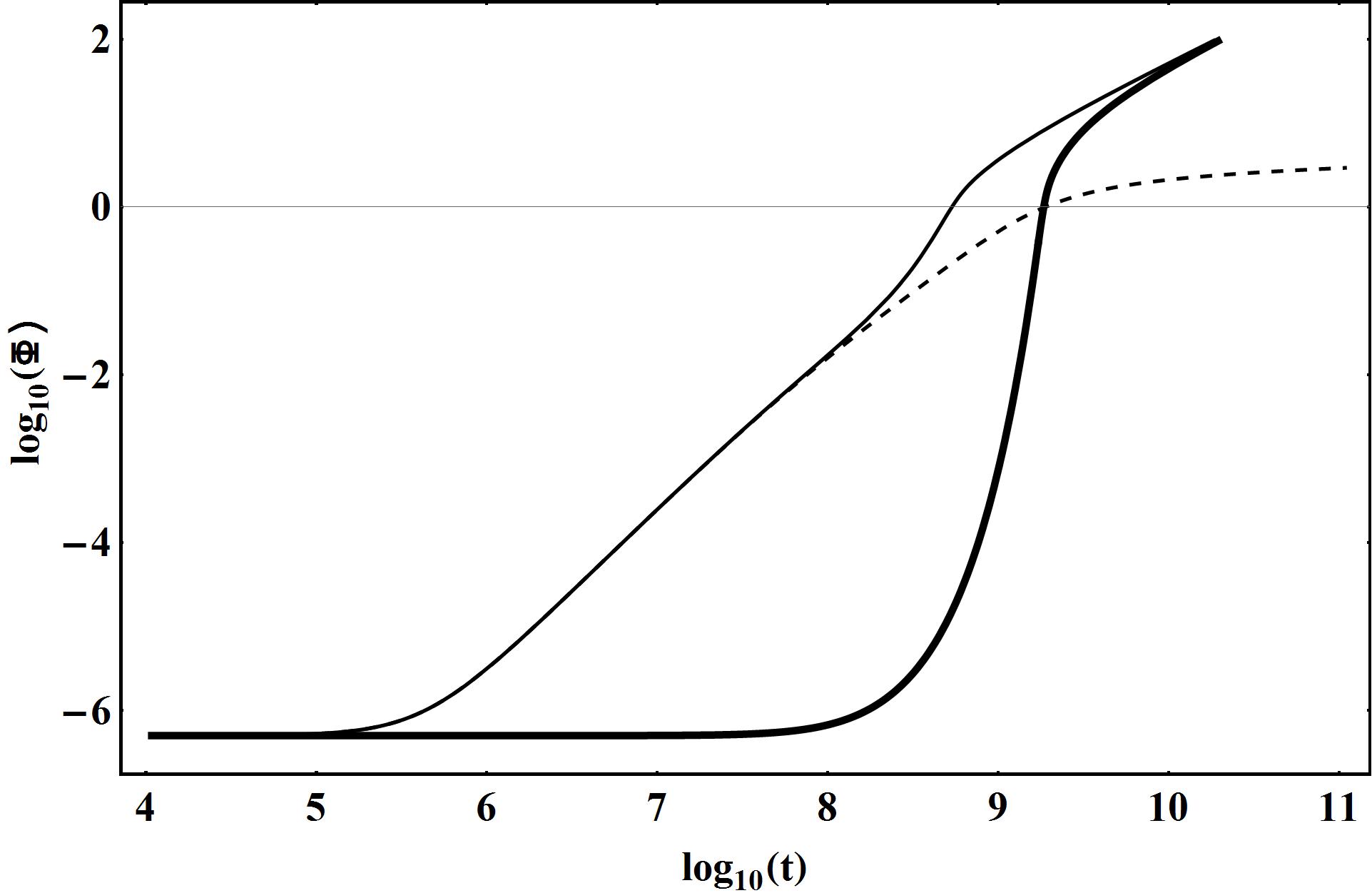}
\end{center}
{\small \textbf{Fig.  \thefigure}. \label{Fig19}The evolution of logarithm of the potential $\log_{10}\Phi$.}\vspace{12pt}

\begin{center}\refstepcounter{figure}
\includegraphics[width=120mm]{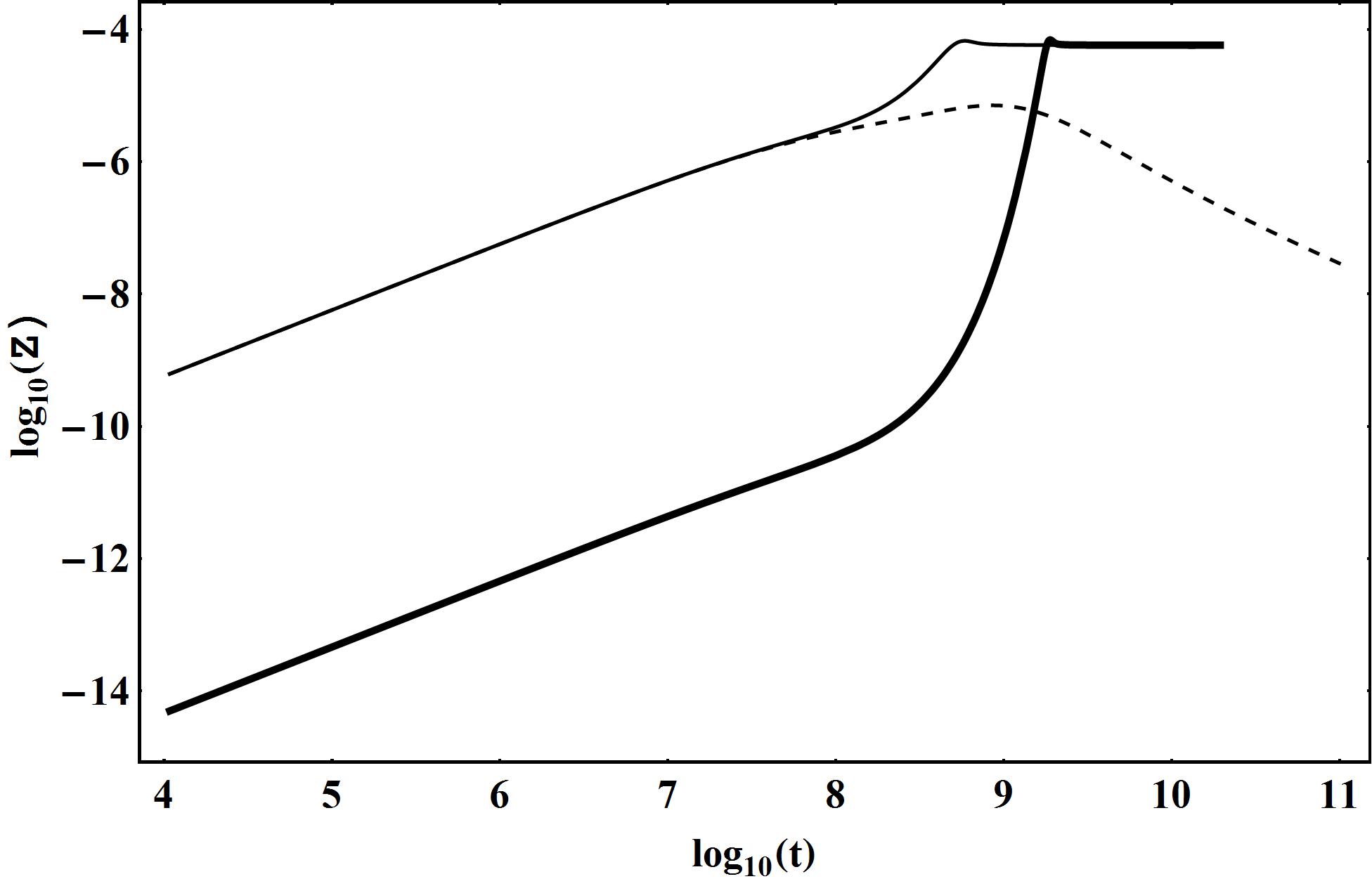}\end{center}
{\small \textbf{Fig.  \thefigure}. \label{Fig20}The evolution of logarithm of the potential's derivative $\log_{10} Z=\log_{10} \dot{\Phi}$.}\vspace{12pt}

\begin{center}\refstepcounter{figure}
\includegraphics[width=120mm]{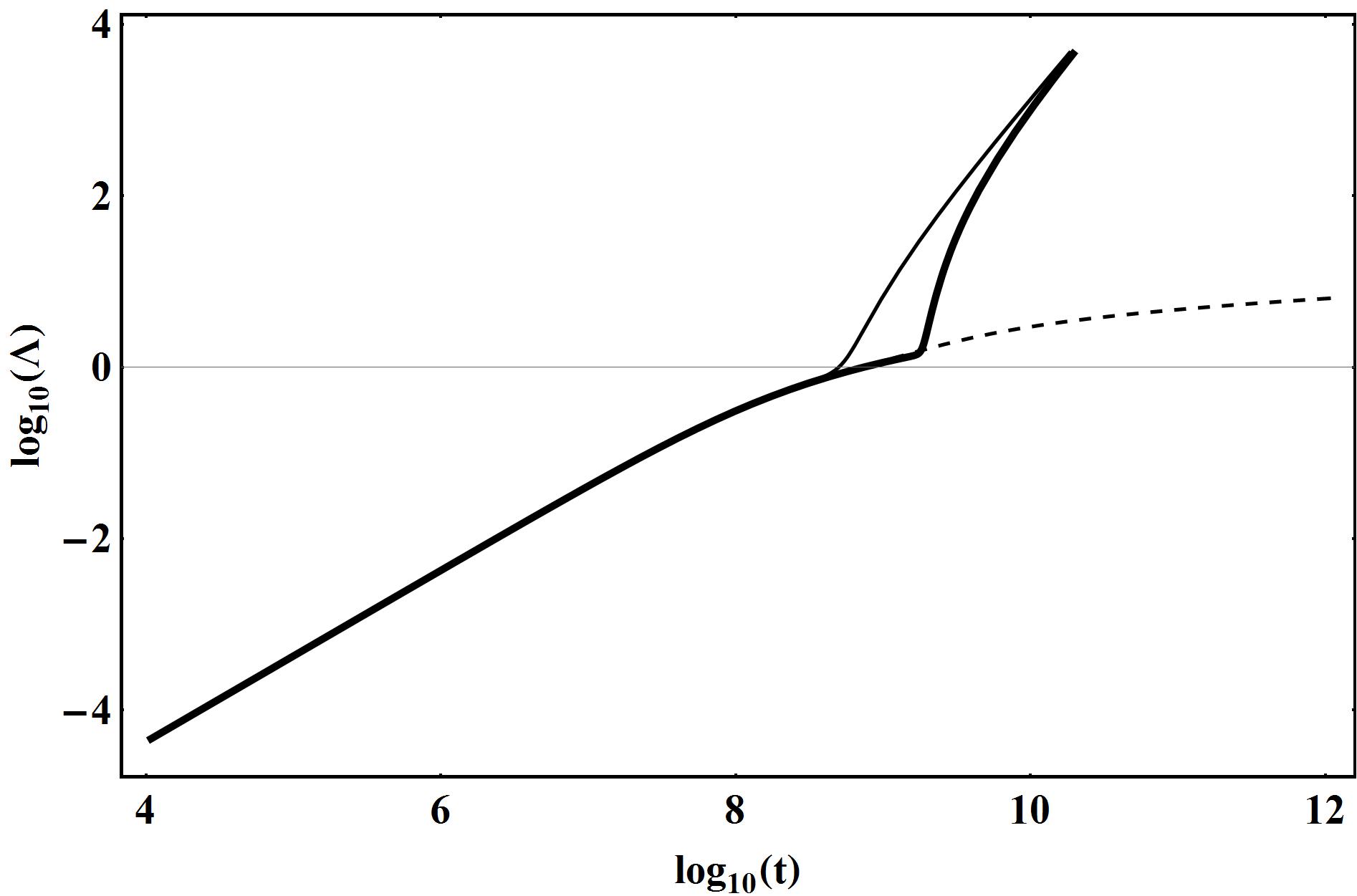}\end{center}
{\small \textbf{Fig.  \thefigure}. \label{Fig21}The evolution of logarithm of the scale function $\log_{10}\Lambda(t)$.}\vspace{12pt}\vspace{12pt}

\begin{center}\refstepcounter{figure}
\includegraphics[width=120mm]{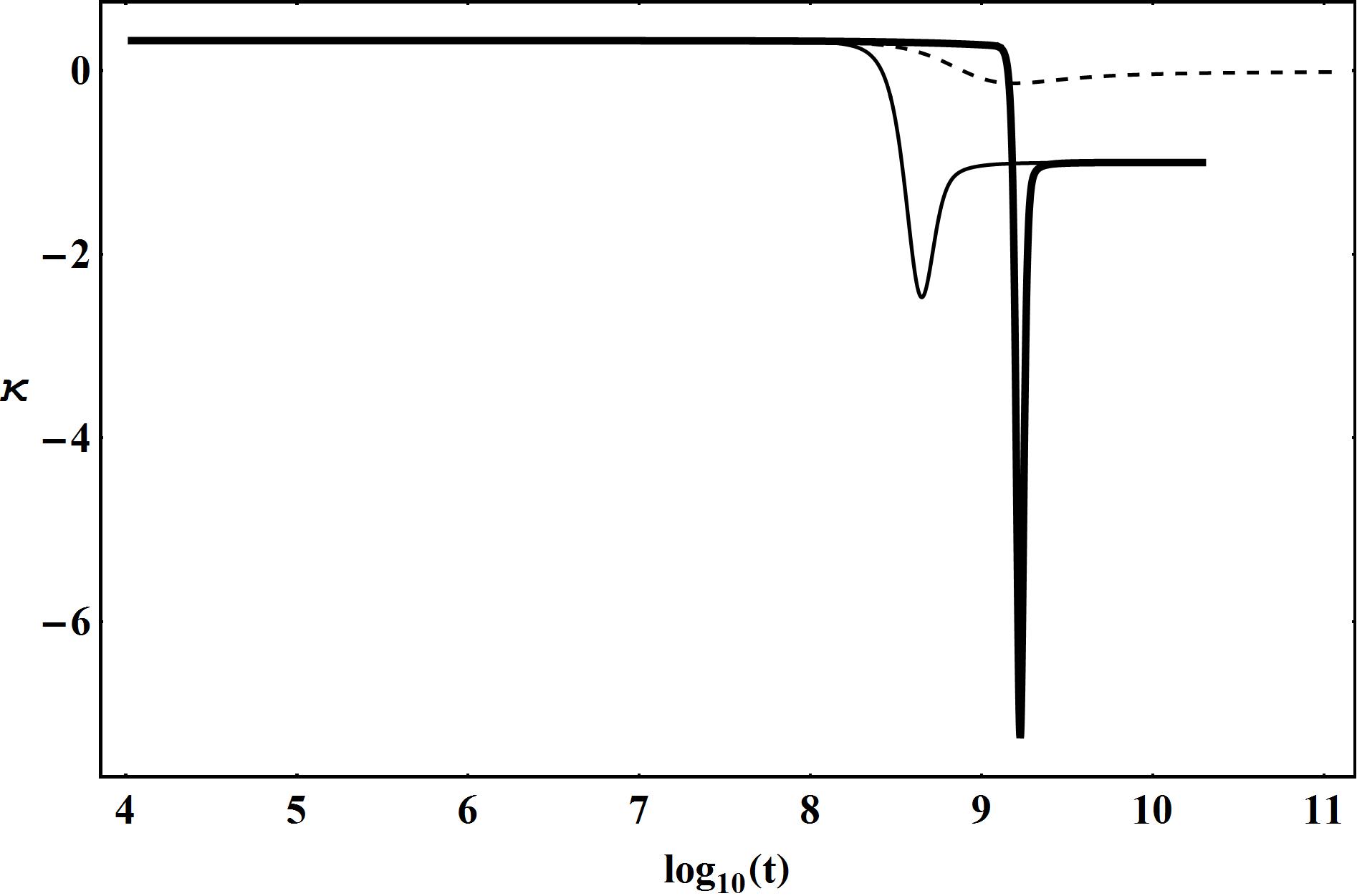}\end{center}
{\small \textbf{Fig.  \thefigure}. \label{Fig22} The evolution of barotropic coefficient $\kappa$.}\vspace{12pt}

\begin{center}\refstepcounter{figure}
\includegraphics[width=120mm]{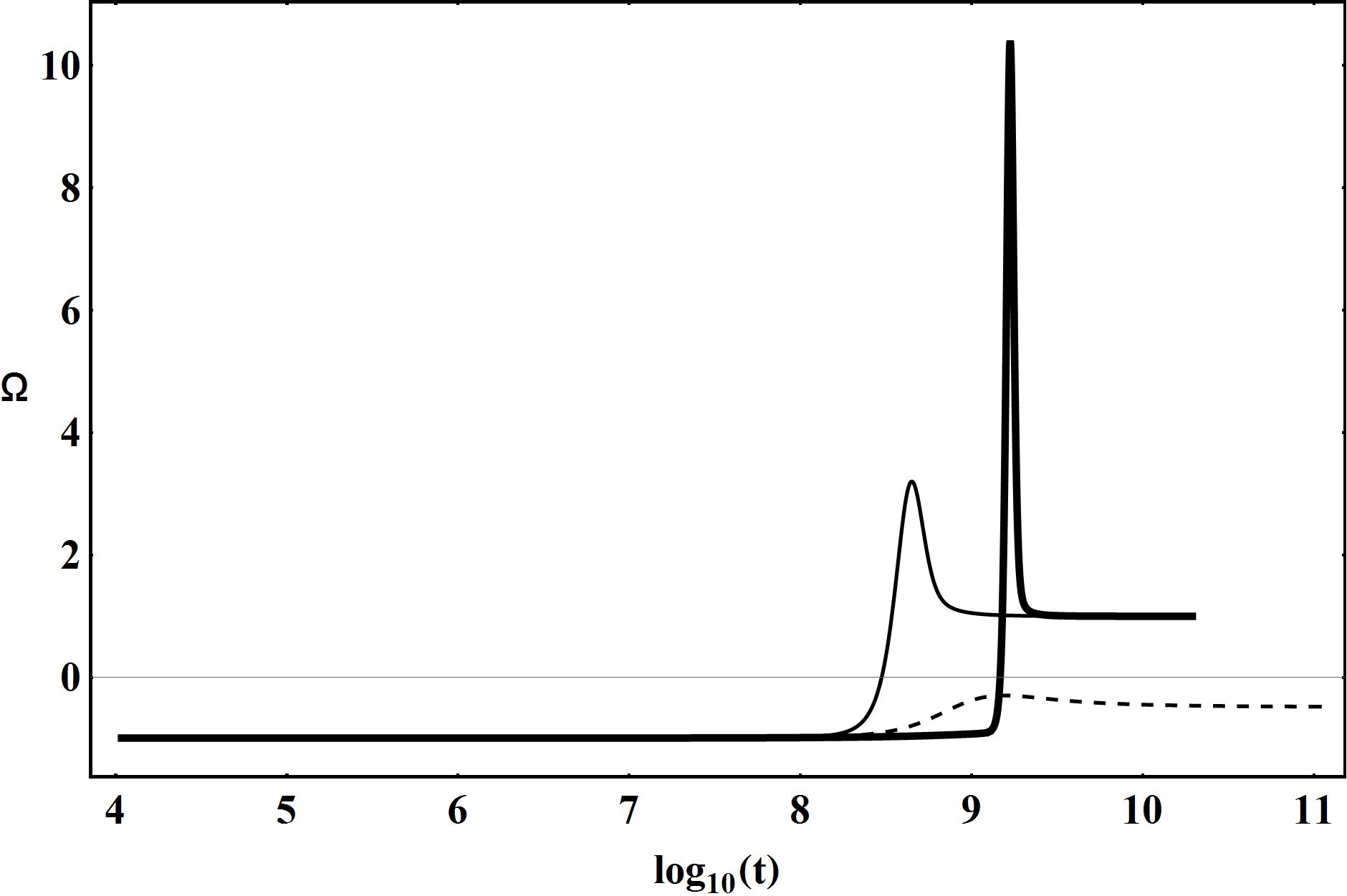}\end{center}
{\small \textbf{Fig.  \thefigure}.\label{Fig23}The evolution of the invariant cosmological acceleration %
$\Omega$.}\vspace{12pt}

\begin{center}\refstepcounter{figure}
\includegraphics[width=120mm]{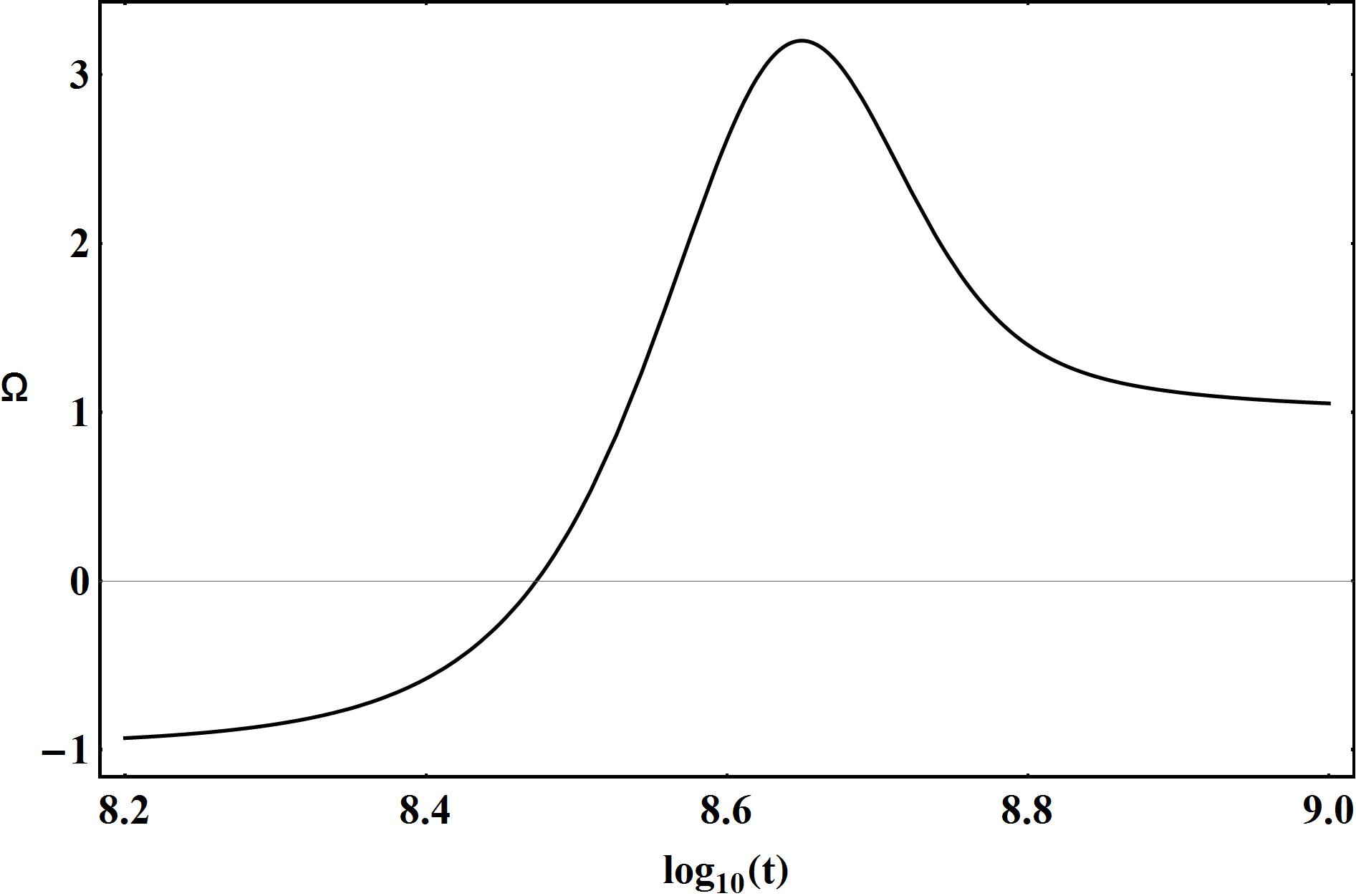}\end{center}
{\small \textbf{Fig.  \thefigure}. \label{Fig24} Detailed structure of the first fantom burst of the invariant cosmological acceleration show on Fig. 23:
$\Omega$; $m_s = 10^{-4}$, $q = 0.1$.\vspace{12pt}\vspace{12pt}

\begin{center}\refstepcounter{figure}
\includegraphics[width=120mm]{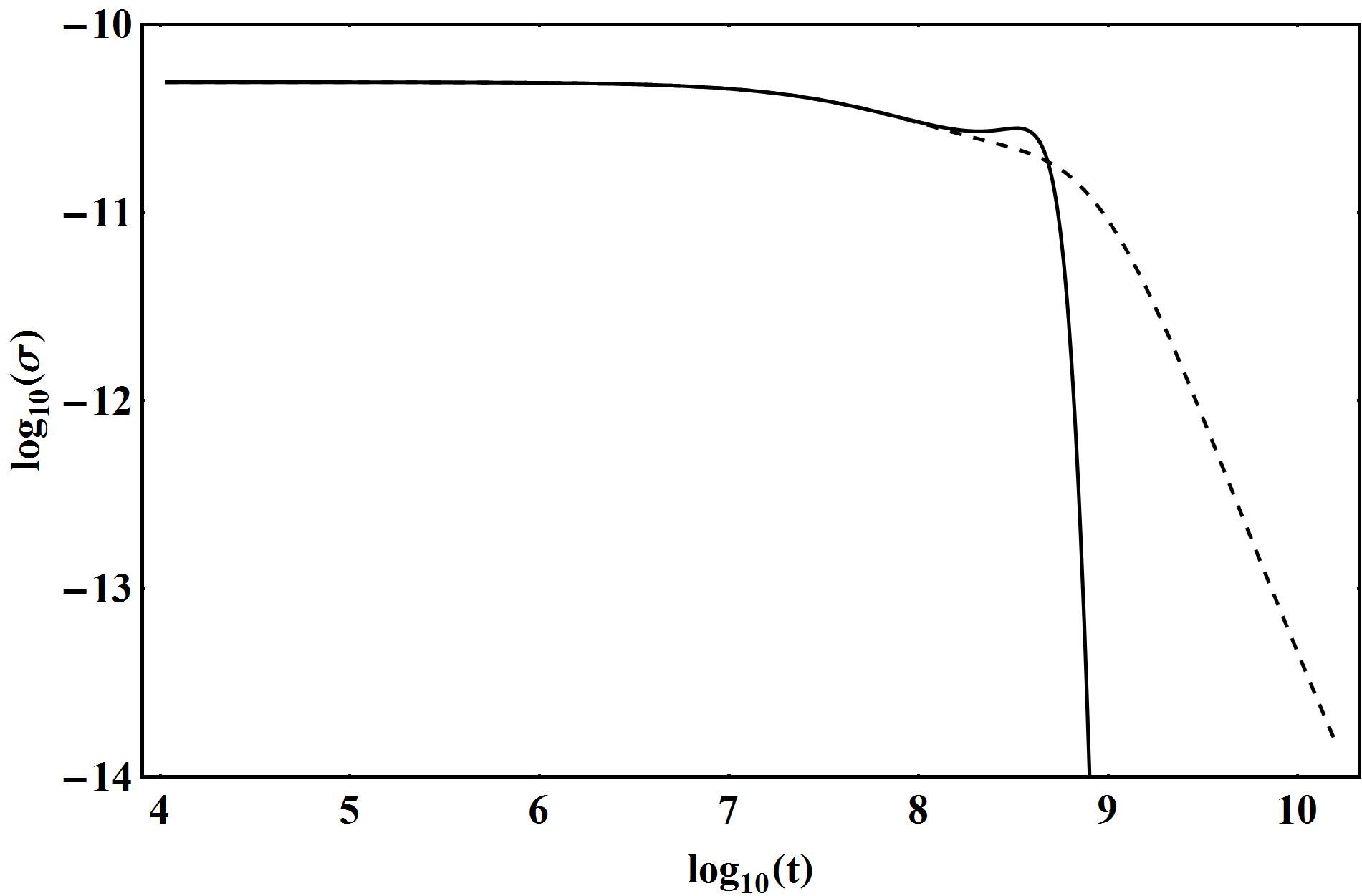}
\end{center}
{\small \textbf{Fig.  \thefigure}. \label{Fig25} The evolution of logarithm of the scalar charge's density $\log_{10}\sigma$.\vspace{12pt}}

\begin{center}\refstepcounter{figure}
\includegraphics[width=120mm]{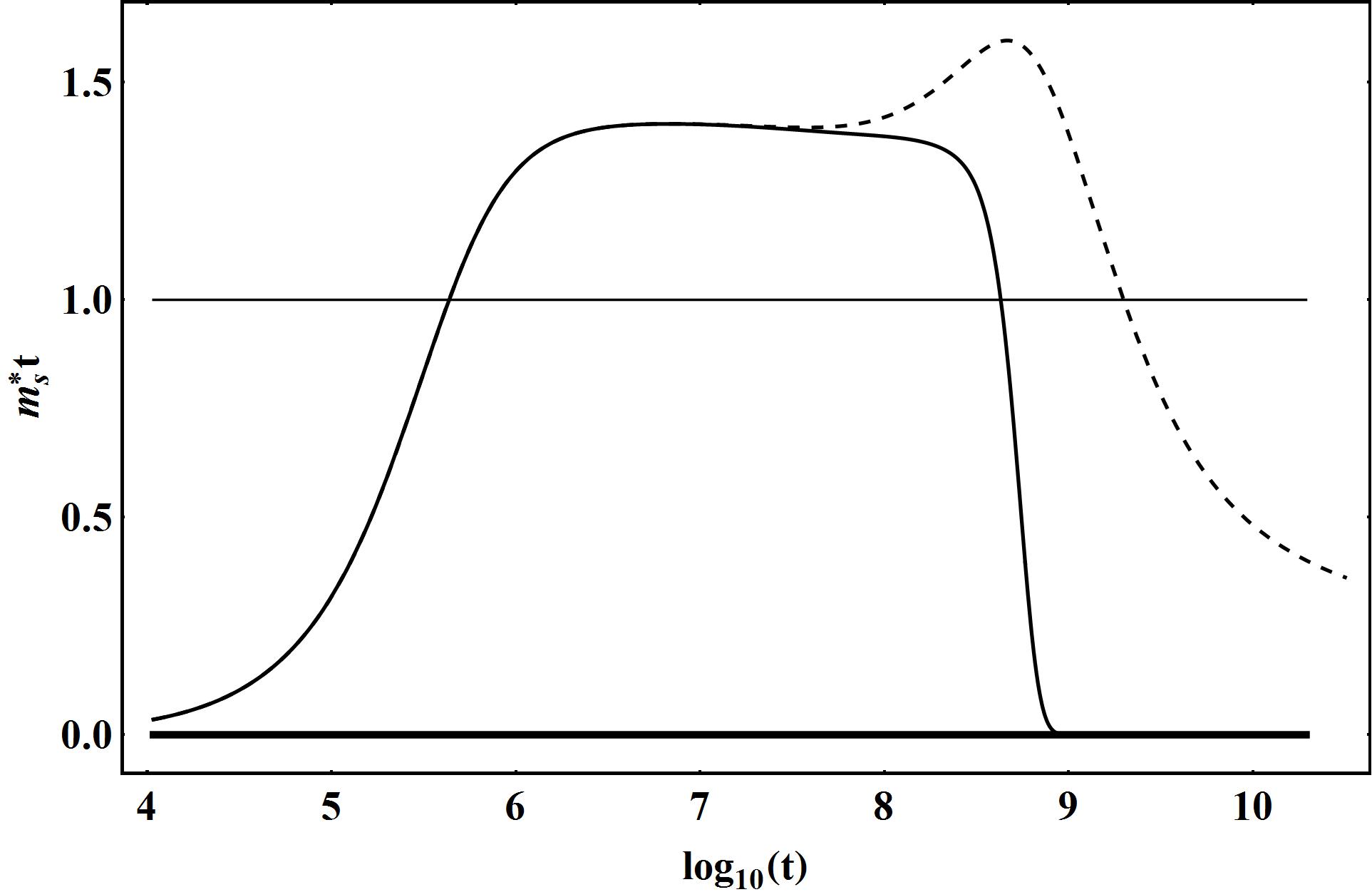}
\end{center}
{\small \textbf{Fig.  \thefigure}. \label{Fig26} The plot of the function $m_s^* t$\vspace{12pt}}

\section{The Characteristic Examples of the Case of Massive Fantom Scalar Field ($m_s\not=0$) with the Source ($\sigma\not=0$)}

\subsection{The Case of Transition from the Relativistic Stage to Inflation Through Non-Relativistic Plateau}

At small values of $m_s$ transition to inflation stage is shifted to late times  and Fermi system has time to become non-relativistic. On the plot of barotropic coefficient a characteristic non-relativistic plateau appears ($\kappa=0$) before fantom burst. At increase of $m_s$ the burst is shifted to early times trimming plateau till its complete vanishing. In this case a transition from relativistic stage to inflation happens through fantom burst.

Let us show the plots of the system's numerical simulation with the following parameters:\\ $p_0 = 1$, $m_0 = 0.1$, $q = 0.001$, $\Phi(0) = 0.05$. Heavy line is $m_s = 10^{-8}$, thin line is $m_s = 10^{-6}$, normal dotted line is $m_s = 10^{-3}$, fine dotted line is $m_s = 10^{-1}$.

\begin{center}\refstepcounter{figure}
\includegraphics[width=120mm]{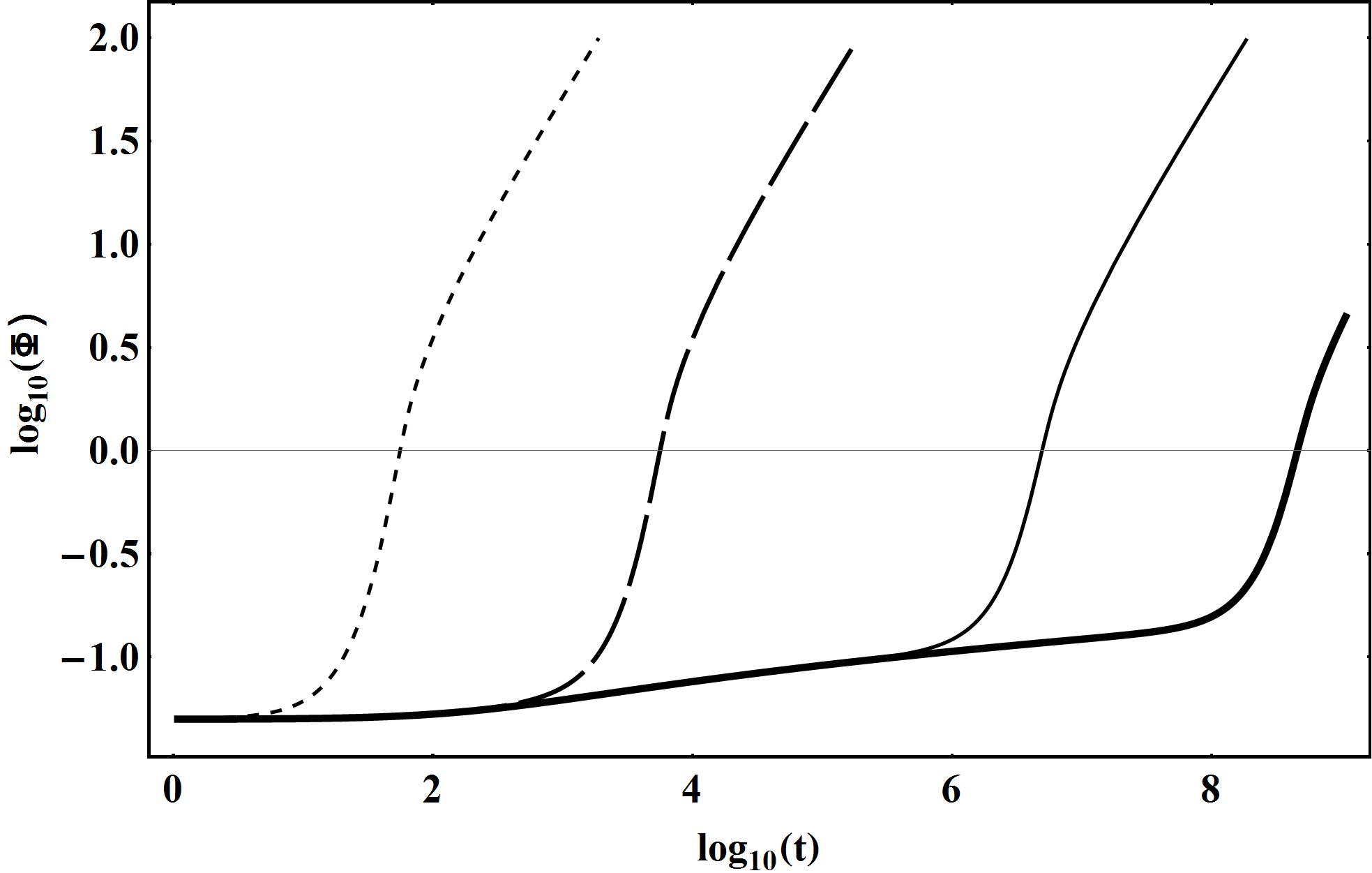}
\end{center}
{\small \textbf{Fig.  \thefigure}. \label{Fig27} The evolution of the potential's logarithm $\log_{10}\Phi$.\vspace{12pt}}

\begin{center}\refstepcounter{figure}
\includegraphics[width=120mm]{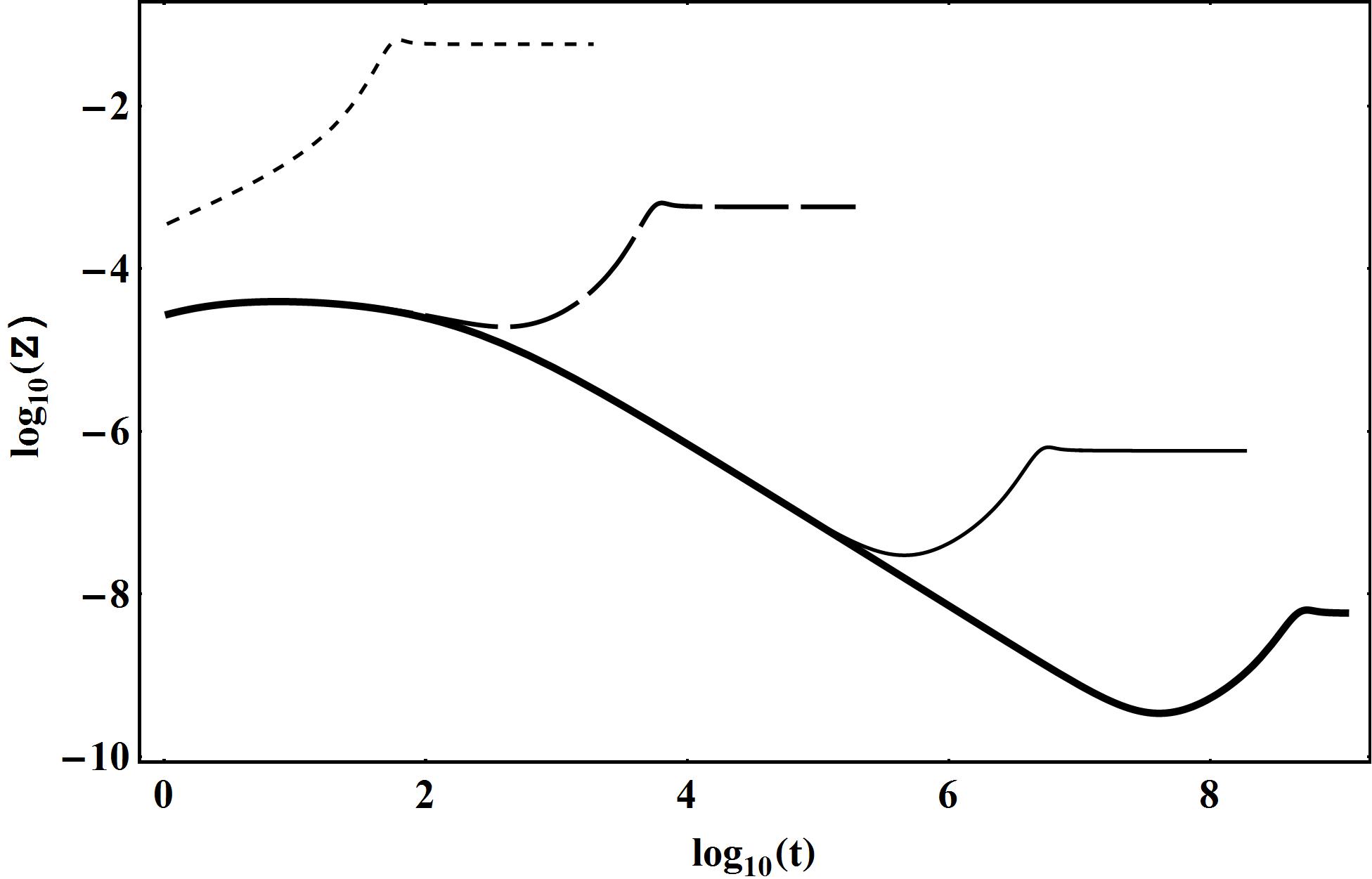}
\end{center}
{\small \textbf{Fig.  \thefigure}.\label{Fig28} The evolution of logarithm of the potential's derivative $\log_{10} Z=\log_{10} \dot{\Phi}$.\vspace{12pt}}

\begin{center}\refstepcounter{figure}
\includegraphics[width=120mm]{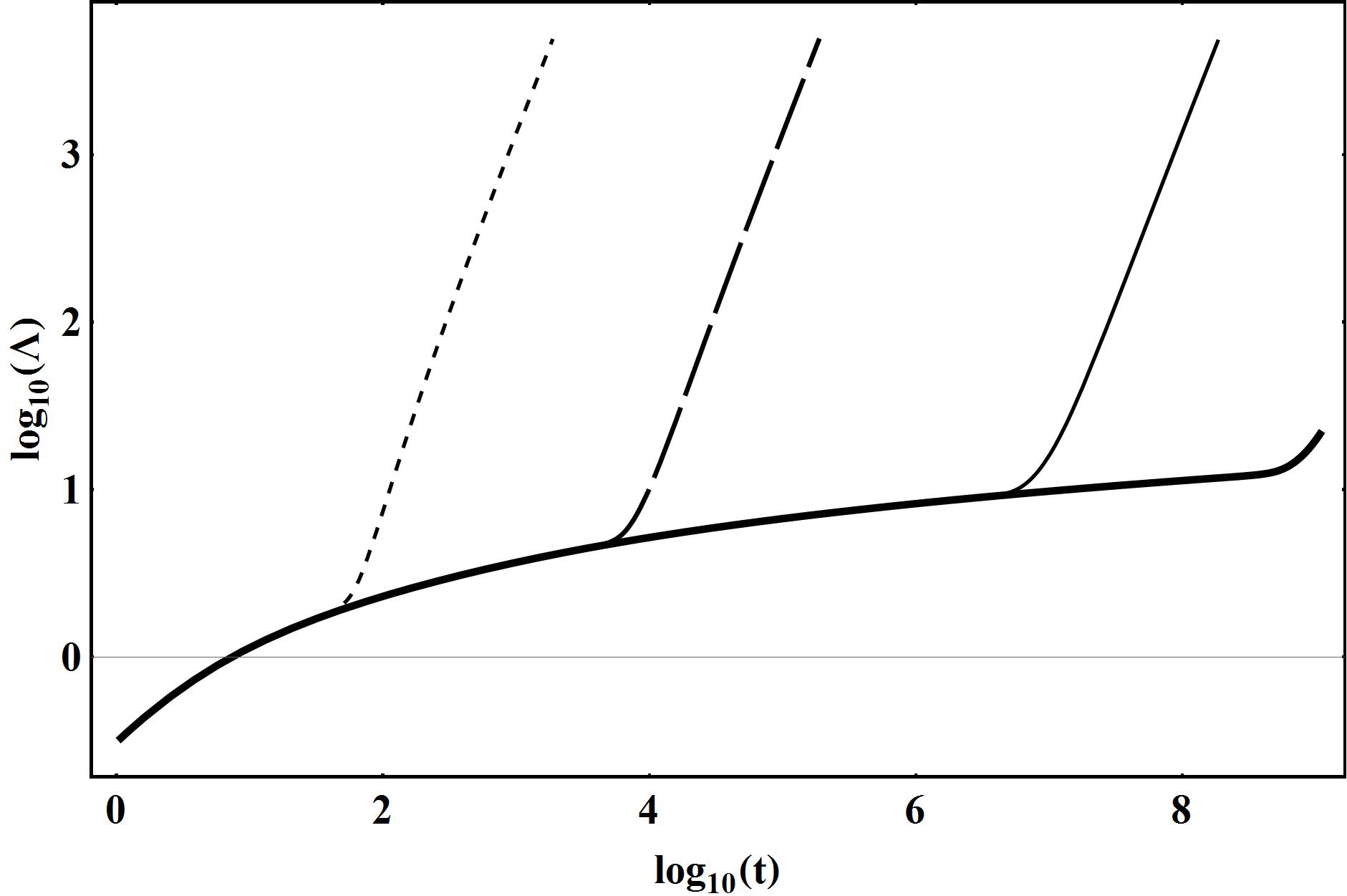}
\end{center}
{\small \textbf{Fig.  \thefigure}.\label{Fig29} The evolution of logarithm of the scale function $\log_{10}\Lambda(t)$.\vspace{12pt}}

\begin{center}\refstepcounter{figure}
\includegraphics[width=120mm]{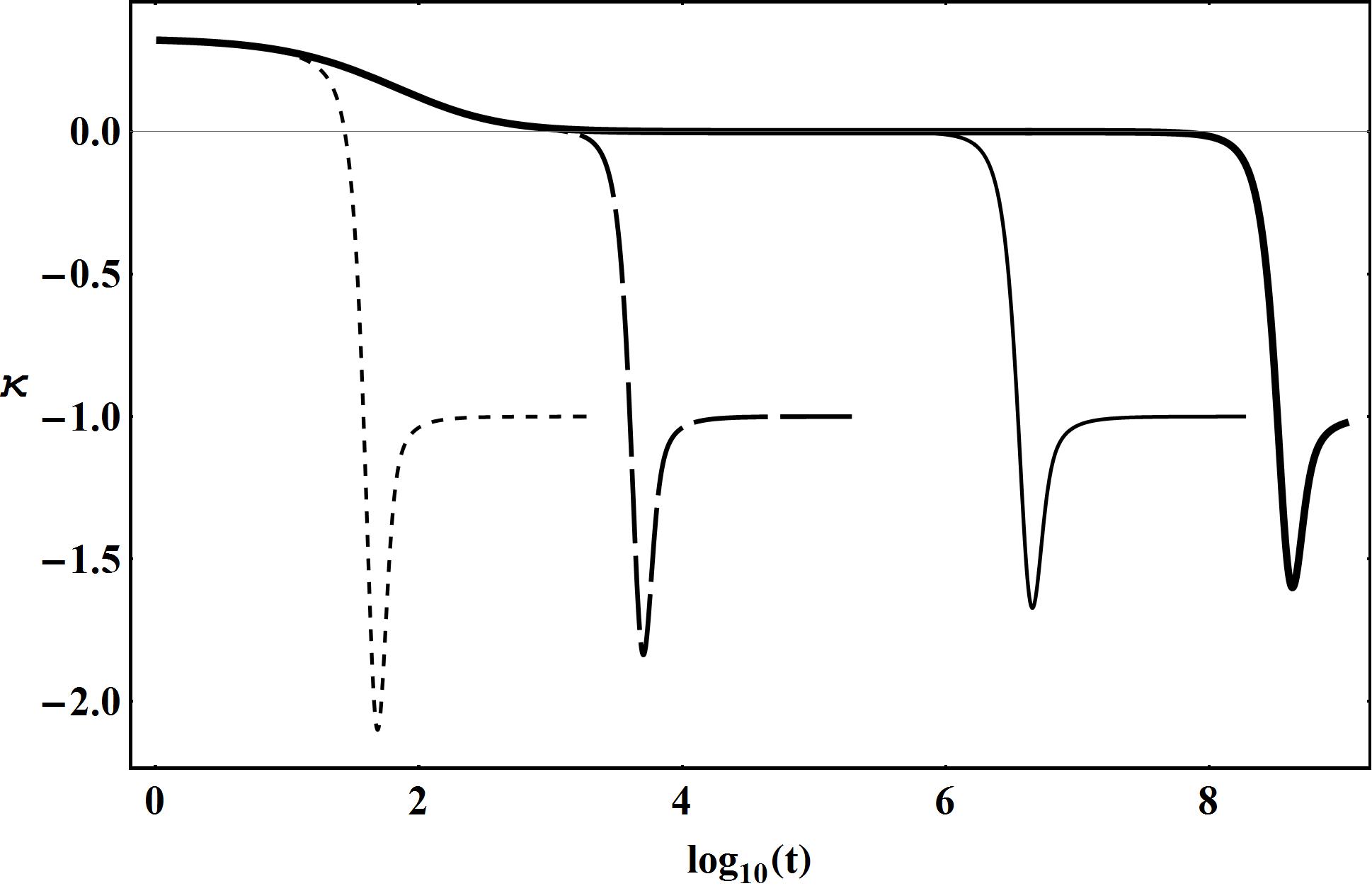}
\end{center}
{\small \textbf{Fig.  \thefigure}.\label{Fig30} The evolution of barotropic coefficient $\kappa$.\vspace{12pt}}

\begin{center}\refstepcounter{figure}
\includegraphics[width=120mm]{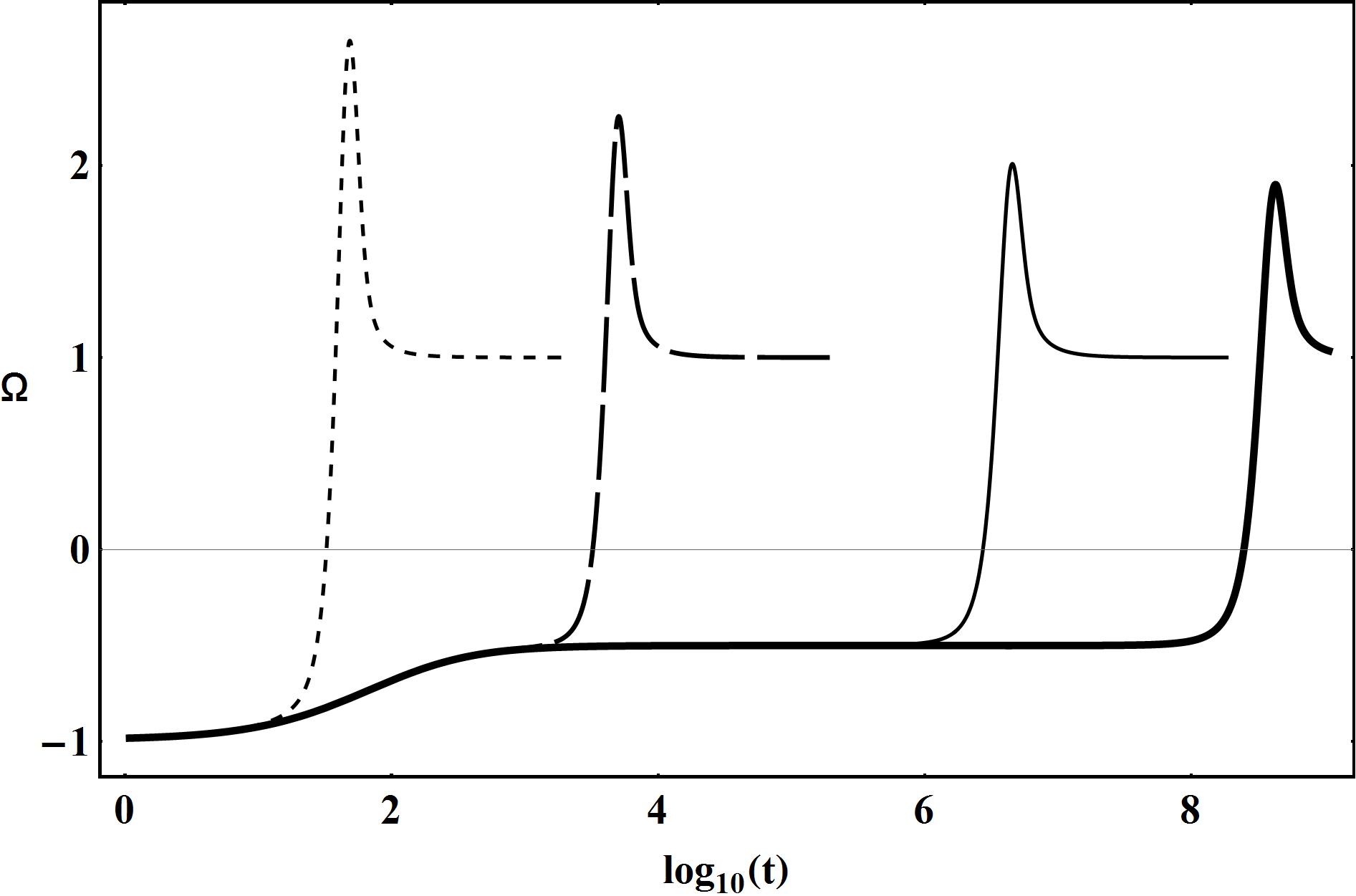}
\end{center}
{\small \textbf{Fig.  \thefigure}.\label{Fig31} The evolution of the invariant cosmological acceleration %
$\Omega$.\vspace{12pt}}

\begin{center}\refstepcounter{figure}
\includegraphics[width=120mm]{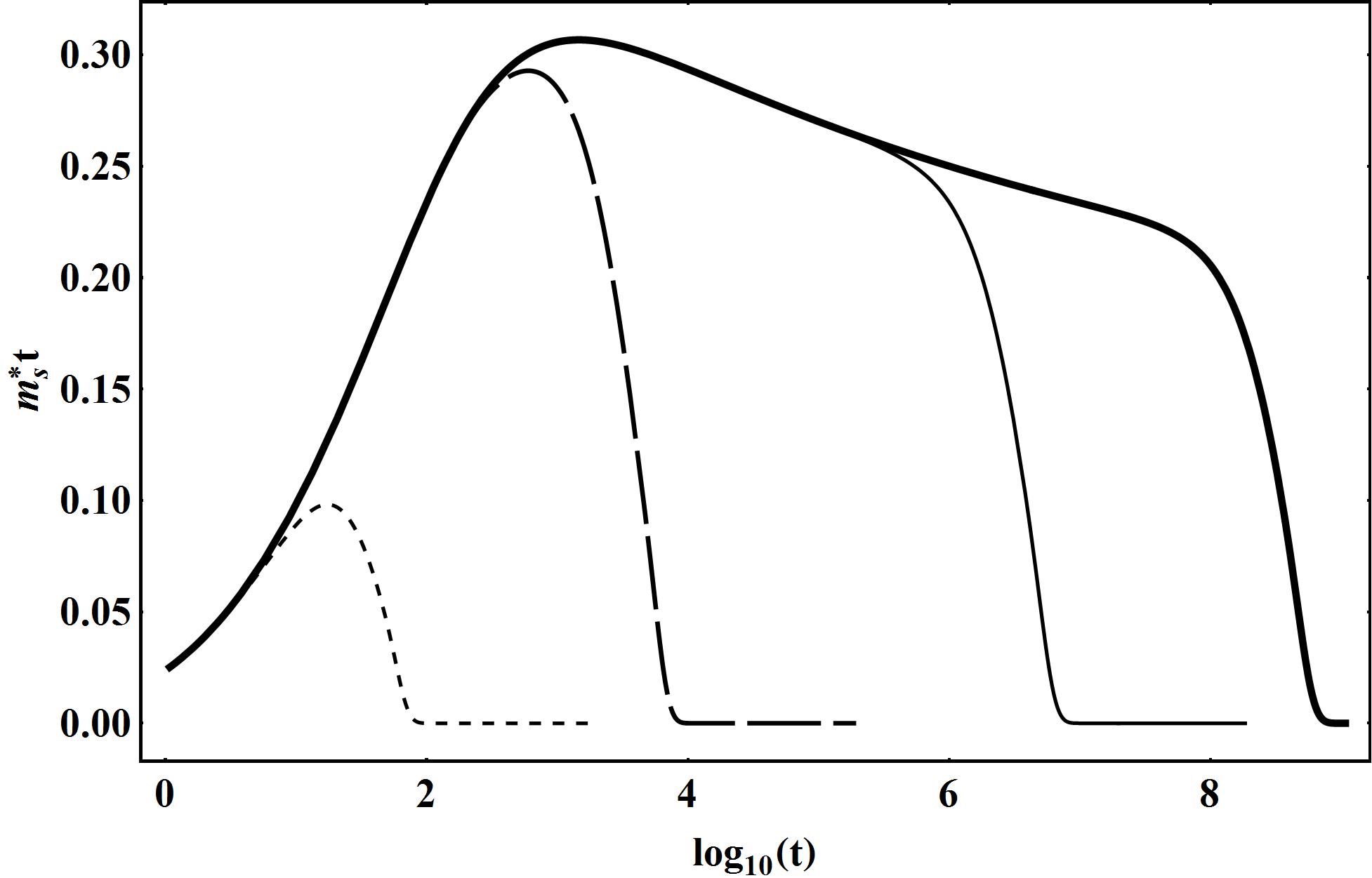}
\end{center}
{\small \textbf{Fig.  \thefigure}.\label{Fig32} The evolution of function $m_s^* t$\vspace{12pt}}

\subsection{The Case of Two Stages of Acceleration with an Intermediate Non-Relativistic Stage}

At certain values of the system parameters we can observe two stages of acceleration: first is a temporary prevalence of the scalar field due to sufficiently large values of the scalar field's mass $m_s^*$ at corresponding times ($m_s^* t > 1$) and second is a stable prevalence of the scalar field at times $m_s t > 1$. There can appear a non-relativistic plateau between inflation stages ($\kappa=0,\ \Omega=-1/2$).

\subsubsection{Dependency on the Scalar Charge}

Changing a value of the scalar charge $q$ at constant value of the scalar field's mass $m_s$, we can observe a shift of the first inflation burst and corresponding change of non-relativistic stage duration.

Let us see plots of the system's numerical simulation with the following parameters:\\
 $p_0 = 1$, $m = 0$, $m_s = 10^{-8}$, $\Phi(0) = 5\cdot 10^{-8}$. Heavy line is $q = 1$, thin line is $q = 0.1$, normal dotted line is $q = 0.01$, fine dotted line is $q = 0.001$.
\vspace{12pt}

\begin{center}\refstepcounter{figure}
\includegraphics[width=120mm]{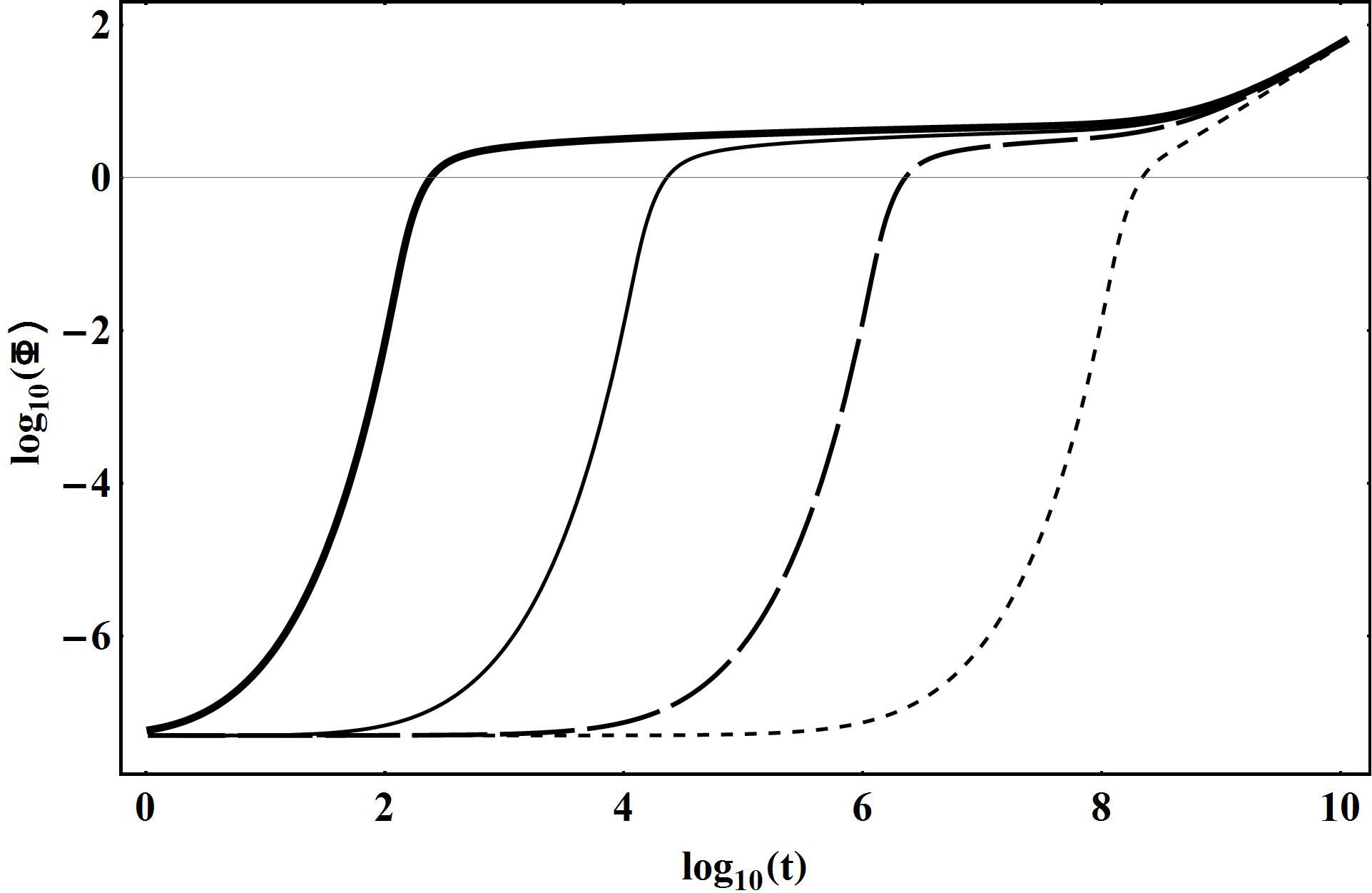}
\end{center}
{\small \textbf{Fig.  \thefigure}.\label{Fig33} The evolution of logarithm of the potential $\log_{10}\Phi$.\vspace{12pt}}

\begin{center}\refstepcounter{figure}
\includegraphics[width=120mm]{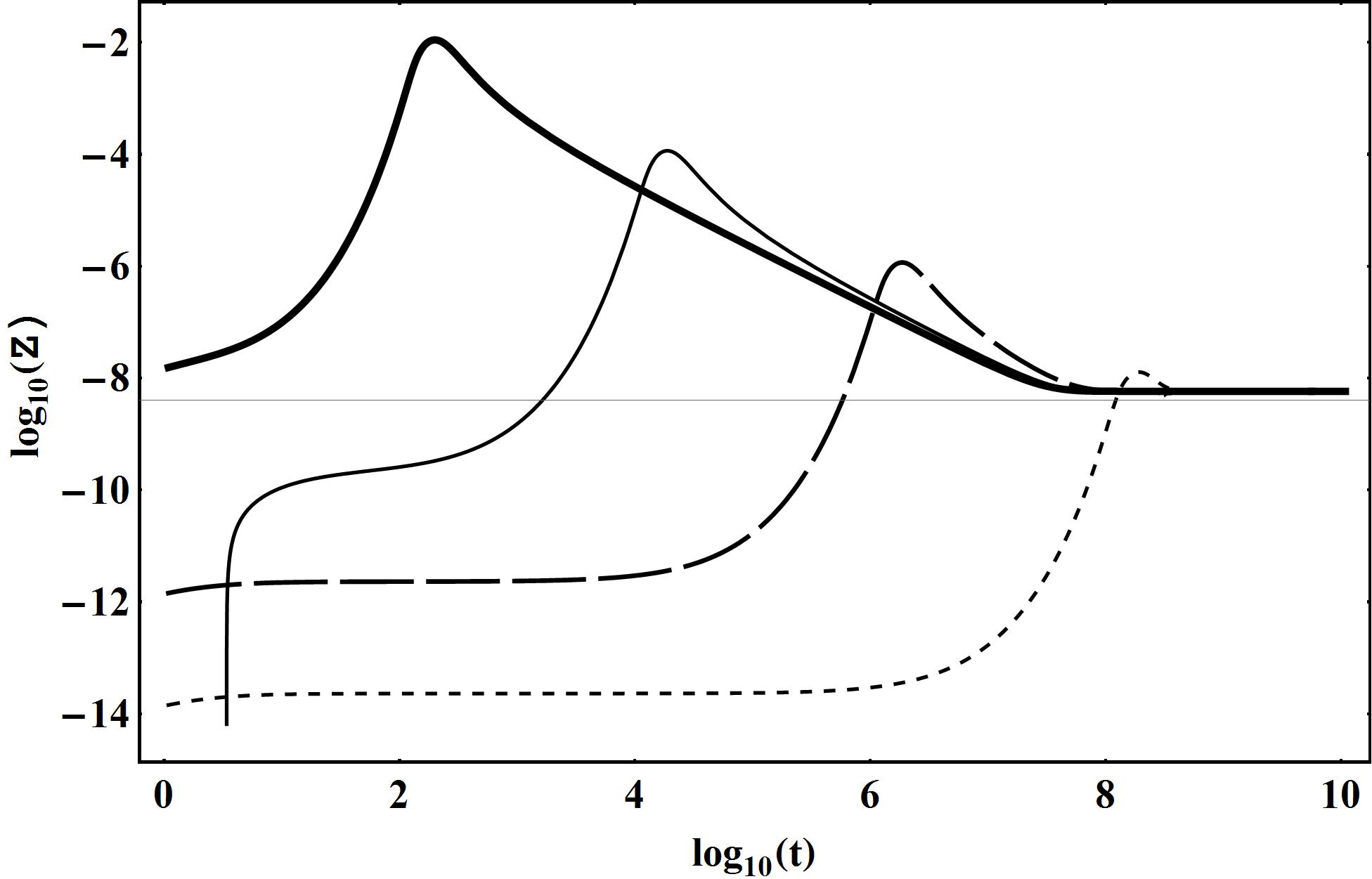}
\end{center}
{\small \textbf{Fig.  \thefigure}.\label{Fig34} The evolution of logarithm of the potential's derivative $\log_{10} Z=\log_{10} \dot{\Phi}$.}\vspace{12pt}}

\begin{center}\refstepcounter{figure}
\includegraphics[width=120mm]{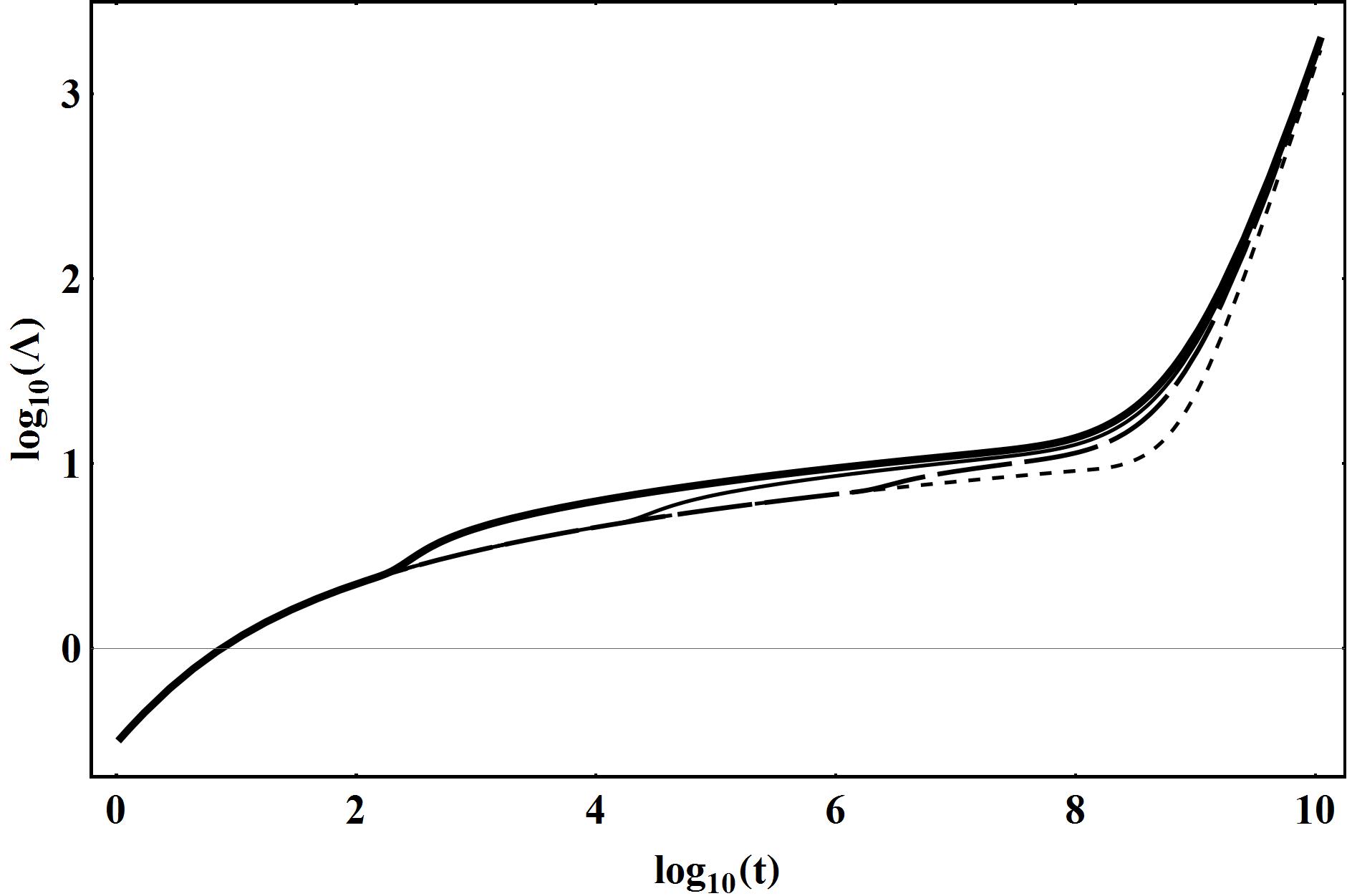}
\end{center}
{\small \textbf{Fig.  \thefigure}.\label{Fig35} The evolution of logarithm of the scale function $\log_{10}\Lambda(t)$.\vspace{12pt}}

\begin{center}\refstepcounter{figure}
\includegraphics[width=120mm]{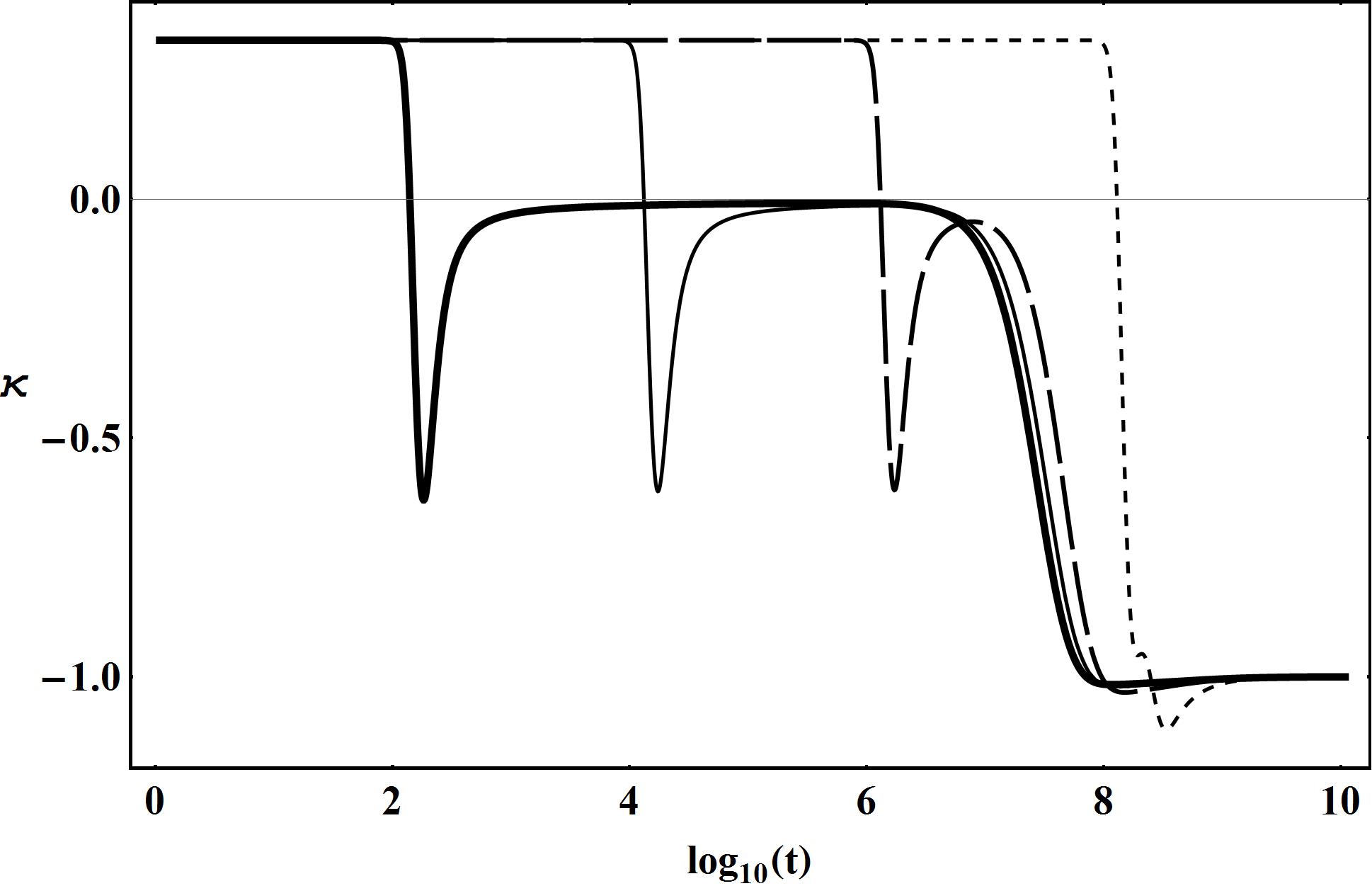}
\end{center}
{\small \textbf{Fig.  \thefigure}.\label{Fig36} The evolution of barotropic coefficient $\kappa$.\vspace{12pt}}

\begin{center}\refstepcounter{figure}
\includegraphics[width=120mm]{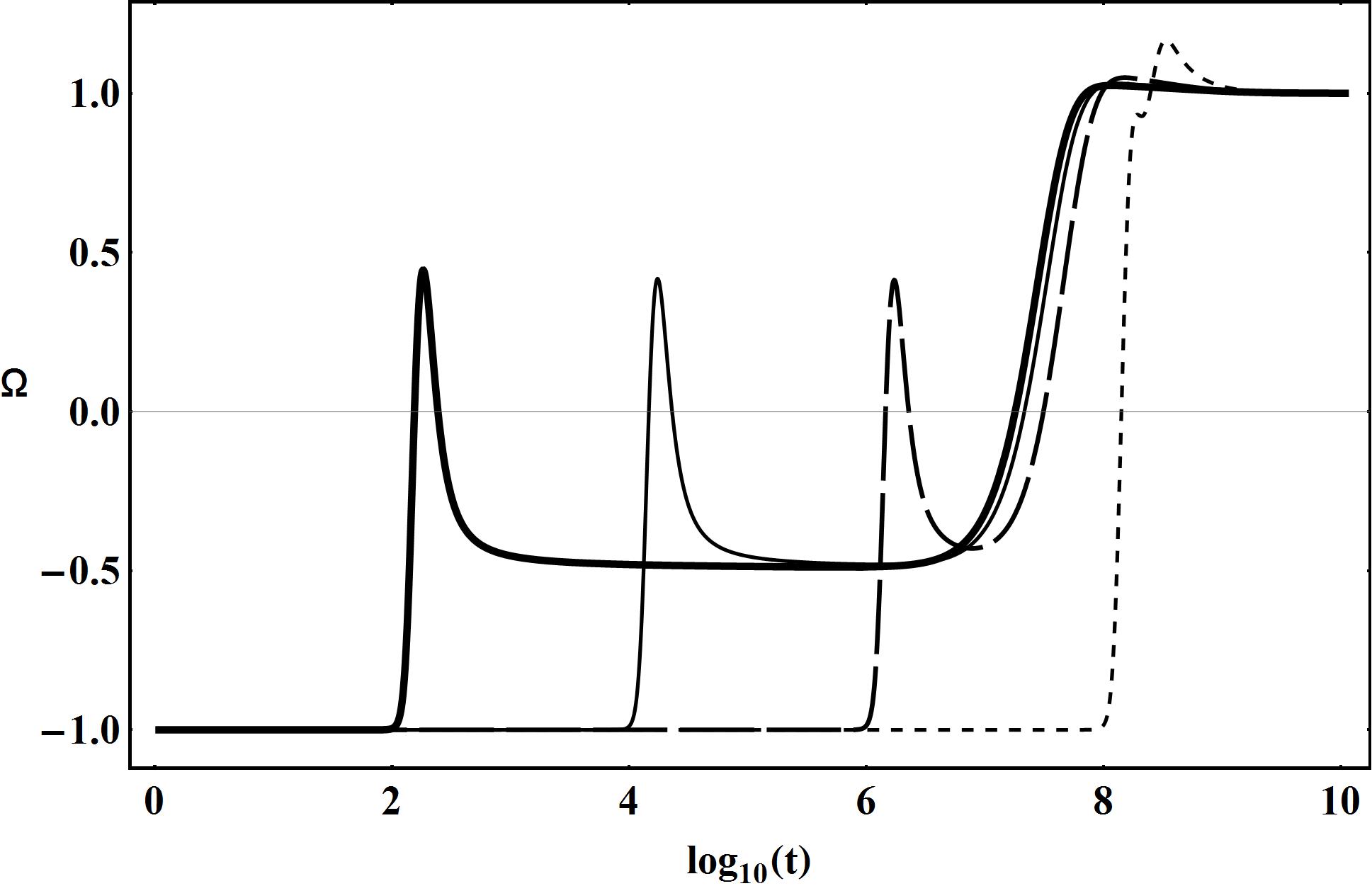}
\end{center}
{\small \textbf{Fig.  \thefigure}.\label{Fig37} The evolution of the invariant cosmological acceleration %
$\Omega$.\vspace{12pt}}

\begin{center}\refstepcounter{figure}
\includegraphics[width=120mm]{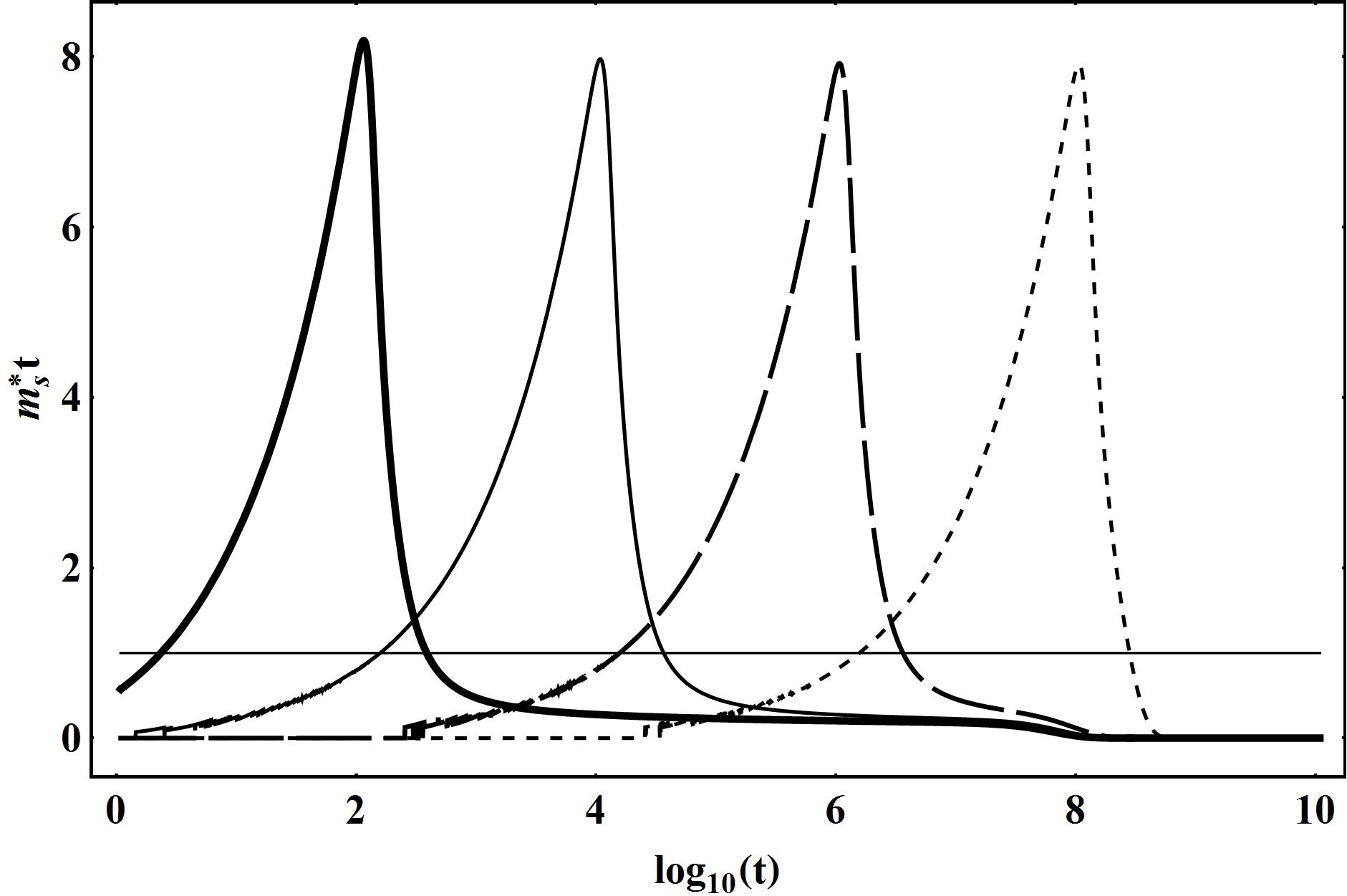}
\end{center}
{\small \textbf{Fig.  \thefigure}.\label{Fig38} The plot of the function $m_s^* t$\vspace{12pt}}

\subsubsection{Dependency of the Cosmological Evolution on Initial Scalar Field's Potential}

Changing the initial value of the scalar field's potential $\Phi_0$, we can observe a shift of the first inflation stage and change of amplitude of the cosmological acceleration's burst.\\[12pt]
\emph{Small initial values of the scalar field potential}$\Phi_0=10^{-4}\div 10^{-14}$.\\
Let us show the plots of numerical simulation of the system with the following parameters:\\
$p_0 = 1$, $m = 0$, $m_s = 10^{-8}$, $q = 1$. Heavy line is $\Phi(0) = 10^{-4}$, thin line is $\Phi(0) = 10^{-8}$, normal dotted line is $\Phi(0) = 10^{-14}$, fine dotted line is $\Phi(0) = 10^{-20}$.

\begin{center}\refstepcounter{figure}
\includegraphics[width=120mm]{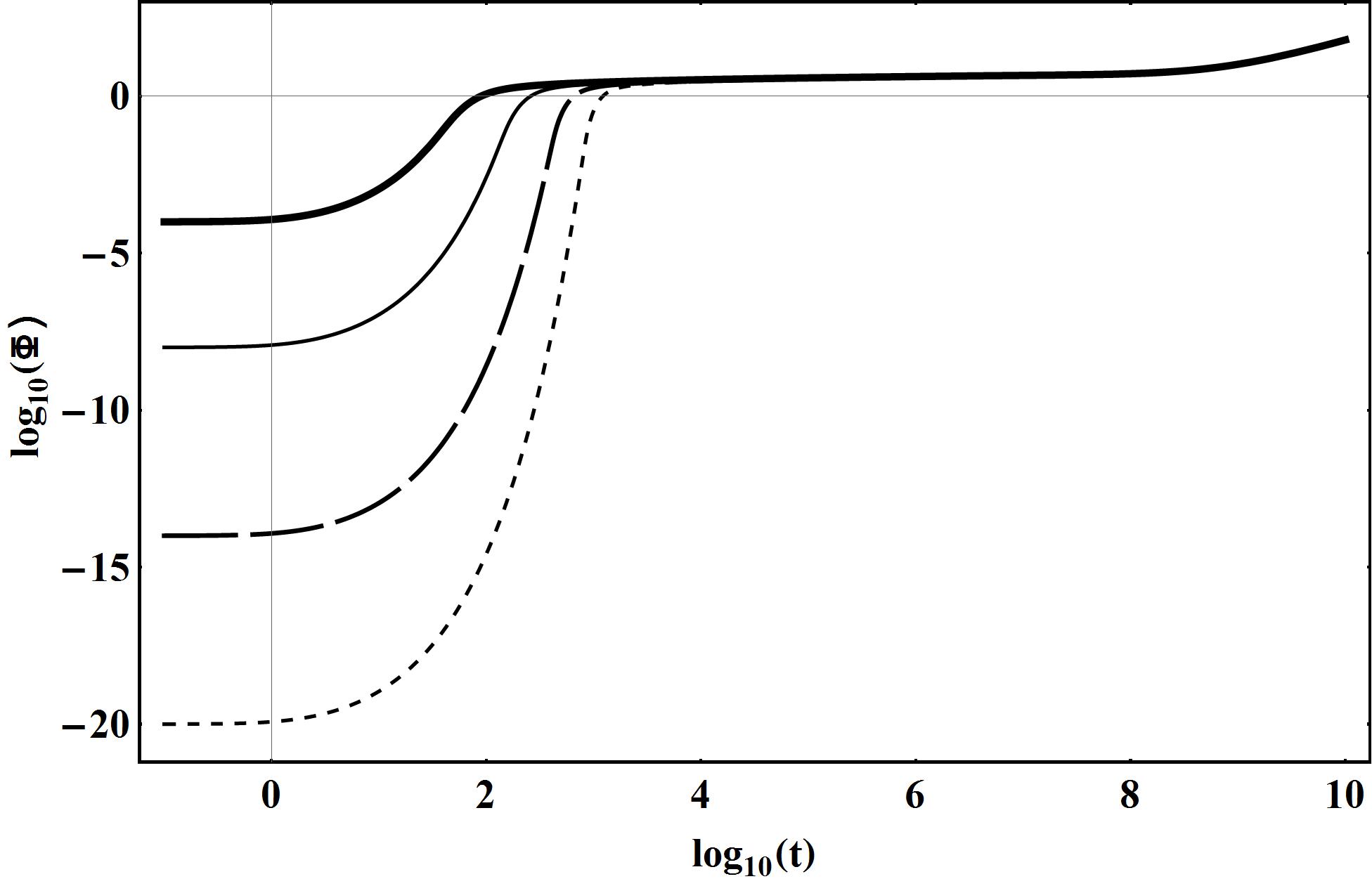}
\end{center}
{\small \textbf{Fig.  \thefigure}.\label{Fig39} The evolution of the potential's logarithm $\log_{10}\Phi$.\vspace{12pt}}

\begin{center}\refstepcounter{figure}
\includegraphics[width=120mm]{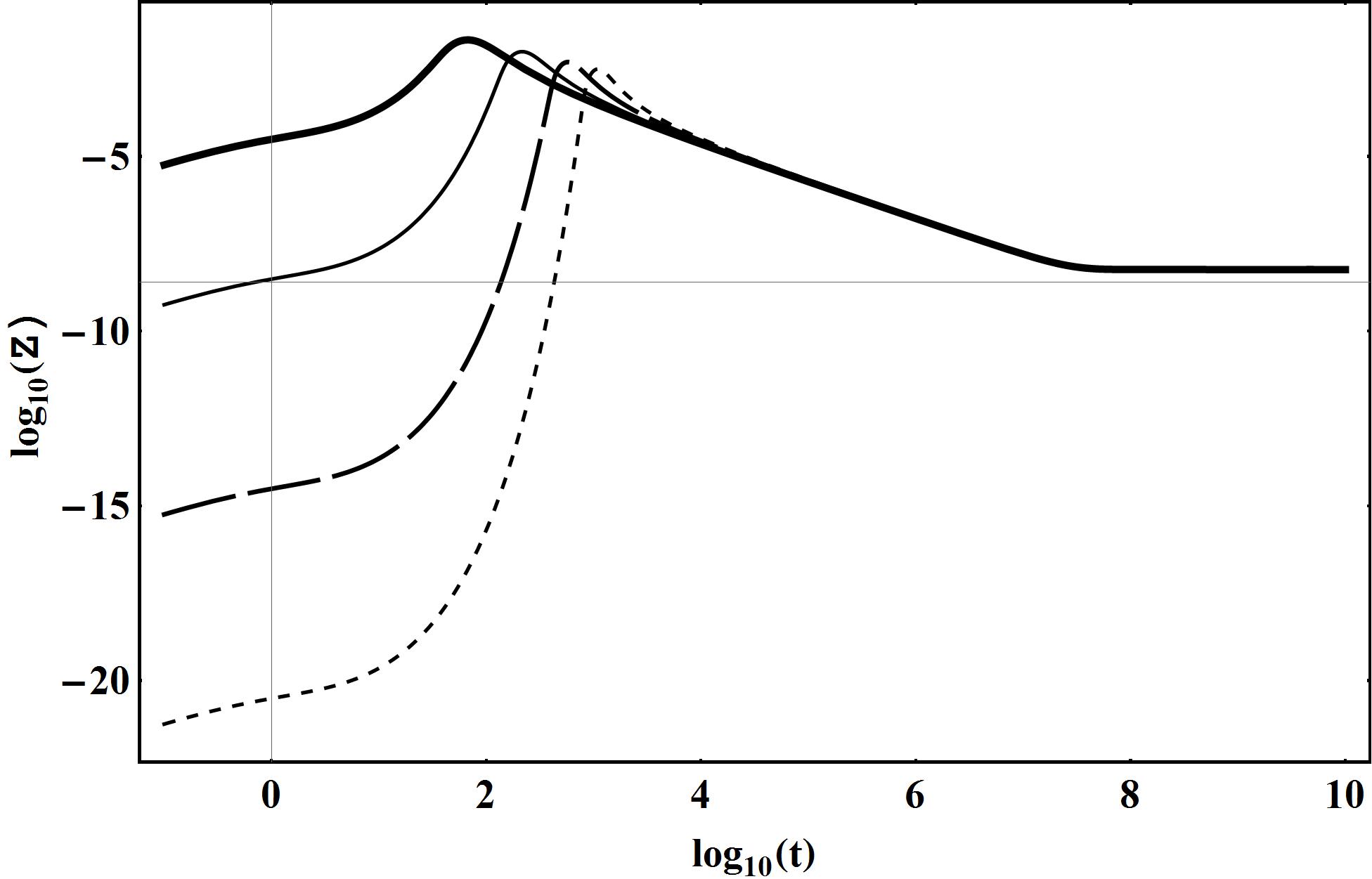}
\end{center}
{\small \textbf{Fig.  \thefigure}.\label{Fig40} The evolution of logarithm of the potential's derivative $\log_{10} Z=\log_{10} \dot{\Phi}$.\vspace{12pt}}

\begin{center}\refstepcounter{figure}
\includegraphics[width=120mm]{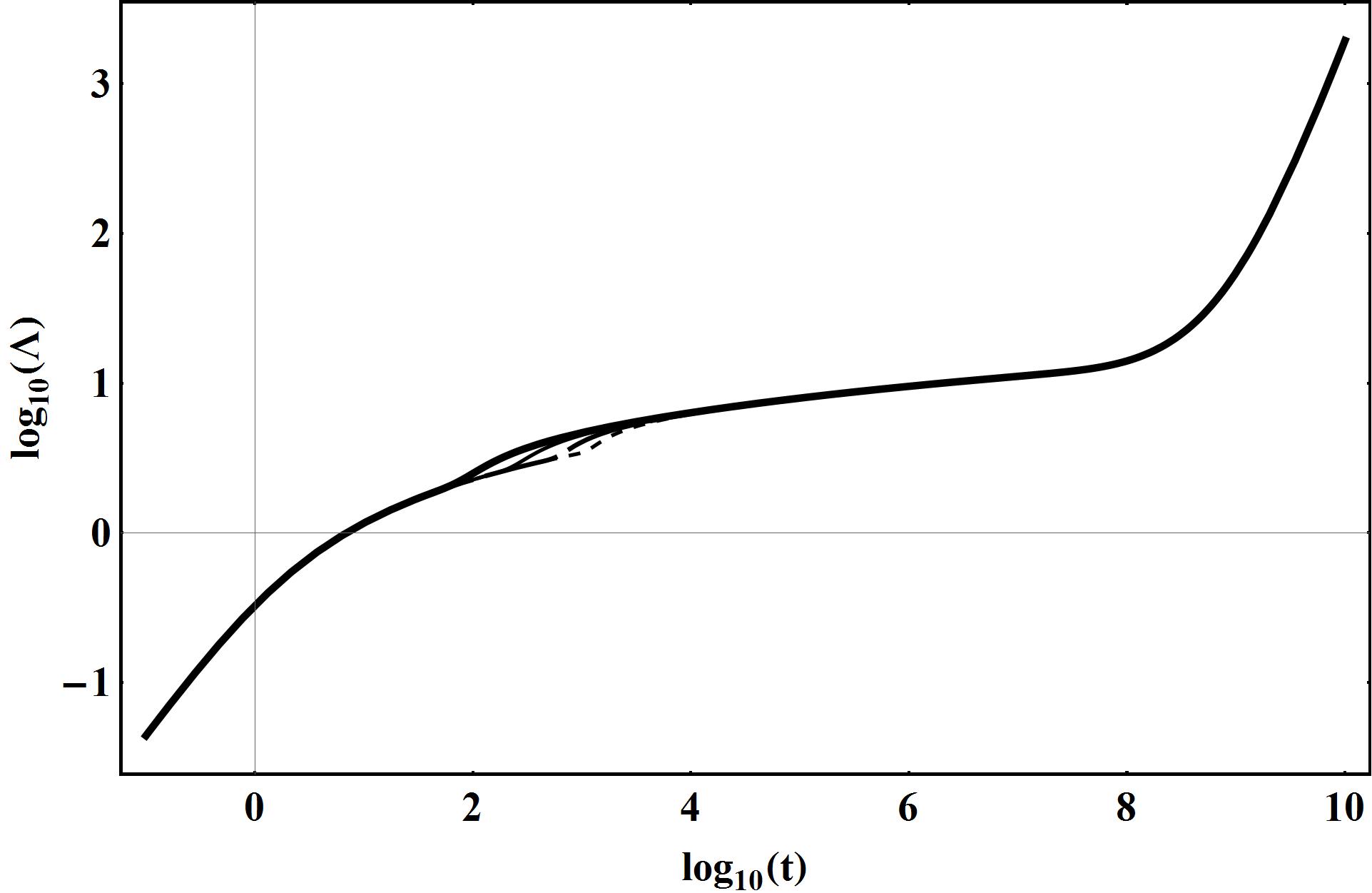}
\end{center}
{\small \textbf{Fig.  \thefigure}.\label{Fig41} The evolution of logarithm of the scale function $\log_{10}\Lambda(t)$.\vspace{12pt}}

\begin{center}\refstepcounter{figure}
\includegraphics[width=120mm]{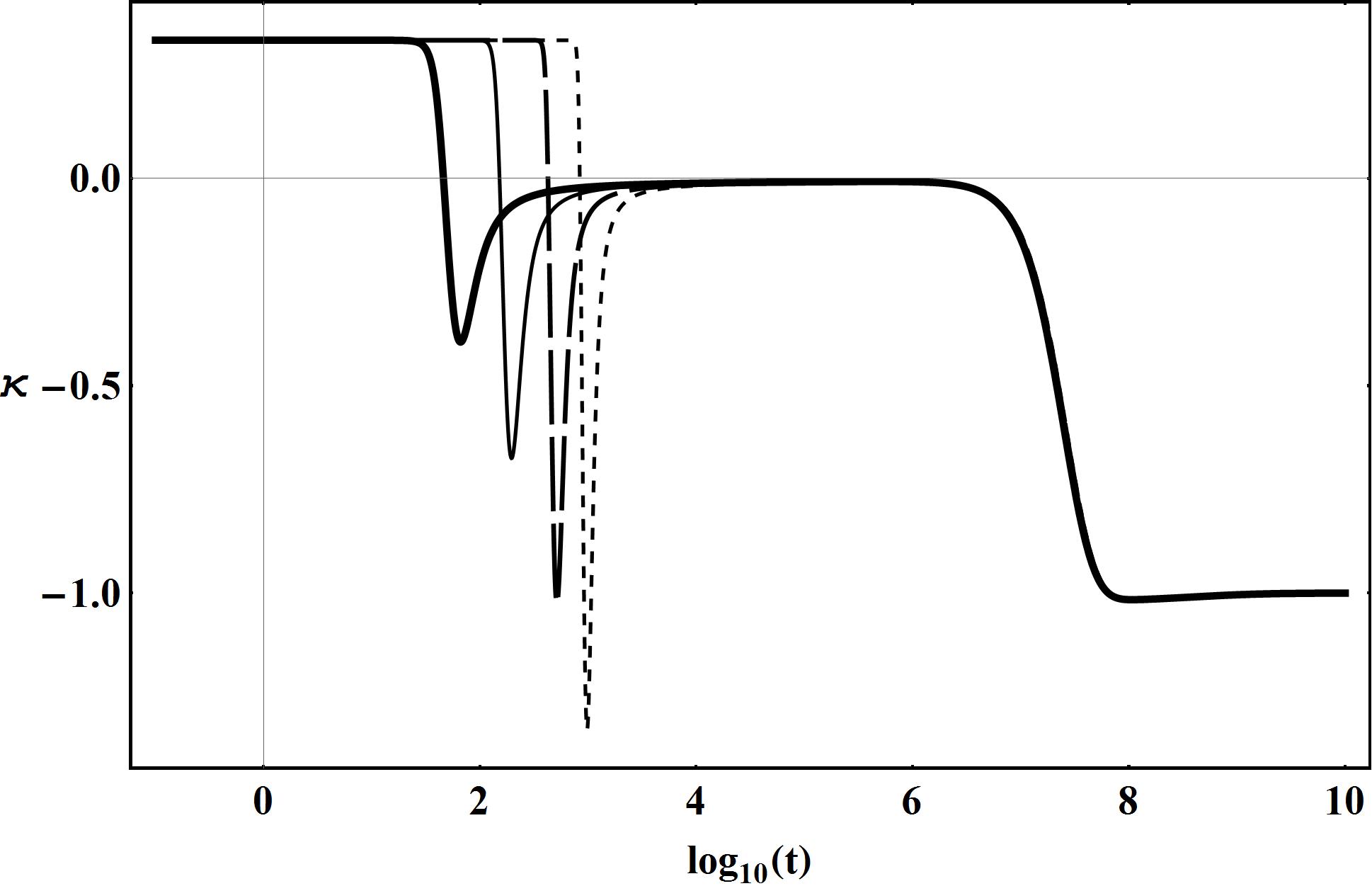}
\end{center}
{\small \textbf{Fig.  \thefigure}.\label{Fig42} The evolution of barotropic coefficient $\kappa$.\vspace{12pt}}

\begin{center}\refstepcounter{figure}
\includegraphics[width=120mm]{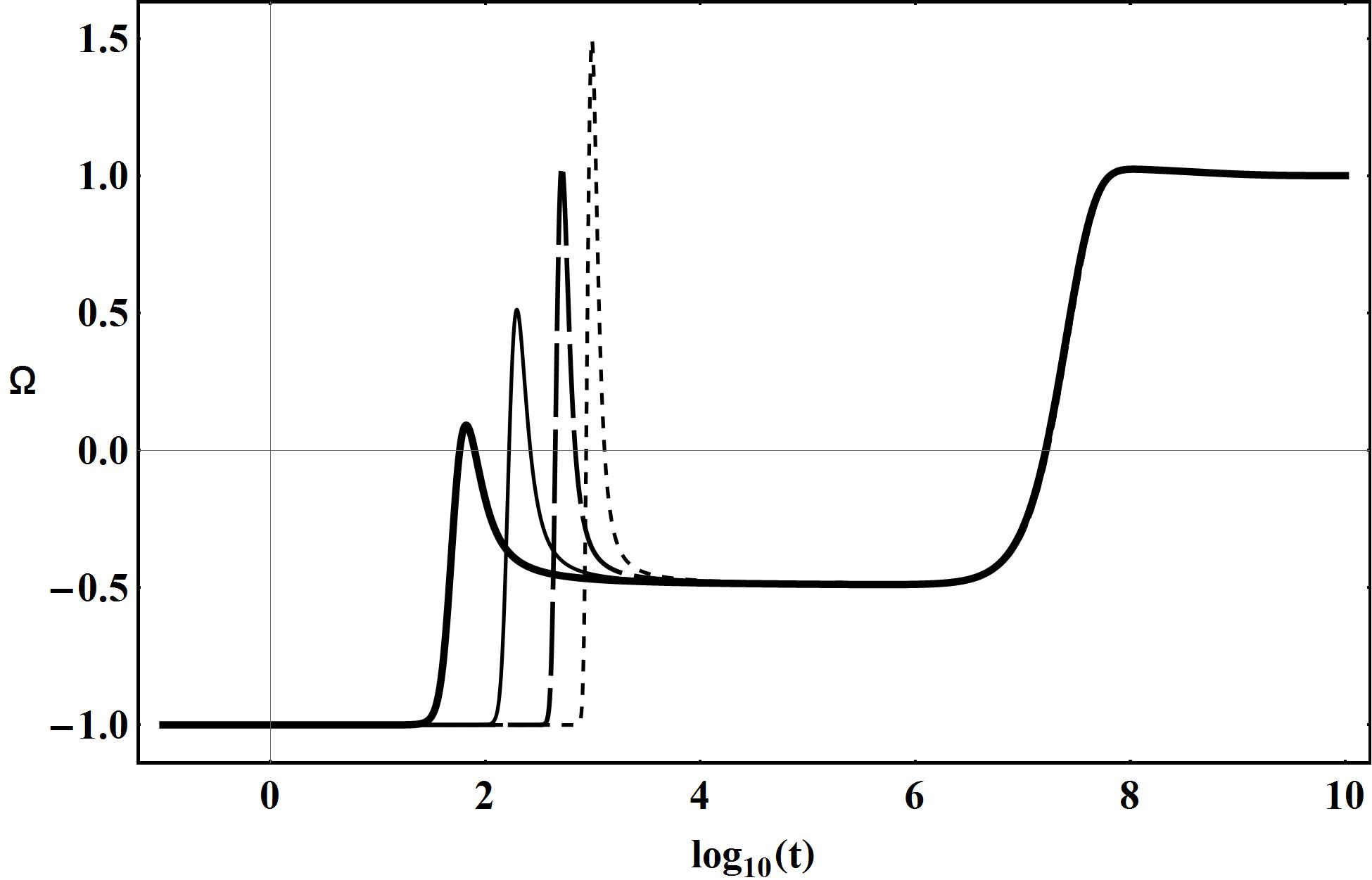}
\end{center}
{\small \textbf{Fig.  \thefigure}.\label{Fig43} The evolution of the invariant cosmological acceleration %
$\Omega$.\vspace{12pt}}

\begin{center}\refstepcounter{figure}
\includegraphics[width=120mm]{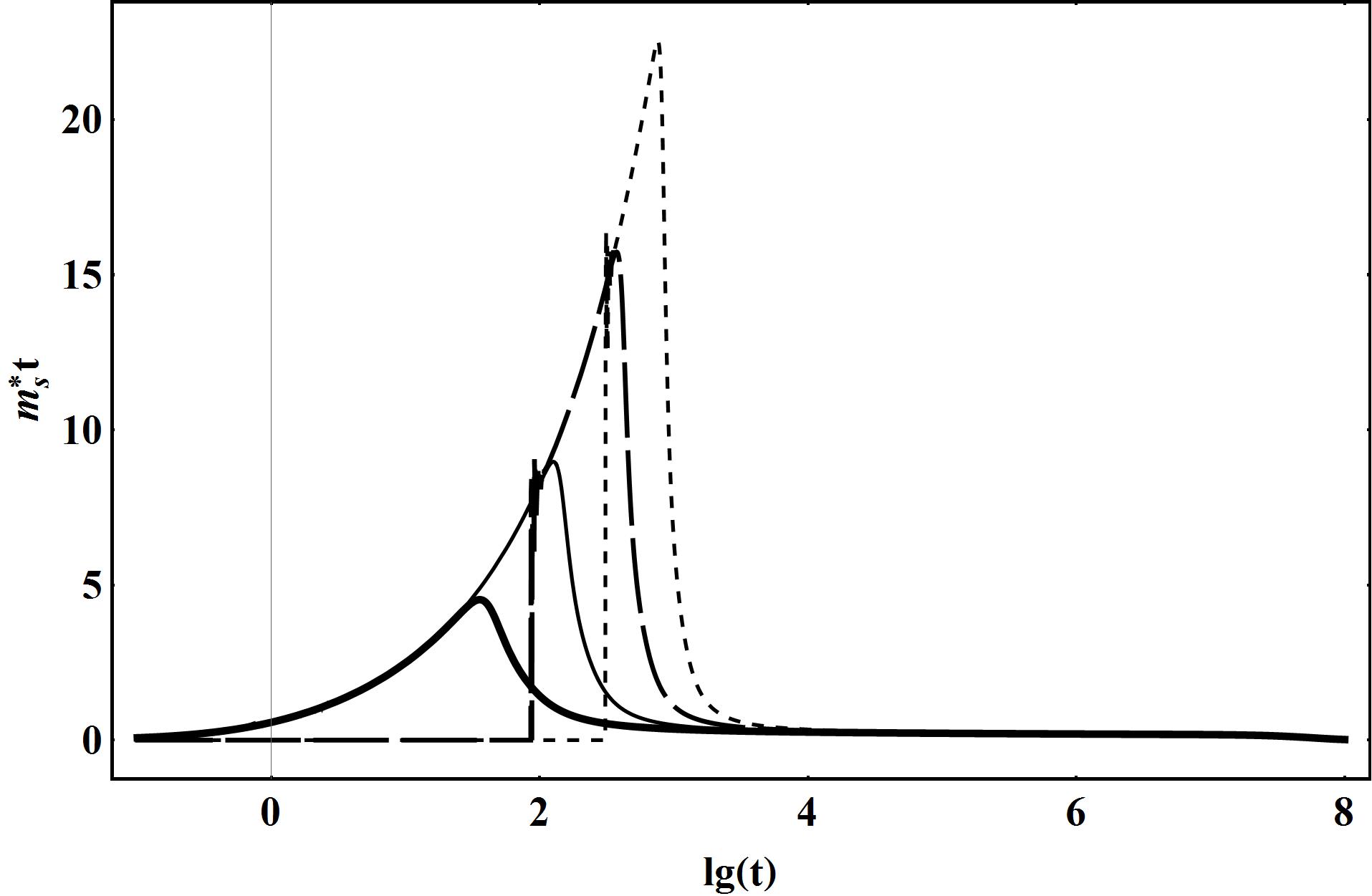}
\end{center}
{\small \textbf{Fig.  \thefigure}.\label{Fig44} The plot of the function $m_s^* t$\vspace{12pt}}

\noindent
\emph{Smaller initial values of the scalar field's potential} $\Phi_0=10^{-50}\div 10^{-300}$.\\[12pt]
Let us show the plots of numerical simulation of the system with the following parameters: $p_0 = .1$, $m = 0$, $m_s = 10^{-10}$, $q = 5$. Heavy black line is $\Phi(0) = 10^{-50}$, thin black line is $\Phi(0) = 10^{-150}$, normal dotted line is $\Phi(0) = 10^{-250}$, fine dotted line is $\Phi(0) = 10^{-300}$.

\begin{center}\refstepcounter{figure}
\includegraphics[width=120mm]{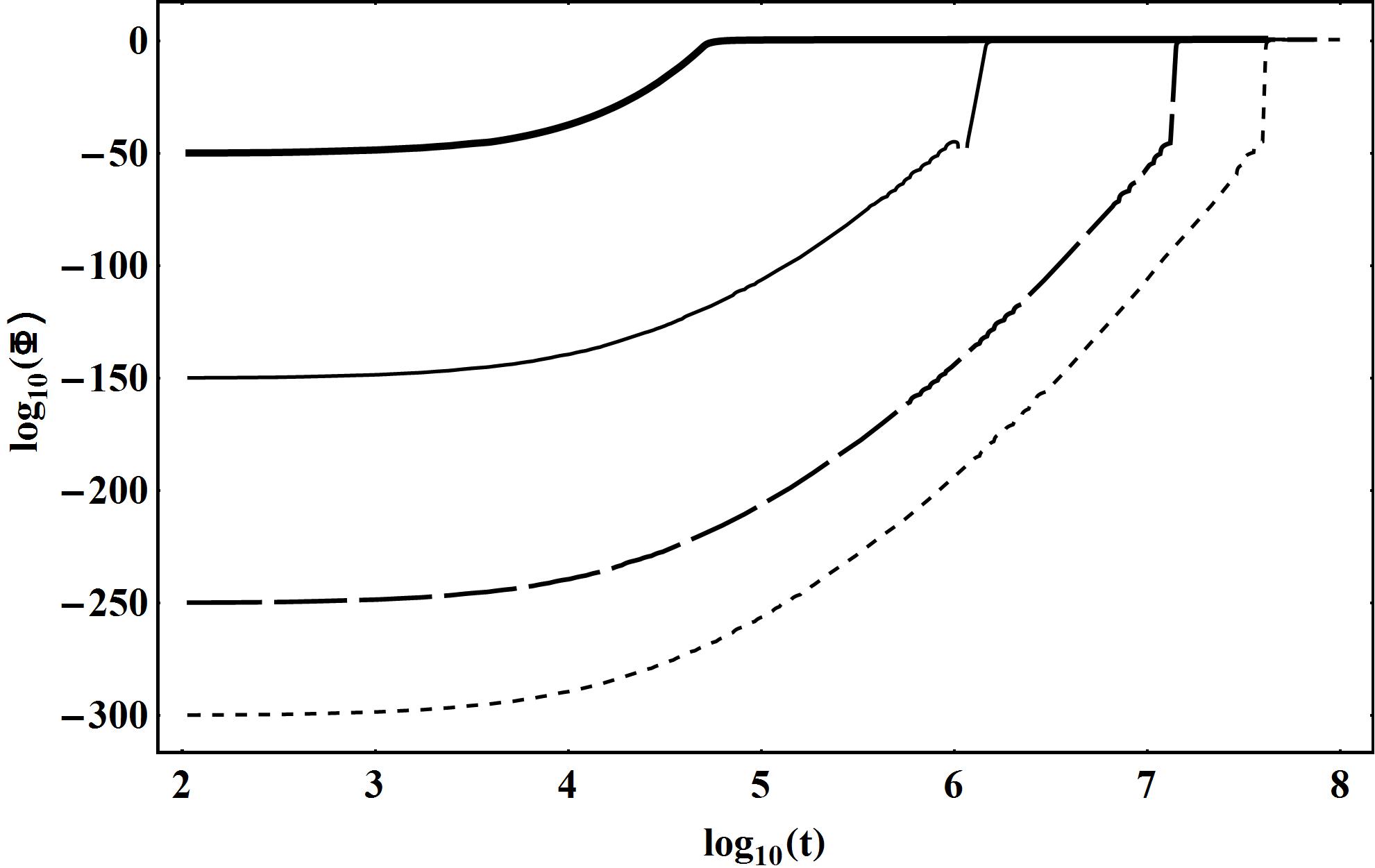}
\end{center}
{\small \textbf{Fig.  \thefigure}.\label{Fig45} The evolution of the potential's logarithm $\log_{10}\Phi$.\vspace{12pt}}

\begin{center}\refstepcounter{figure}
\includegraphics[width=120mm]{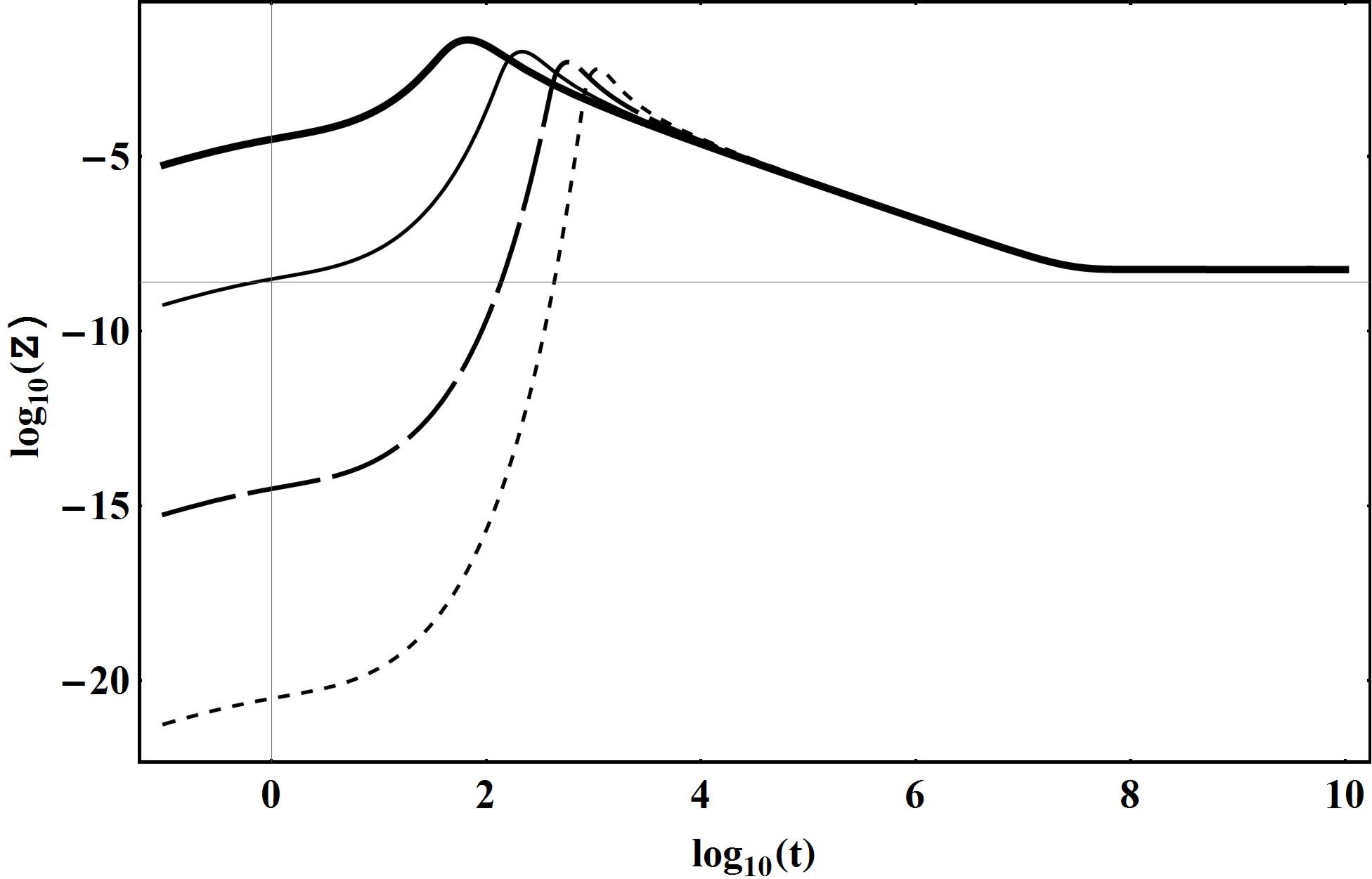}
\end{center}
{\small \textbf{Fig.  \thefigure}.\label{Fig46} The evolution of logarithm of the potential's derivative $\log_{10} Z=\log_{10} \dot{\Phi}$.\vspace{12pt}}

\begin{center}\refstepcounter{figure}
\includegraphics[width=120mm]{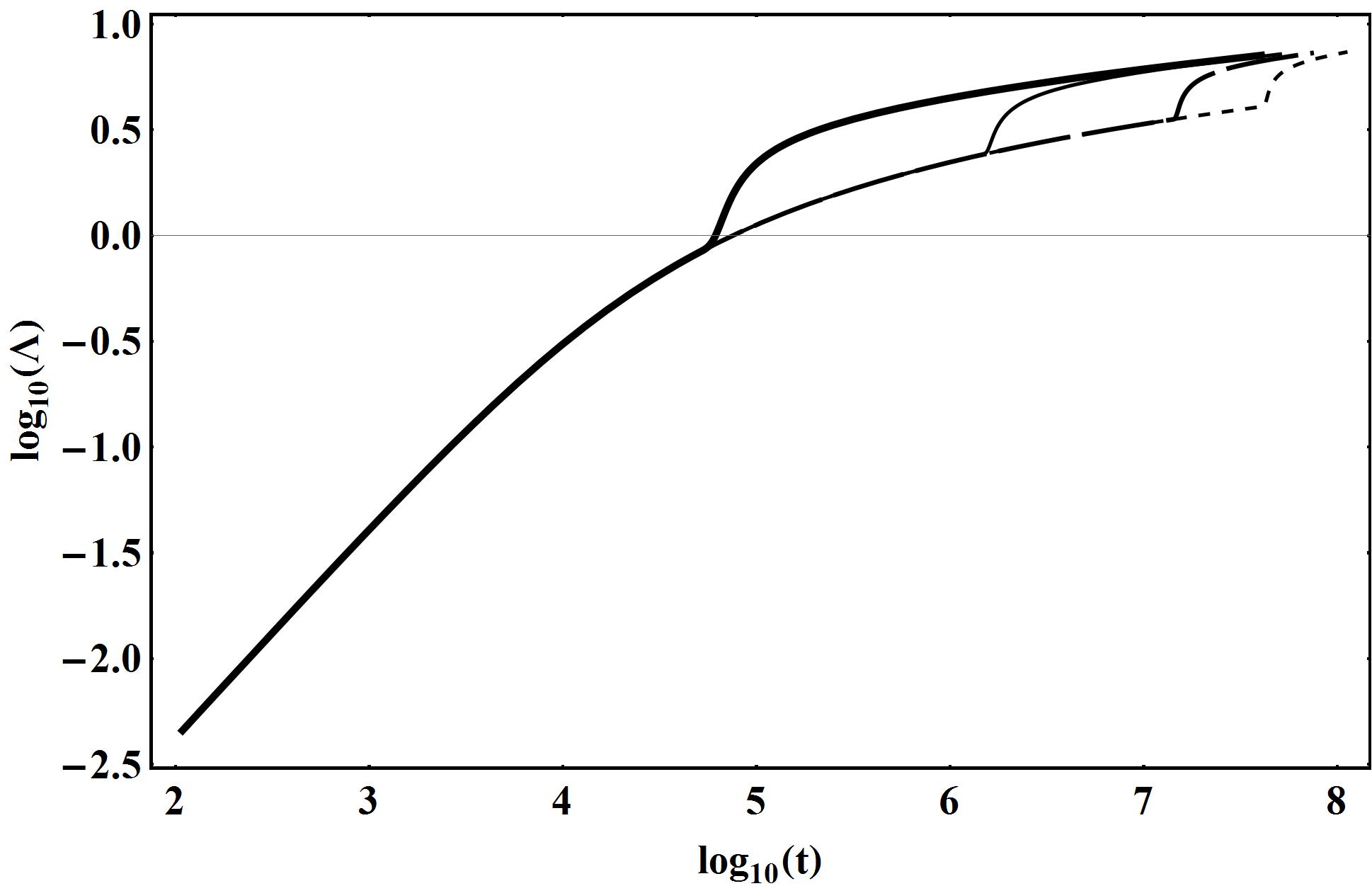}
\end{center}
{\small \textbf{Fig.  \thefigure}.\label{Fig47} The evolution of logarithm of the scale function $\log_{10}\Lambda(t)$.\vspace{12pt}}

\begin{center}\refstepcounter{figure}
\includegraphics[width=120mm]{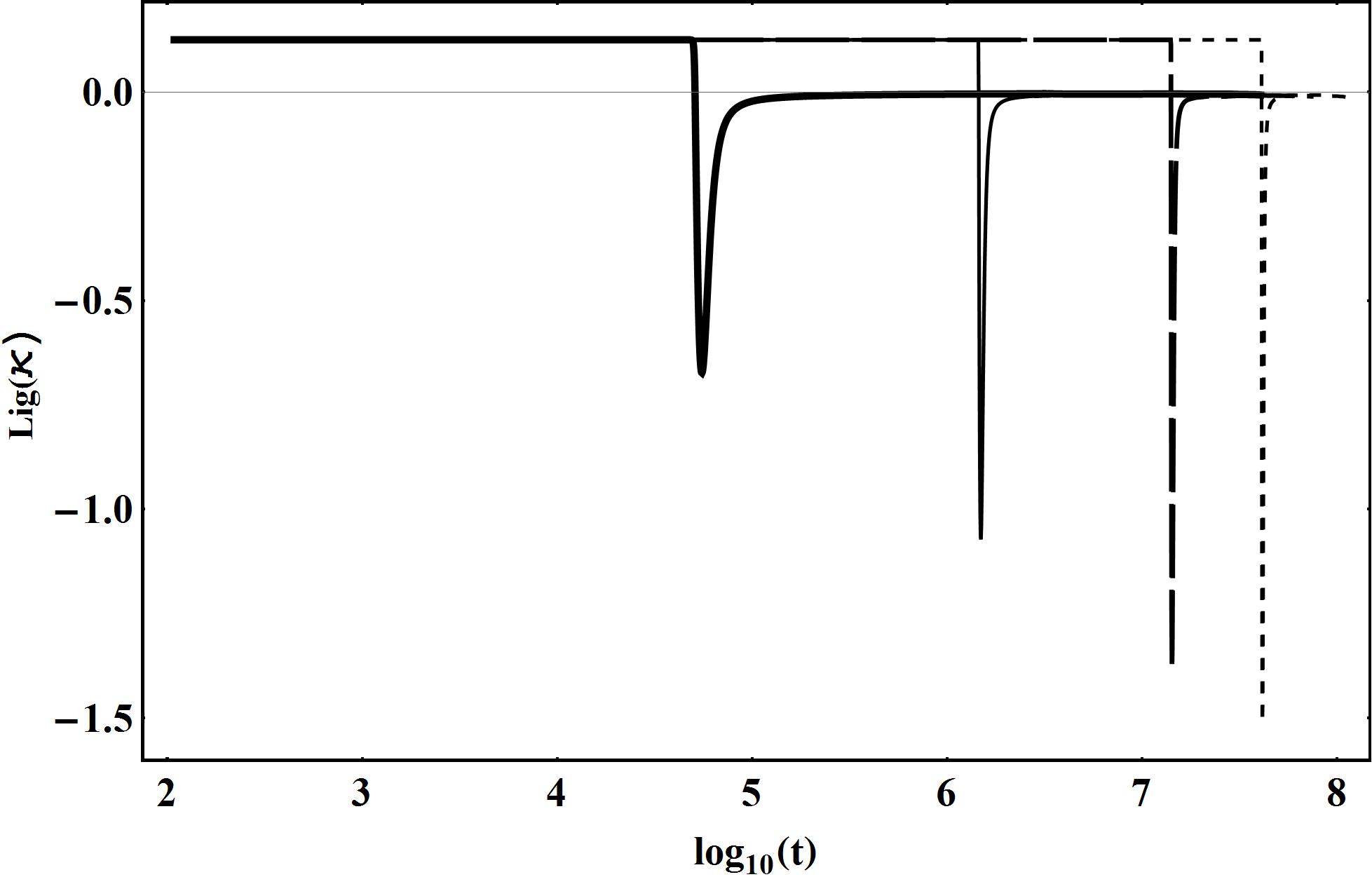}
\end{center}
{\small \textbf{Fig.  \thefigure}.\label{Fig48} The evolution of barotropic coefficient $\kappa$.\vspace{12pt}}

\begin{center}\refstepcounter{figure}
\includegraphics[width=120mm]{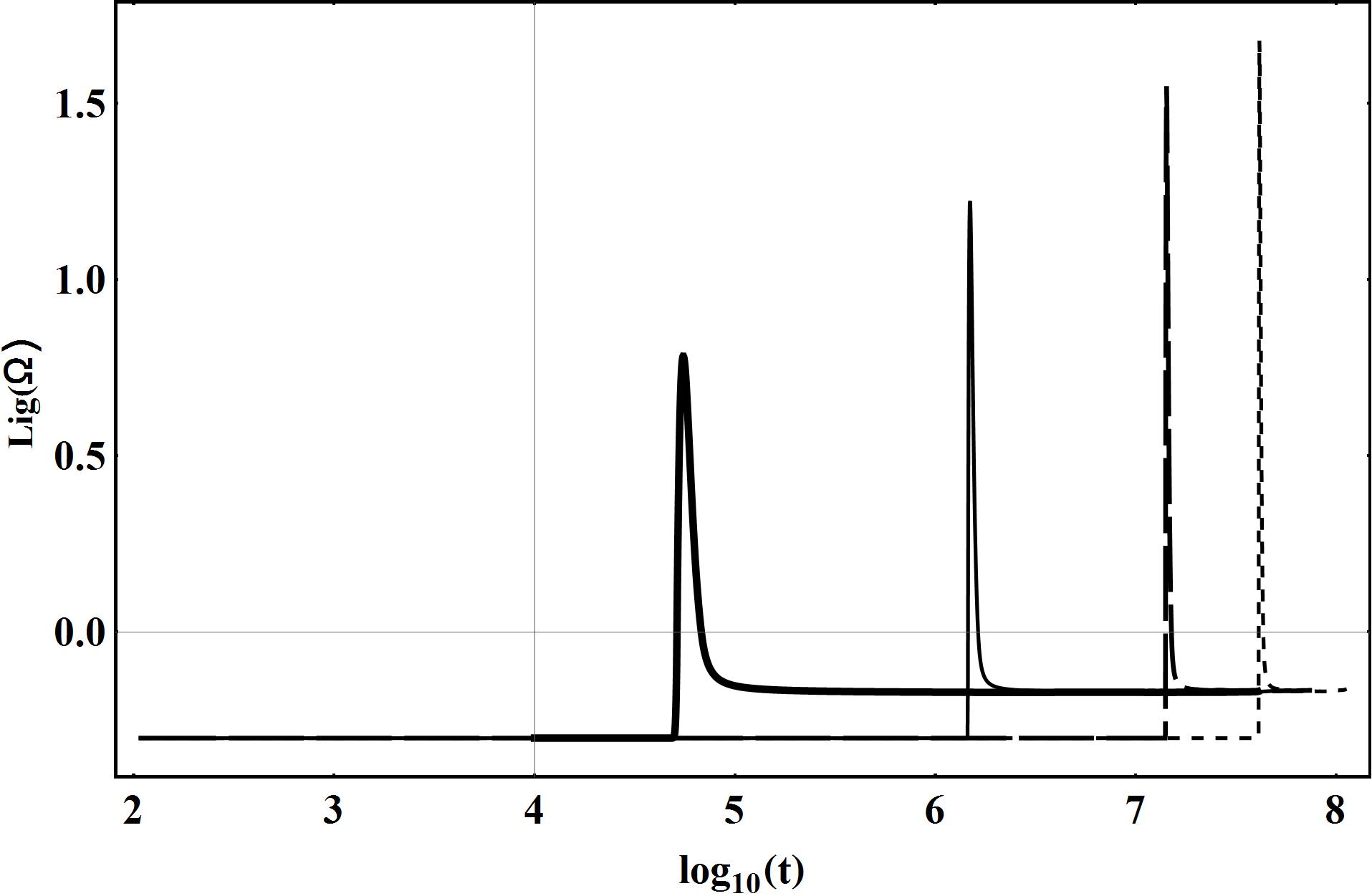}
\end{center}
{\small \textbf{Fig.  \thefigure}.\label{Fig49} The evolution of the invariant cosmological acceleration %
$\Omega$.\vspace{12pt}}

\begin{center}\refstepcounter{figure}
\includegraphics[width=120mm]{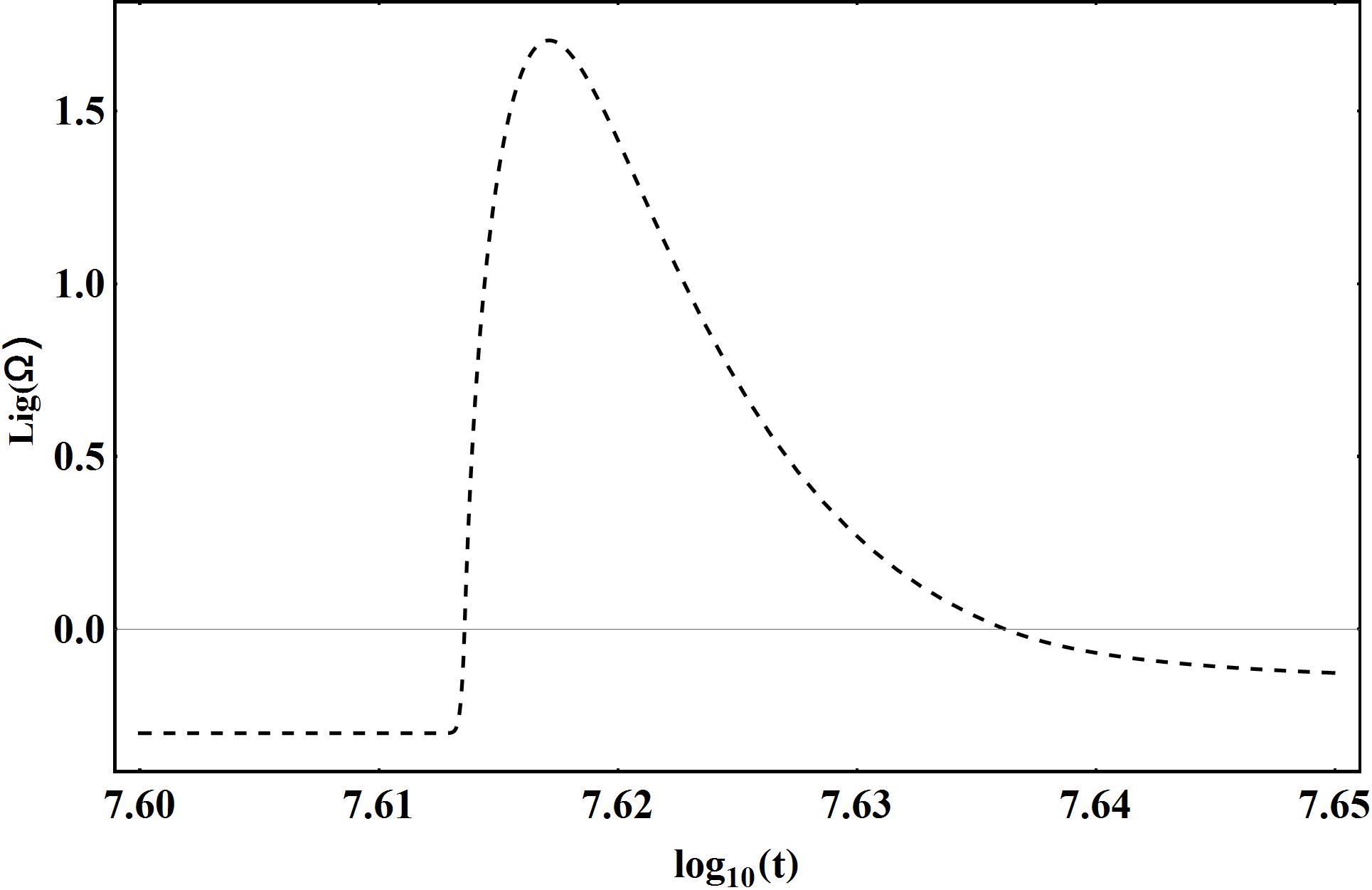}
\end{center}
{\small \textbf{Fig.  \thefigure}.\label{Fig50} Detailed structure of the fourth fantom burst of the invariant cosmological acceleration shown on Fig. 49}

\section{The Cosmological Evolution of Thermodynamic Values}
Let us investigate now thermal corrections on almost degenerated Ferm system with fantom scalar interaction.
In this case in the field and Einstein equations (\ref{Z,lambda}) -- (\ref{dL}) instead of (\ref{3a}) -- (\ref{3c}) for macroscopic scalars of the completely degenerated Fermi system it is necessary to use expressions for macroscopic scalars of almost degenerated Fermi system (\ref{E_p_theta}) -- (\ref{sigma_theta}). The initial value of the inverse temperature would be added to the initial data of the Cauchy problem $\lambda=m_*/\theta$.
\begin{eqnarray}\label{Coshe_theta}
\Lambda(0)=0; \; p_f(0)=p_0;\;\lambda(0)=\lambda_0;\nonumber\\
\Phi(0)=\Phi_0;\; ñ(0)=Z_0.
\end{eqnarray}
Herewith dimensionless chemical potential $\gamma$ is defined through two dimensionless functions $\lambda=m_*/\theta$ (\ref{gamma})  è $\psi = p_f/m_*$ (\ref{psi}) êàê $\gamma = \lambda\sqrt{1+\psi^2}$.

Thus, we have a completely defined system of five differential equations of the first order that can be used to find four unknown scalar functions $\Lambda(t)$, $\Phi$, $Z(t)$ $\psi$, $\lambda$:
%
\begin{eqnarray}
\label{Deltan}
\Delta \dot{n} +3\dot{\Lambda}\Delta n = 0;\\
\label{Z1}
\dot{\Phi}=Z;\\
\label{dZ1}
\dot{Z}=-3\dot{\Lambda}Z+m^2\Phi+4\pi\sigma(\Lambda,\Phi);\\
\label{dL1}
\dot{\Lambda}=8\pi\Eps_p-Z^2+m_s^2\Phi^2;\\
\dot{\mathcal{E}}_{p}+3\dot{\Lambda}(\mathcal{E}_{p}+\mathcal{P}_{p})=\sigma\dot{\Phi}.
\end{eqnarray}
Let us proceed to the results of numerical integration of this system.
\subsection{The Case of Massless Fantom Scalar Field ($m_s = 0$) with the Source $\sigma\not=0$}
Fig. 51 -- Fig. 59 show the results of numerical modelling of the system with the following parameters:\\
$\lambda_0 = 10^{-6}$; $\Phi(0) = 5\cdot 10^{-7}$; $q = 0.01$; $\psi_0 = 2\cdot 10^6$.\\
\begin{center}\refstepcounter{figure}
\includegraphics[width=120mm]{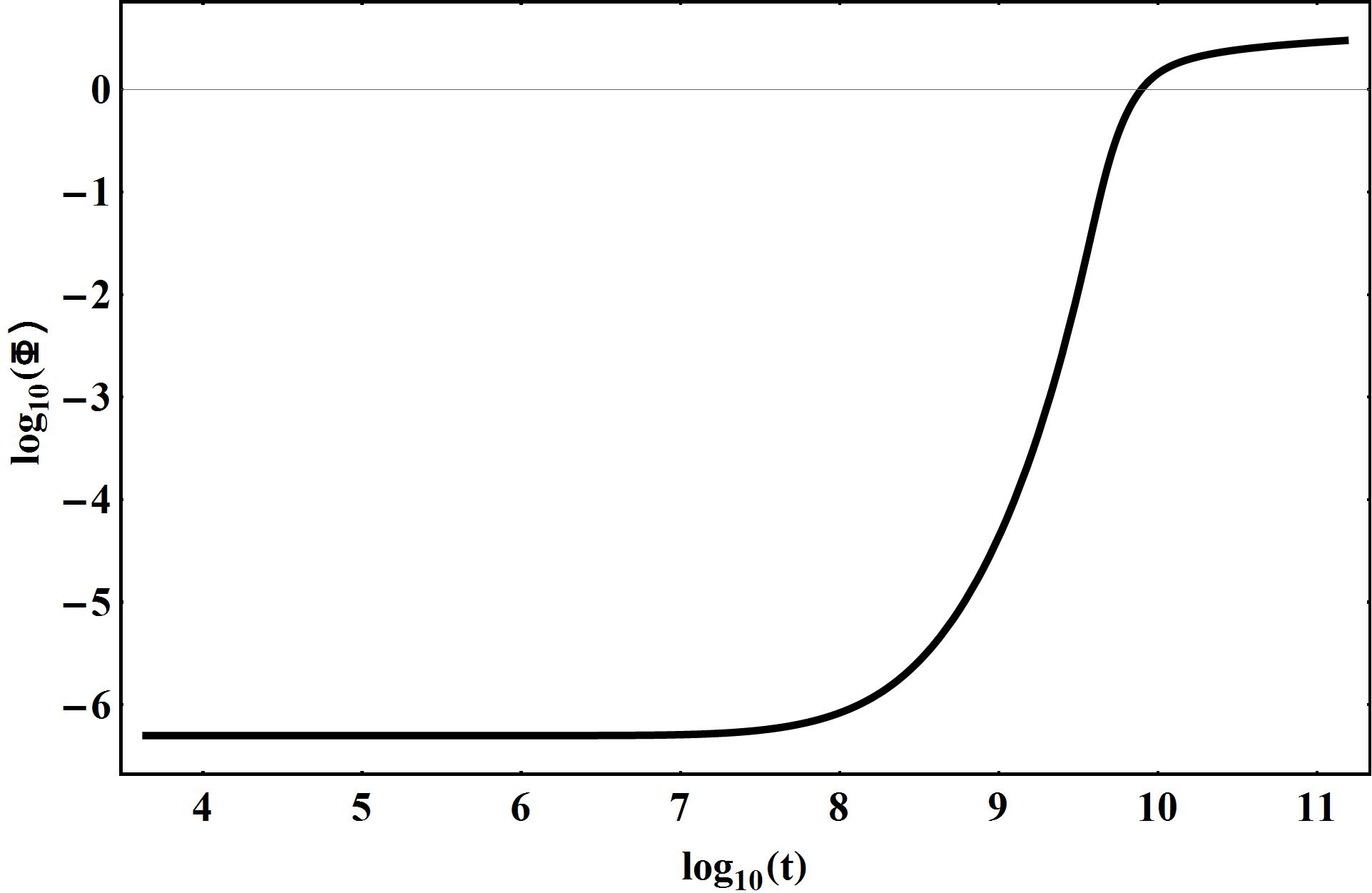}
\end{center}
{\small \textbf{Fig.  \thefigure}.\label{Fig51} The evolution of the potential's logarithm $\log_{10}\Phi$.}
\begin{center}\refstepcounter{figure}
\includegraphics[width=120mm]{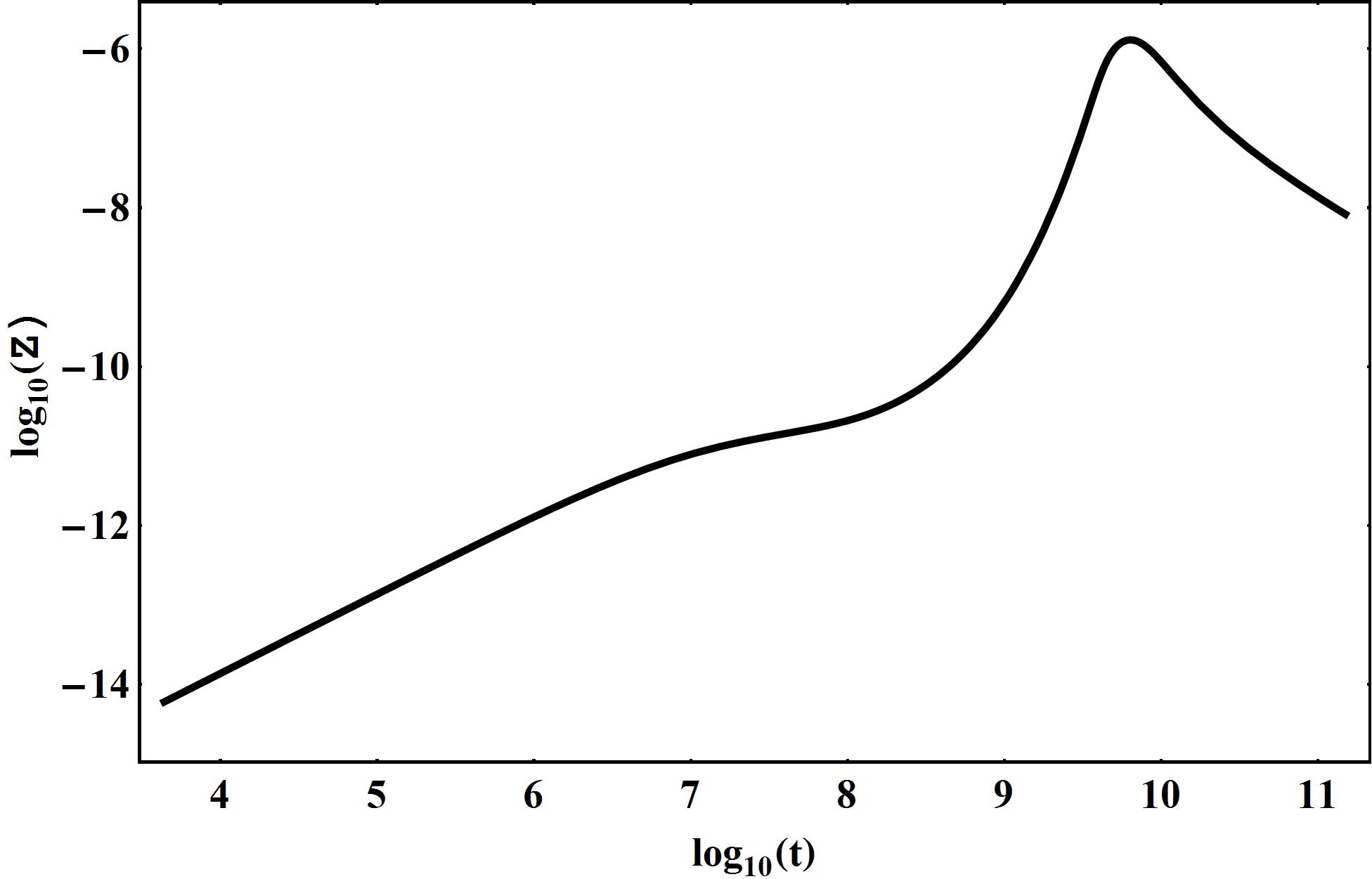}
\end{center}
{\small \textbf{Fig.  \thefigure}.\label{Fig52} The evolution of logarithm of the potential's derivative $\log_{10} Z=\log_{10} \dot{\Phi}$.\vspace{8pt}}
\begin{center}\refstepcounter{figure}
\includegraphics[width=120mm]{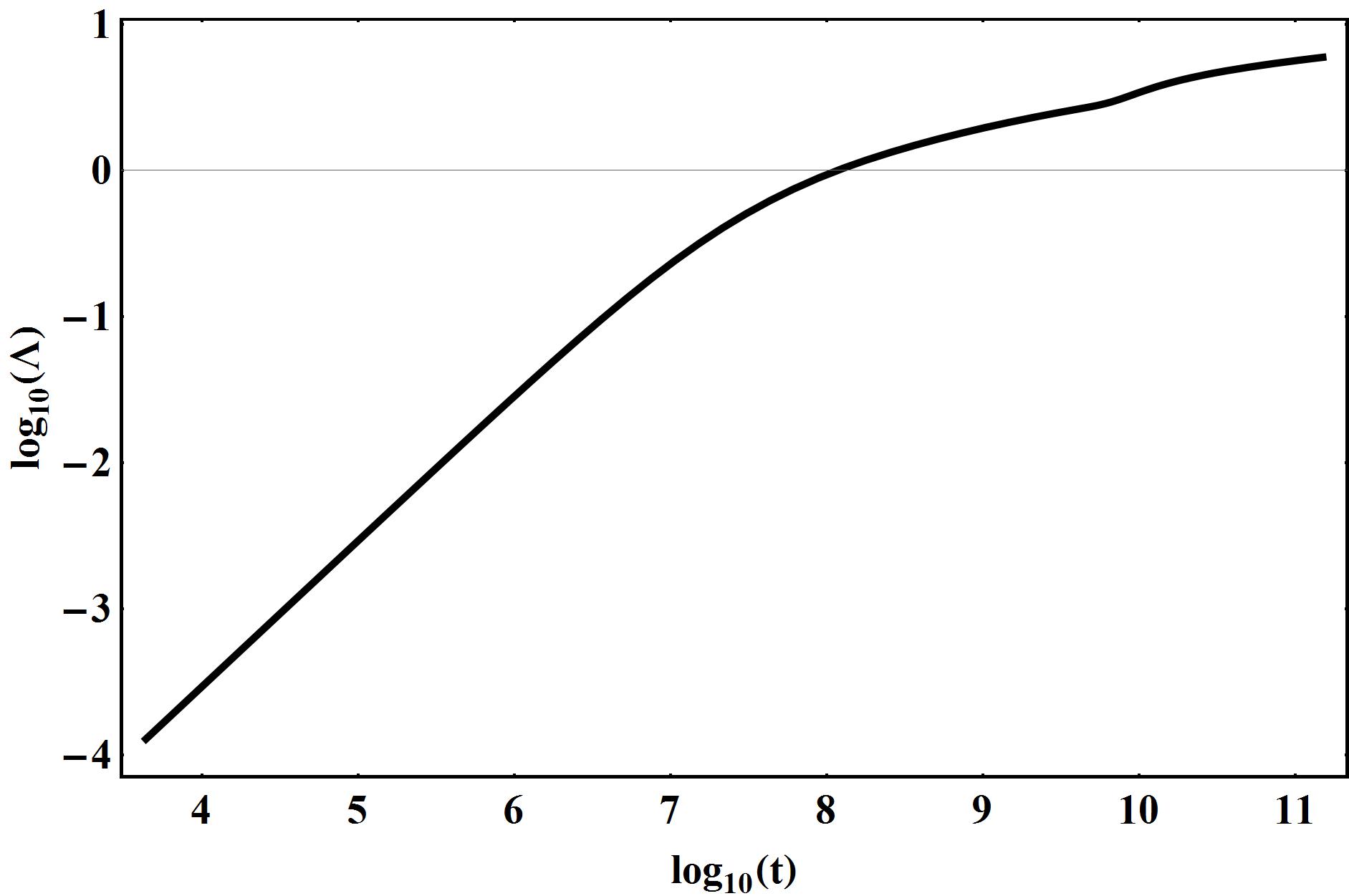}
\end{center}
{\small \textbf{Fig.  \thefigure}.\label{Fig53} The evolution of logarithm of the scale function $\log_{10}\Lambda(t)$.\vspace{12pt}}

\begin{center}\refstepcounter{figure}
\includegraphics[width=120mm]{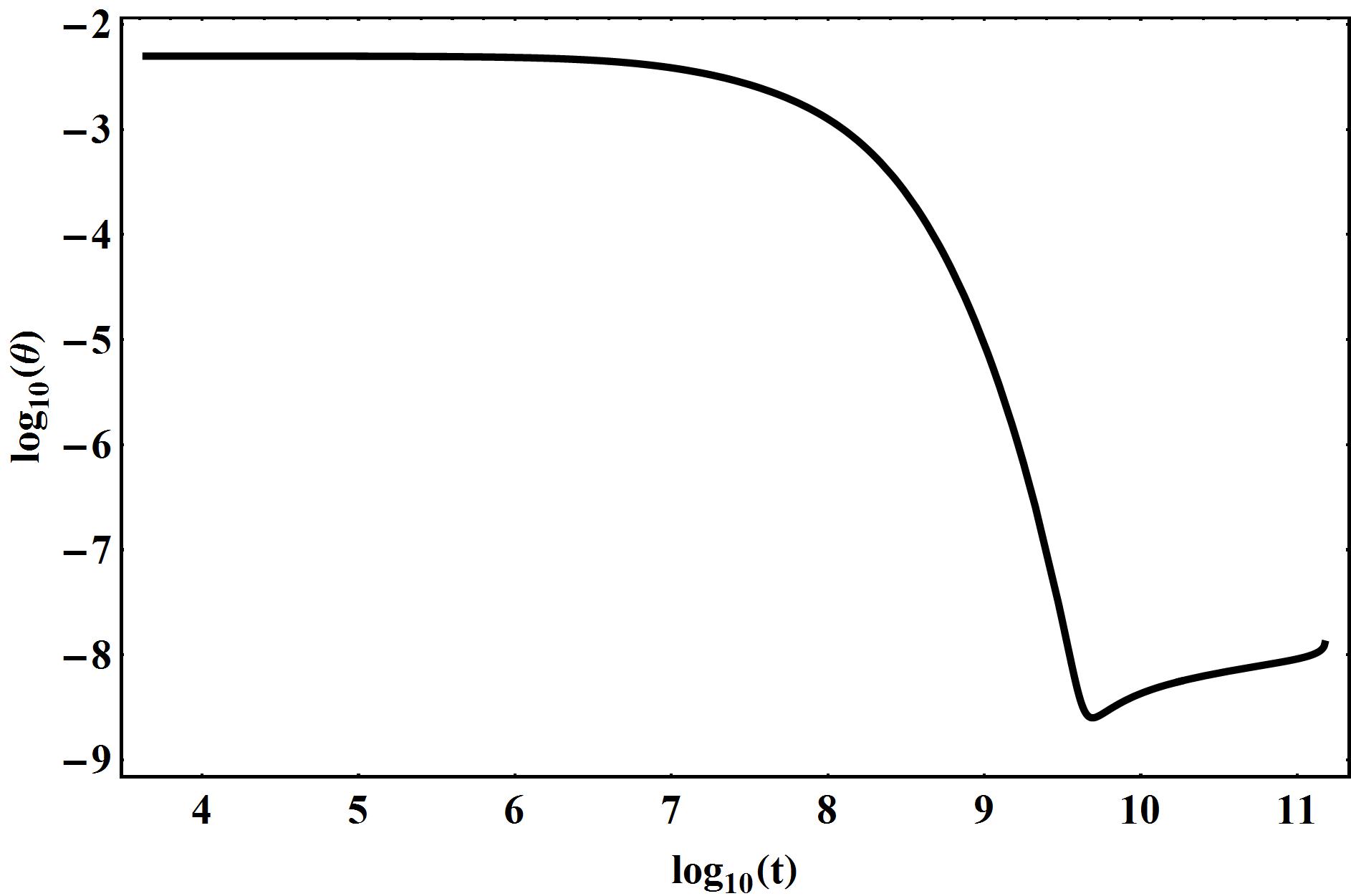}
\end{center}
{\small \textbf{Fig.  \thefigure}.\label{Fig54} The evolution of logarithm of the temperature $\log_{10}\theta(t)$.\vspace{12pt}}

\begin{center}\refstepcounter{figure}
\includegraphics[width=120mm]{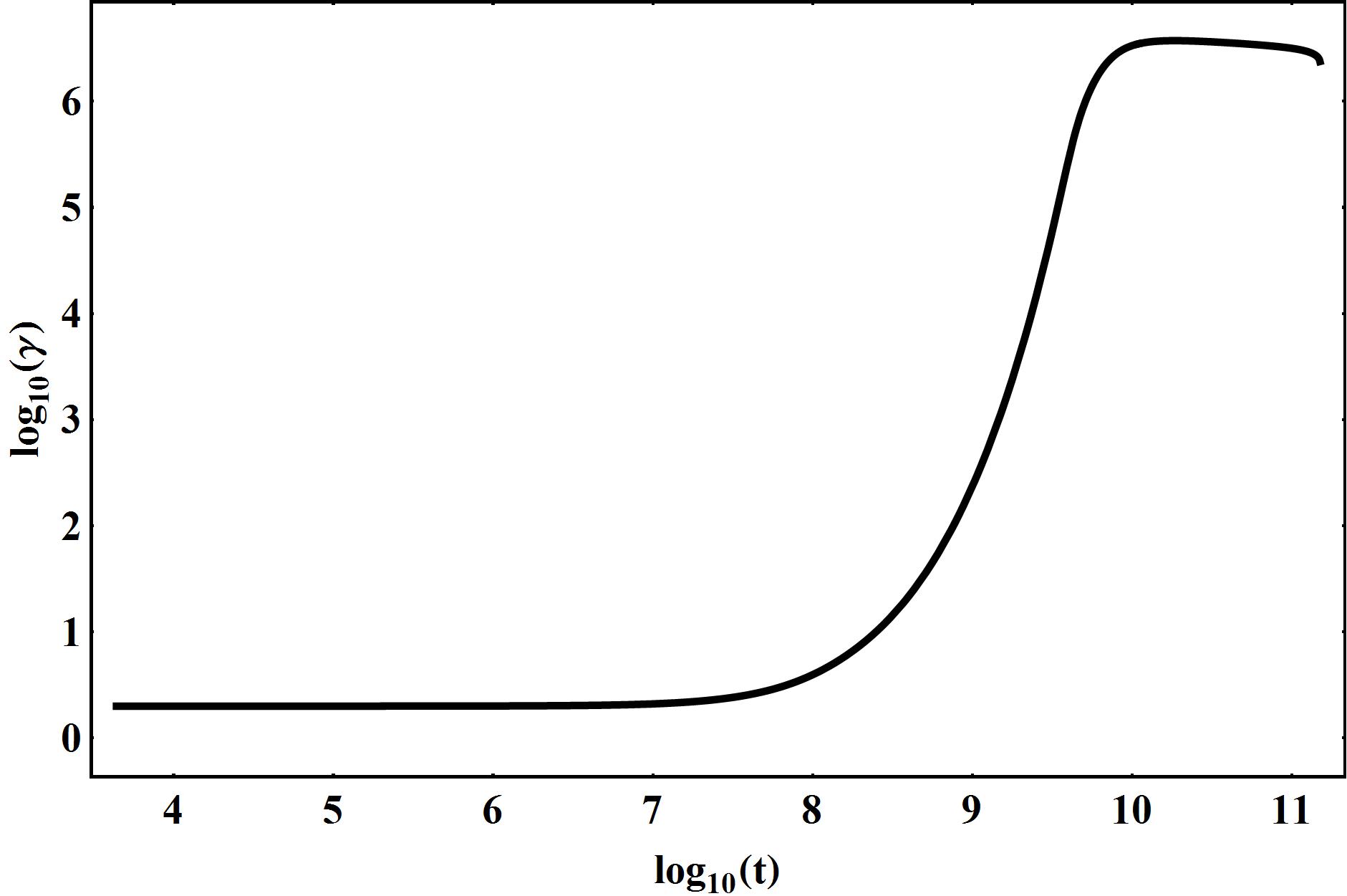}
\end{center}
{\small \textbf{Fig.  \thefigure}.\label{Fig55} The evolution of the chemical potential's logarithm $\log_{10}\mu(t)$.\vspace{12pt}}

\begin{center}\refstepcounter{figure}
\includegraphics[width=120mm]{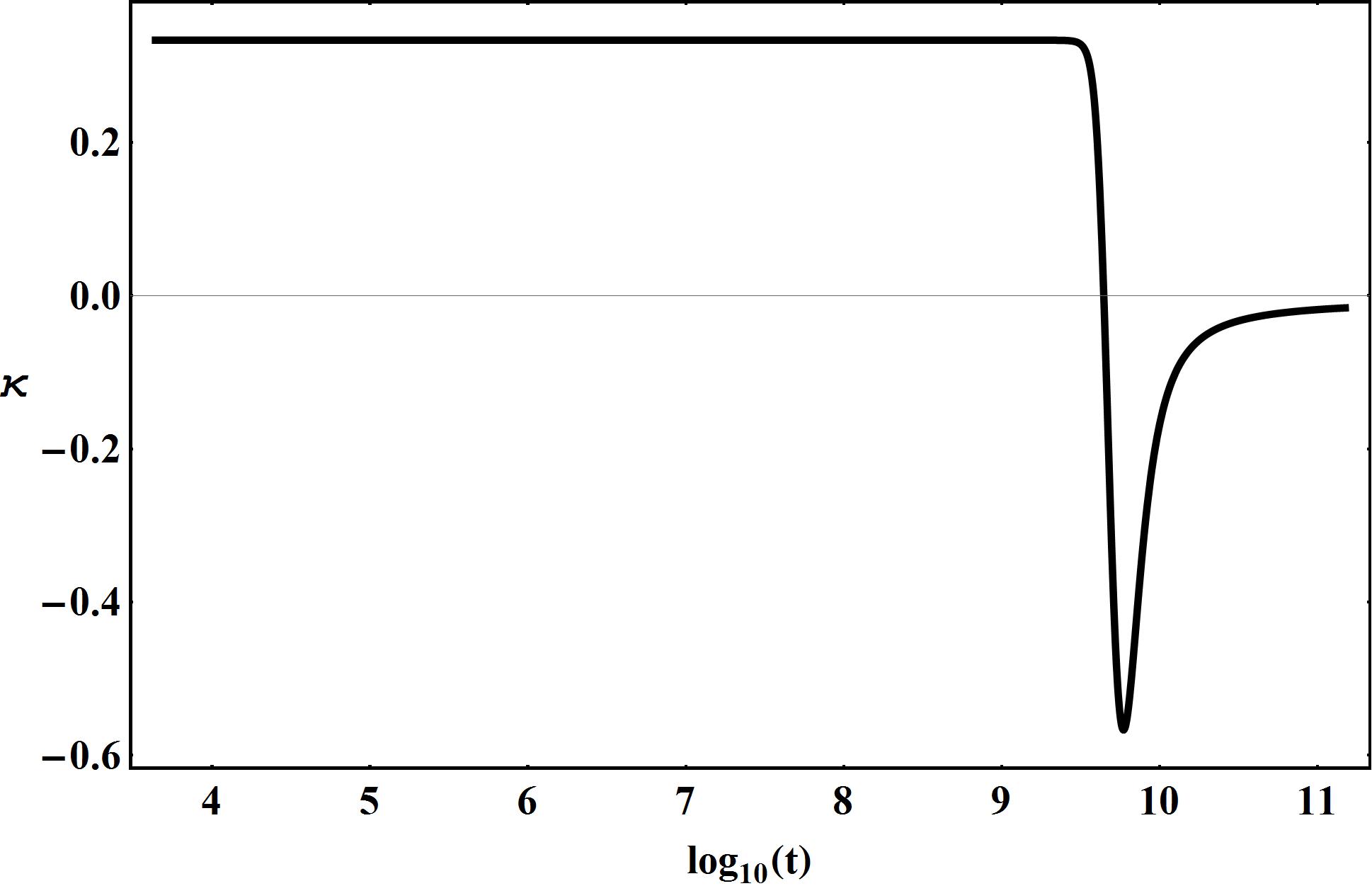}
\end{center}
{\small \textbf{Fig.  \thefigure}.\label{Fig56} The evolution of barotropic coefficient $\kappa$.\vspace{12pt}}

\begin{center}\refstepcounter{figure}
\includegraphics[width=120mm]{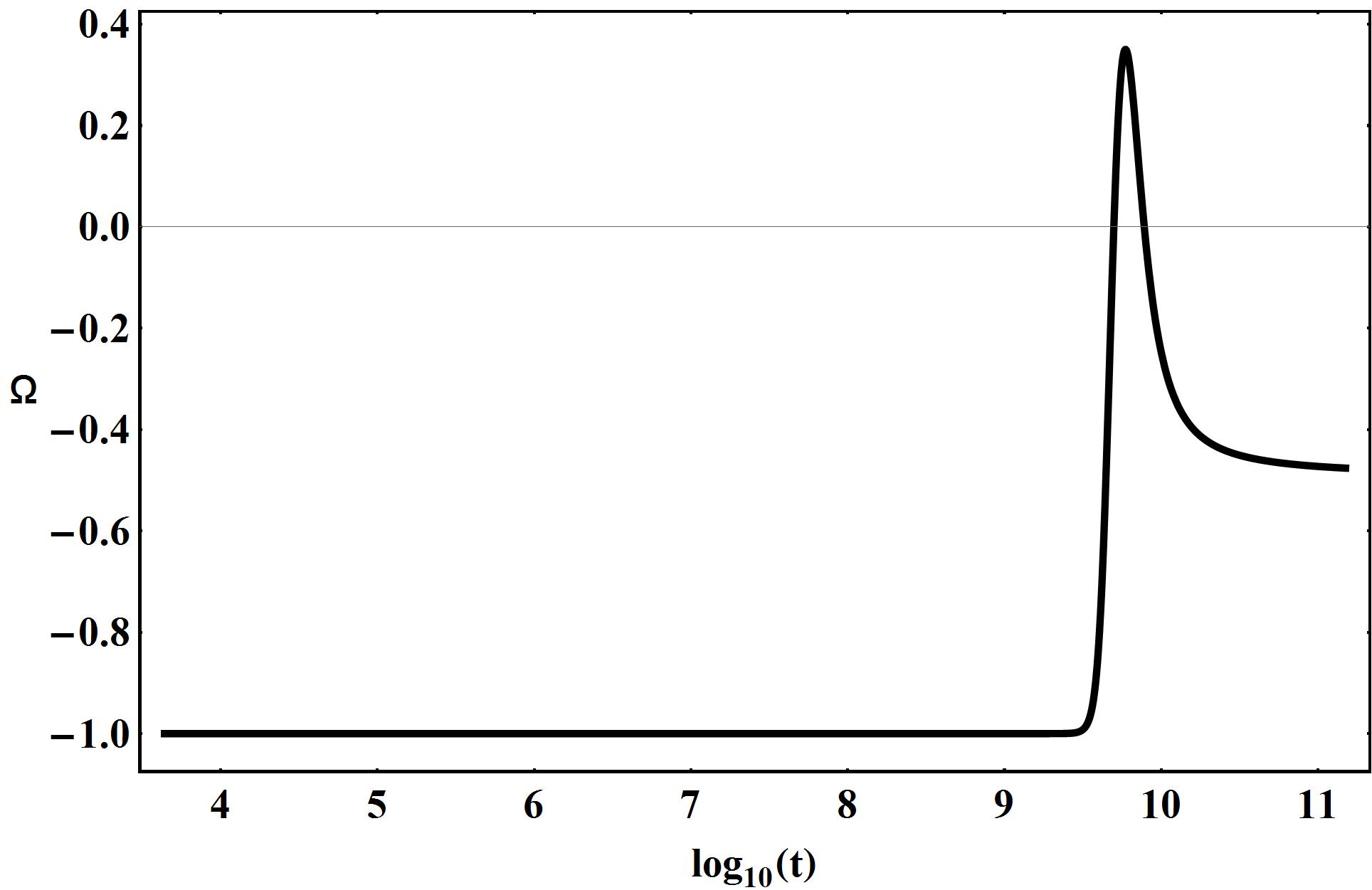}
\end{center}
{\small \textbf{Fig.  \thefigure}.\label{Fig57} The evolution of the invariant cosmological acceleration $\Omega$.\vspace{12pt}}

\begin{center}\refstepcounter{figure}
\includegraphics[width=120mm]{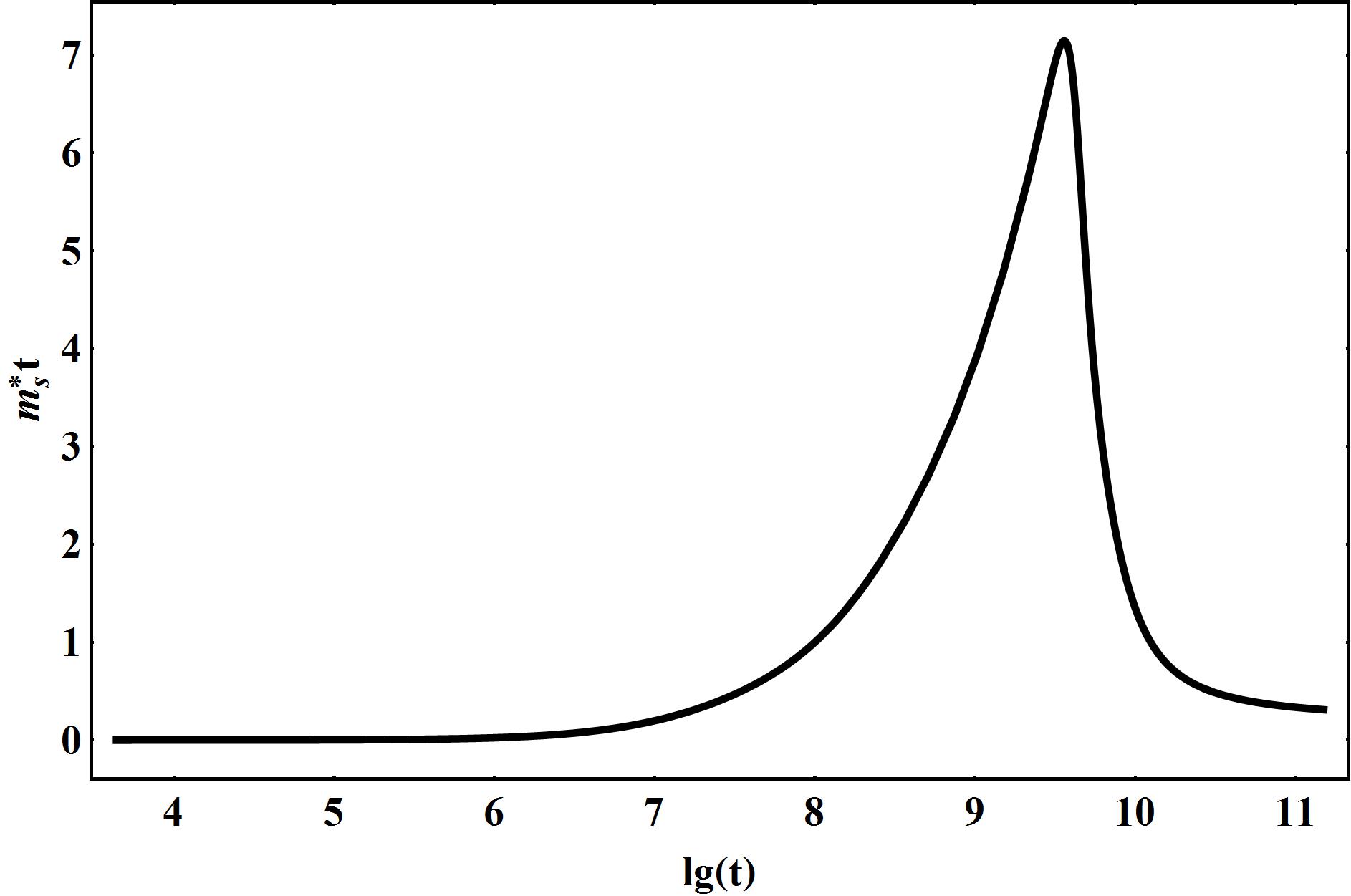}
\end{center}
{\small \textbf{Fig.  \thefigure}.\label{Fig58} The plot of the function $m_s^* t$\vspace{12pt}}

\begin{center}\refstepcounter{figure}
\includegraphics[width=120mm]{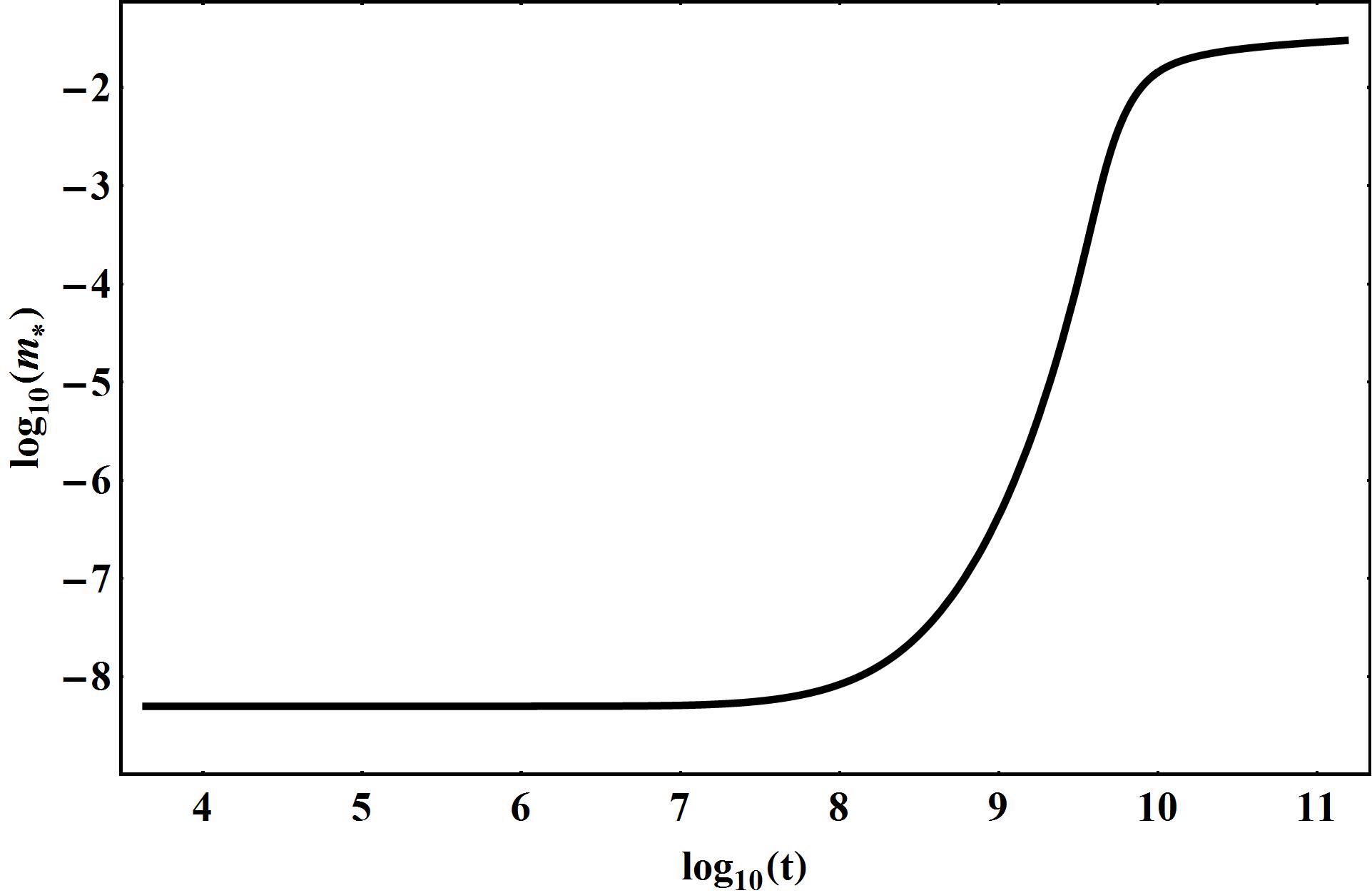}
\end{center}
{\small \textbf{Fig.  \thefigure}.\label{Fig59} The evolution of logarithm of the effective mass\\ $\log_{10} m_*$.\vspace{12pt}}

Plots above reveal the next regularities:
\begin{enumerate}
\item At the stage of growth $Z=\dot\Phi$ ($t>10^8$) one can observe a simultaneous growth of chemical potential $\mu$ and fall of temperature $\theta$
\item Starting from instant of maximum $Z$ and, simultaneously, maximum of acceleration $\Omega$ and the scalar field's effective mass $m^*_s$ ($t\approx 3\cdot 10^9$)
the chemical potential becomes approximately constant one while temperature start to grow nevertheless staying small.
\item Thus, generally, the level of the Fermi system's degeneracy grows with time.
\end{enumerate}

\subsection{The case of Massive Fantom Scalar Field ($m_s\not=0$) with the Source $\sigma\not=0$}
Fig. 60 -- Fig. 67 show the results of numerical simulation of the system with the following parameters:
$m_s = 10^{-6}$, $\lambda_0 = 10^{-7}$, $\Phi(0) = 5\cdot 10^{-8}$. Heavy line is $q = 1$, $\psi_0 = 2\cdot 10^7$; thin line is $q = .1$, $\psi_0 = 2\cdot 10^8$; normal dotted line is $q = .01$, $\psi_0 = 2\cdot 10^9$; fine dotted line is $q = .001$, $\psi_0 = 2\cdot 10^{10}$.
\begin{center}\refstepcounter{figure}
\includegraphics[width=120mm]{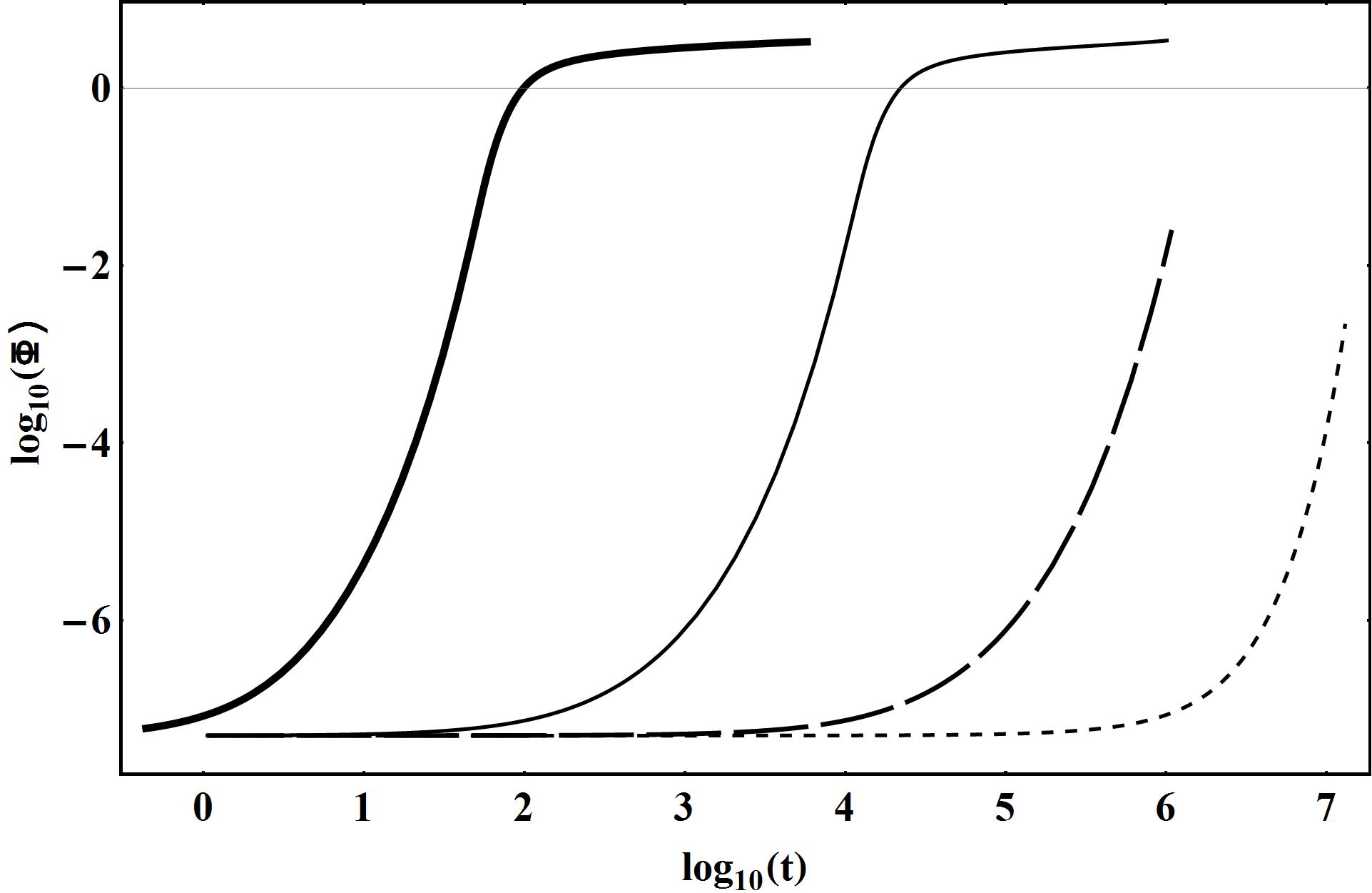}
\end{center}
{\small \textbf{Fig.  \thefigure}.\label{Fig60} The evolution of the potential's logarithm $\log_{10}\Phi$.\vspace{12pt}}

\begin{center}\refstepcounter{figure}
\includegraphics[width=120mm]{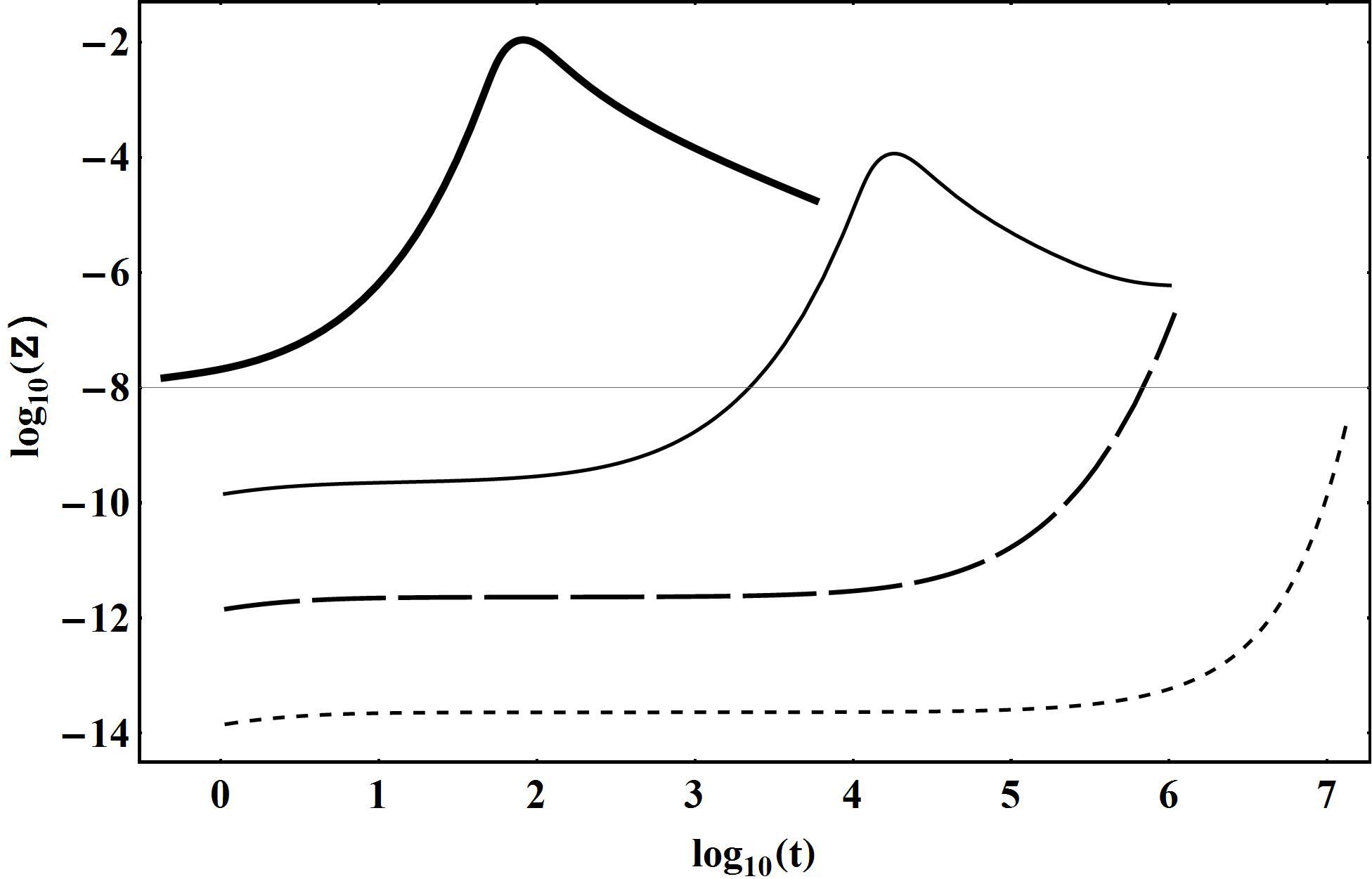}
\end{center}
{\small \textbf{Fig.  \thefigure}.\label{Fig61} The evolution of logarithm of the potential's derivative $\log_{10} Z=\log_{10} \dot{\Phi}$.\vspace{12pt}}

\begin{center}\refstepcounter{figure}
\includegraphics[width=120mm]{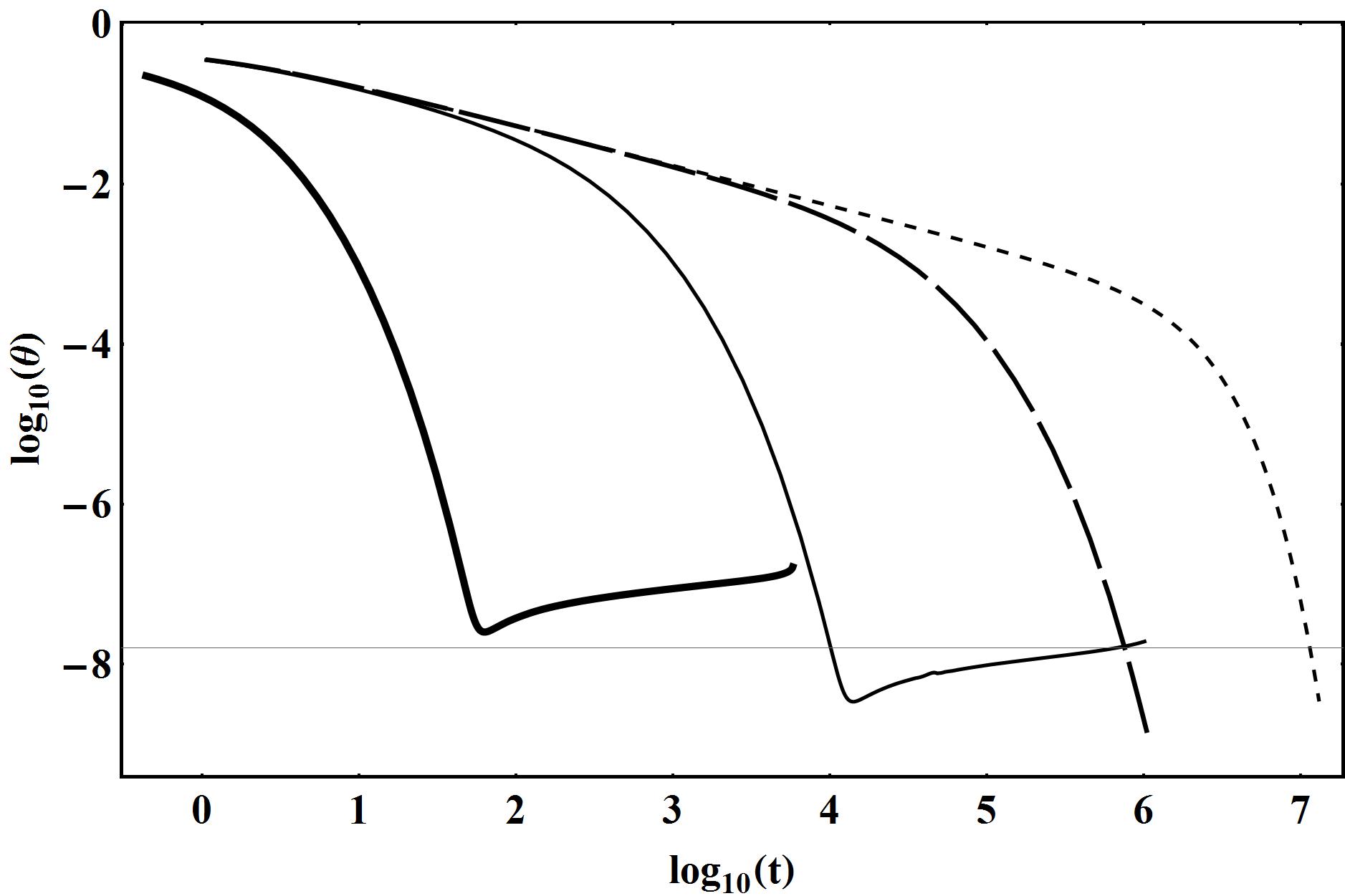}
\end{center}
{\small \textbf{Fig.  \thefigure}.\label{Fig62} The evolution of logarithm of the temperature $\log_{10}\theta(t)$.\vspace{12pt}}

\begin{center}\refstepcounter{figure}
\includegraphics[width=120mm]{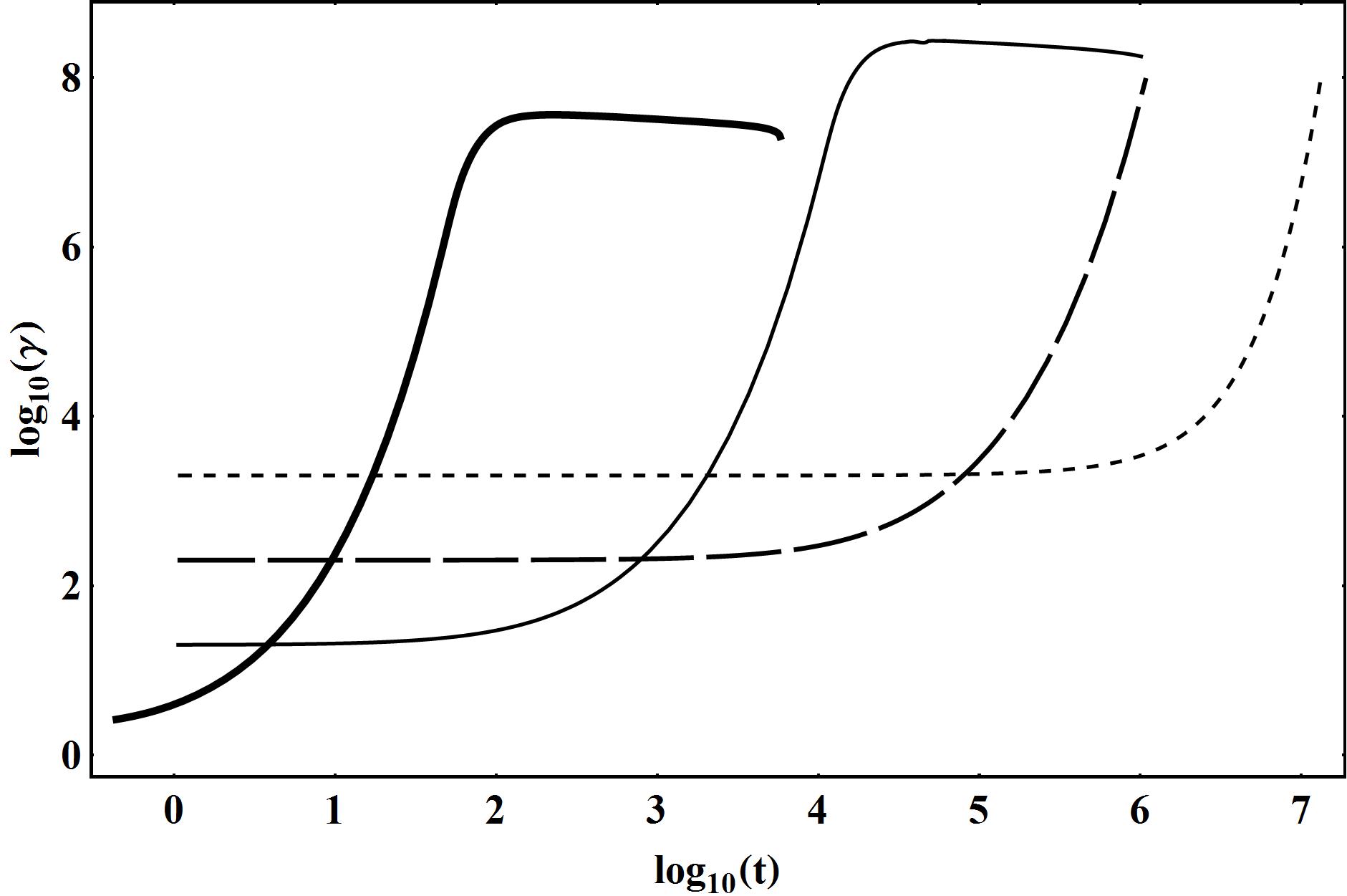}
\end{center}
{\small \textbf{Fig.  \thefigure}.\label{Fig63} The evolution of the chemical potential's logarithm $\log_{10}\mu(t)$.\vspace{12pt}}

\begin{center}\refstepcounter{figure}
\includegraphics[width=120mm]{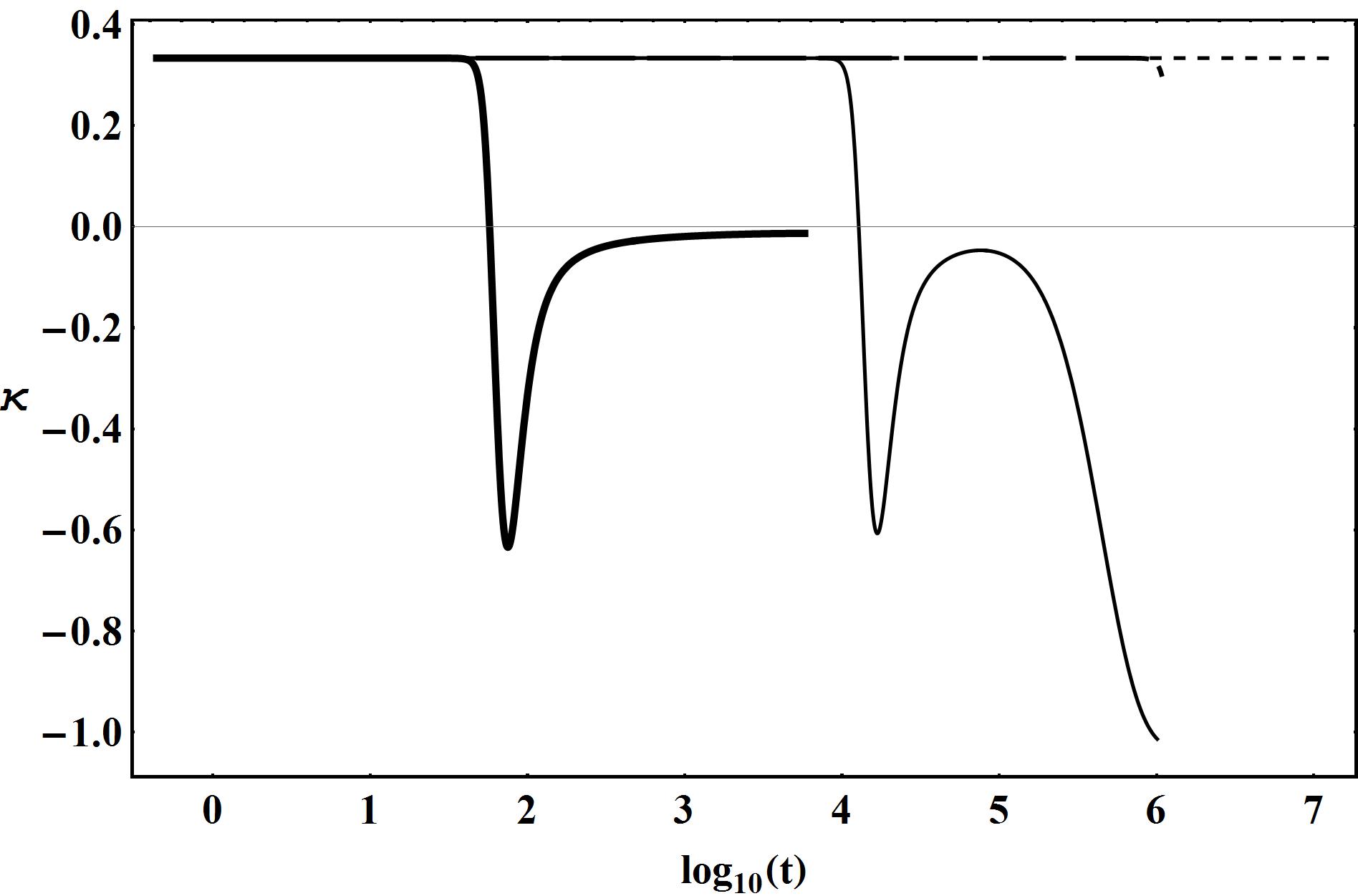}
\end{center}
{\small \textbf{Fig.  \thefigure}.\label{Fig64} The evolution of barotropic coefficient $\kappa$.\vspace{12pt}}

\begin{center}\refstepcounter{figure}
\includegraphics[width=120mm]{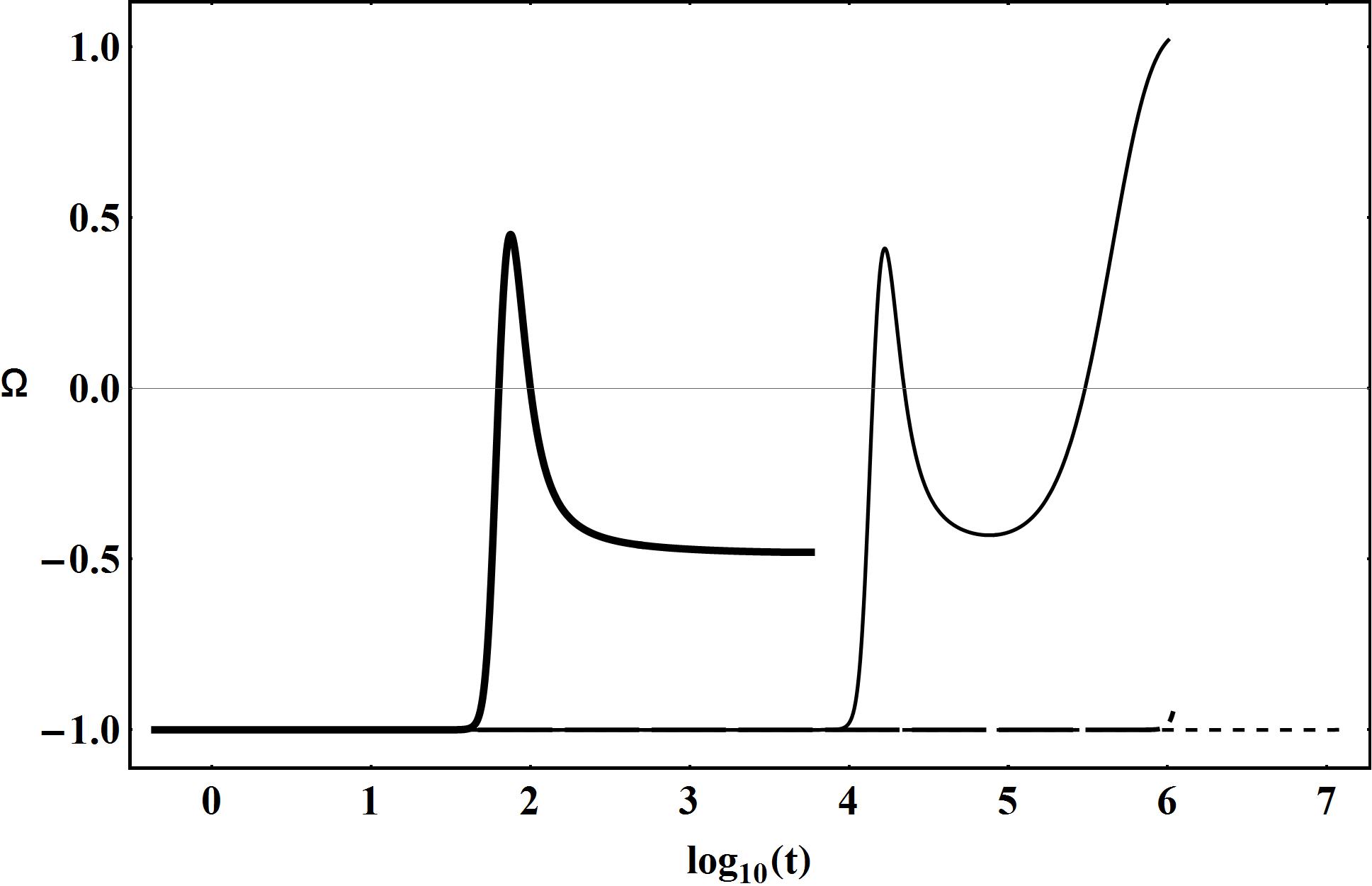}
\end{center}
{\small \textbf{Fig.  \thefigure}.\label{Fig65} The evolution of the invariant cosmological acceleration %
$\Omega$.\vspace{12pt}}

\begin{center}\refstepcounter{figure}
\includegraphics[width=120mm]{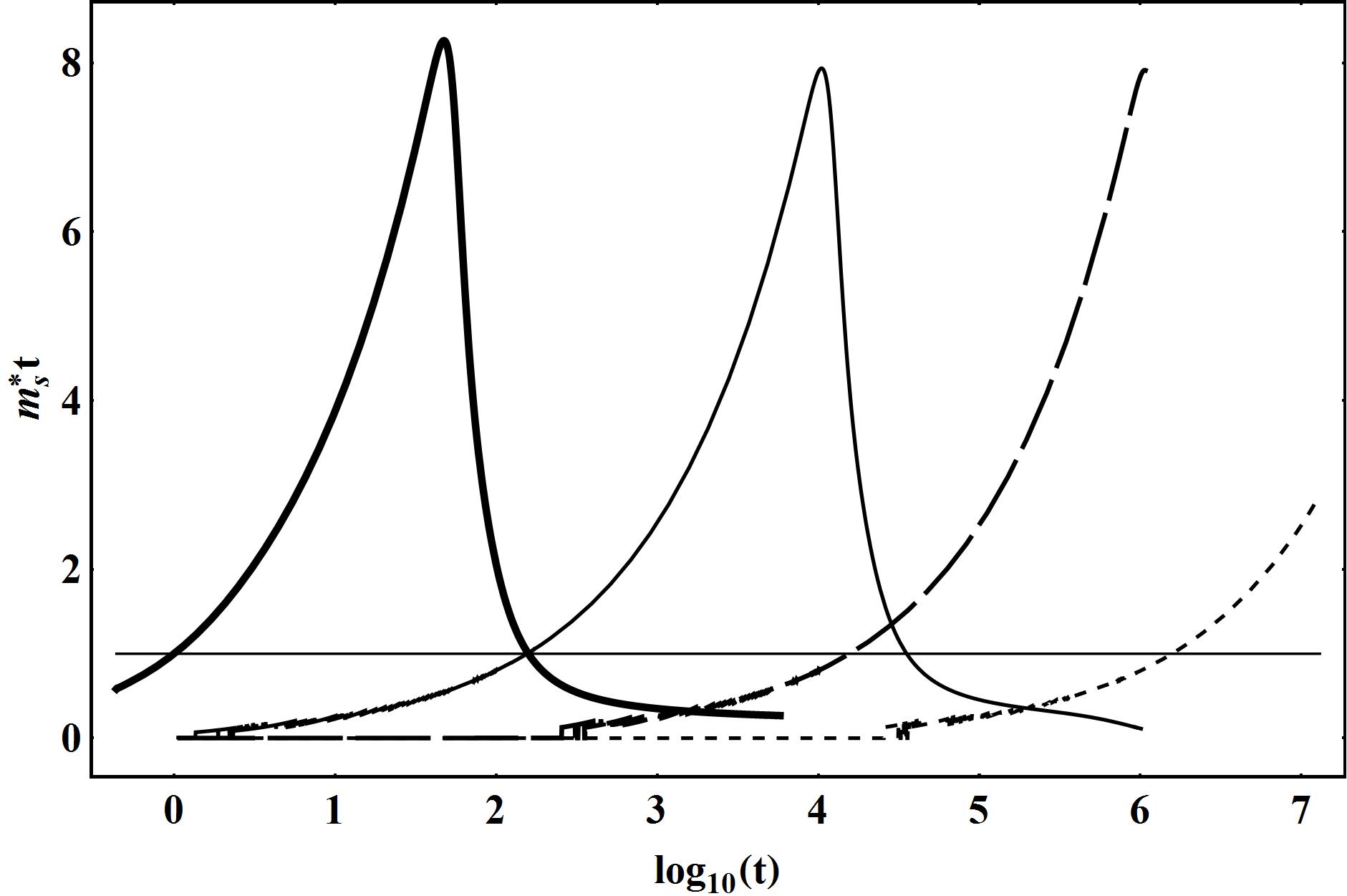}
\end{center}
{\small \textbf{Fig.  \thefigure}.\label{Fig66} The plot of the function $m_s^* t$\vspace{12pt}}

\begin{center}\refstepcounter{figure}
\includegraphics[width=120mm]{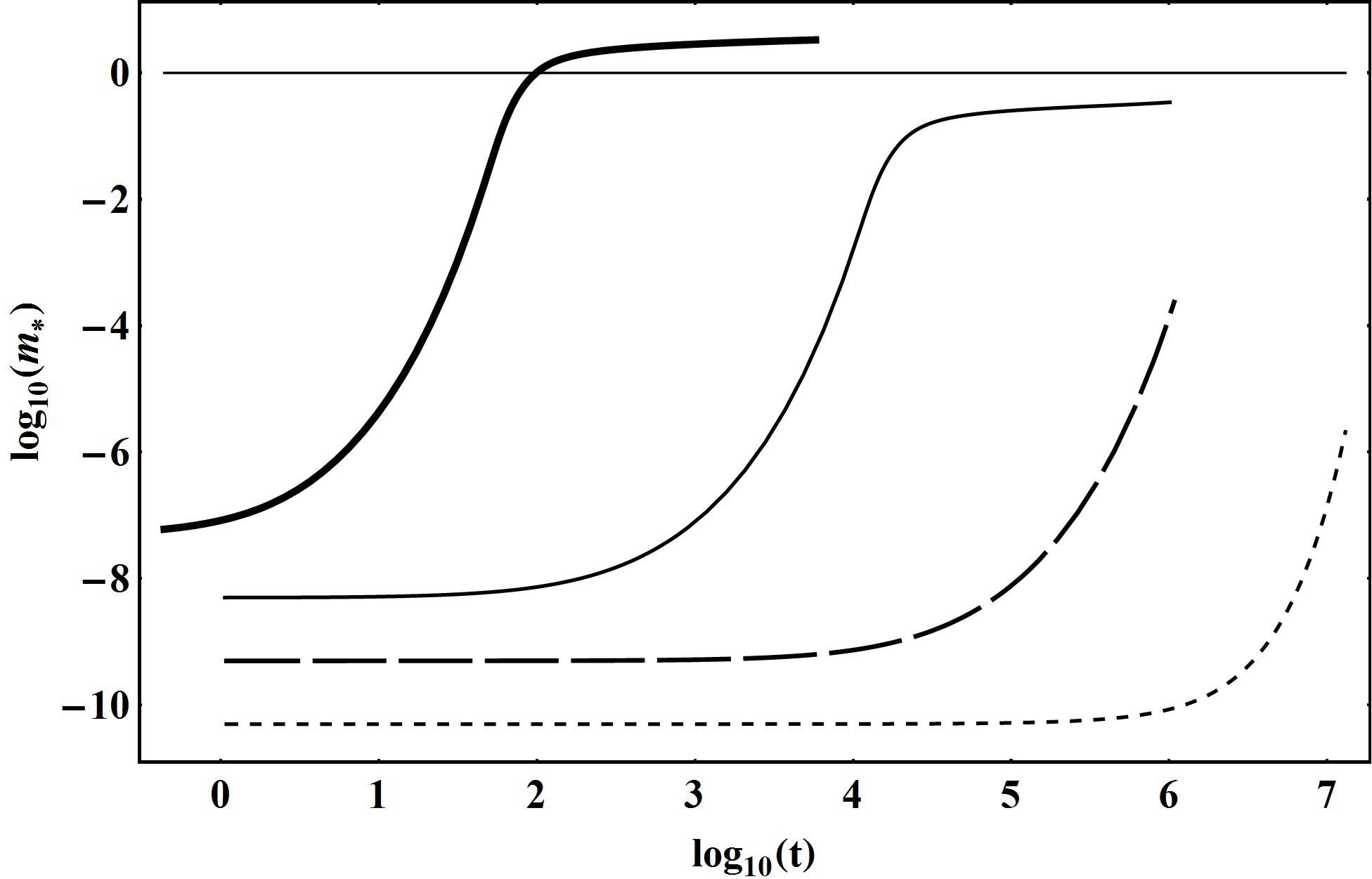}
\end{center}
{\small \textbf{Fig.  \thefigure}.\label{Fig67} The evolution of the effective mass'es logarithm $\log_{10} m_*$.\vspace{12pt}}

\section{Discussion of the Results}
Thus, the made research revealed the following regularities of the cosmological evolution of statistical systems of fermions with fantom scalar interaction.

\begin{enumerate}

\item During the process of cosmological evolution in the statistical systems of fermions with fantom scalar interaction there by all means occur acceleration bursts $\Omega$, which can be characterized by time parameters: instant of time of the burst's maximum $t_m$, half-width of the burst $\Delta t$ and height $h$ of the burst in its maximum.
\item The statistical systems of fermions with fantom scalar interaction reveal a tendency for generation of stable modes with constant acceleration (stages of the cosmological evolution) $\Omega=-1$ ($\kappa=1/3$, ultrarelativistic state), $\Omega=-1/2$ ($\kappa=0$, non-relativistic state) and $\Omega=1$ ($\kappa=-1$, inflation, vacuum state).
\item The dynamic properties of the statistical system with fantom scalar interaction weekly depend on a type of the particle systems' statistics
\cite{Ignatev_AAS}.
\item There are 4 clearly distinct in kind cosmological scenarios for the statistical systems of particles with fantom scalar interaction depending on the system's parameters (fundamental constants and initial conditions):
\begin{itemize}
\item \emph{1st type}.\; Ultrarelativistic start $\rightarrow$\\ acceleration burst $\rightarrow$ inflation stage (Fig. 68). This scenario is realized for a case of minimal interaction ($\sigma=0\rightarrow q=0$) of massive scalar field ($m_s\not=0$).
\begin{center}\refstepcounter{figure}
\includegraphics[width=0.8120mm]{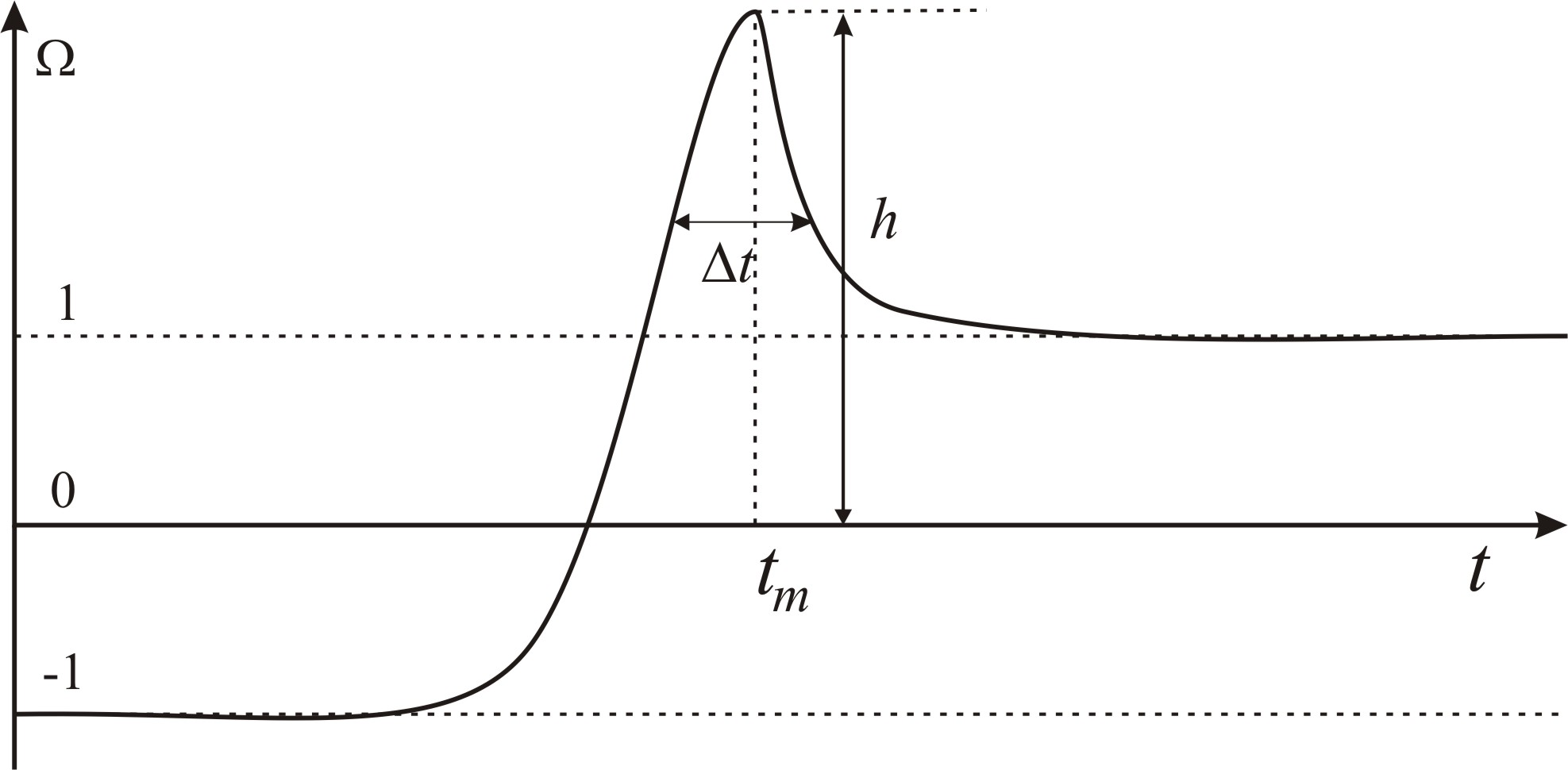}
\end{center}
{\small \textbf{Fig.  \thefigure}.\label{Fig68} The first type of the cosmological acceleration.\vspace{12pt}}
The following characteristic parameters meet this scenario: $t_m\sim 10^{-1}\div 10^3$; $\Delta t \sim 2$; $h\sim 10$, i.e. early acceleration burst and early transition to inflation stage.
\item \emph{2nd type}.\; Ultrarelaticistic start $\rightarrow$\\ acceleration burst $\rightarrow$ non-relativistic stage (Fig. 69). This scenario is realized for a case of non-minimal interaction ($\sigma\not=0\rightarrow q\not=0$) and massless scalar field ($m_s=0$).
\begin{center}\refstepcounter{figure}
\includegraphics[width=0.8120mm]{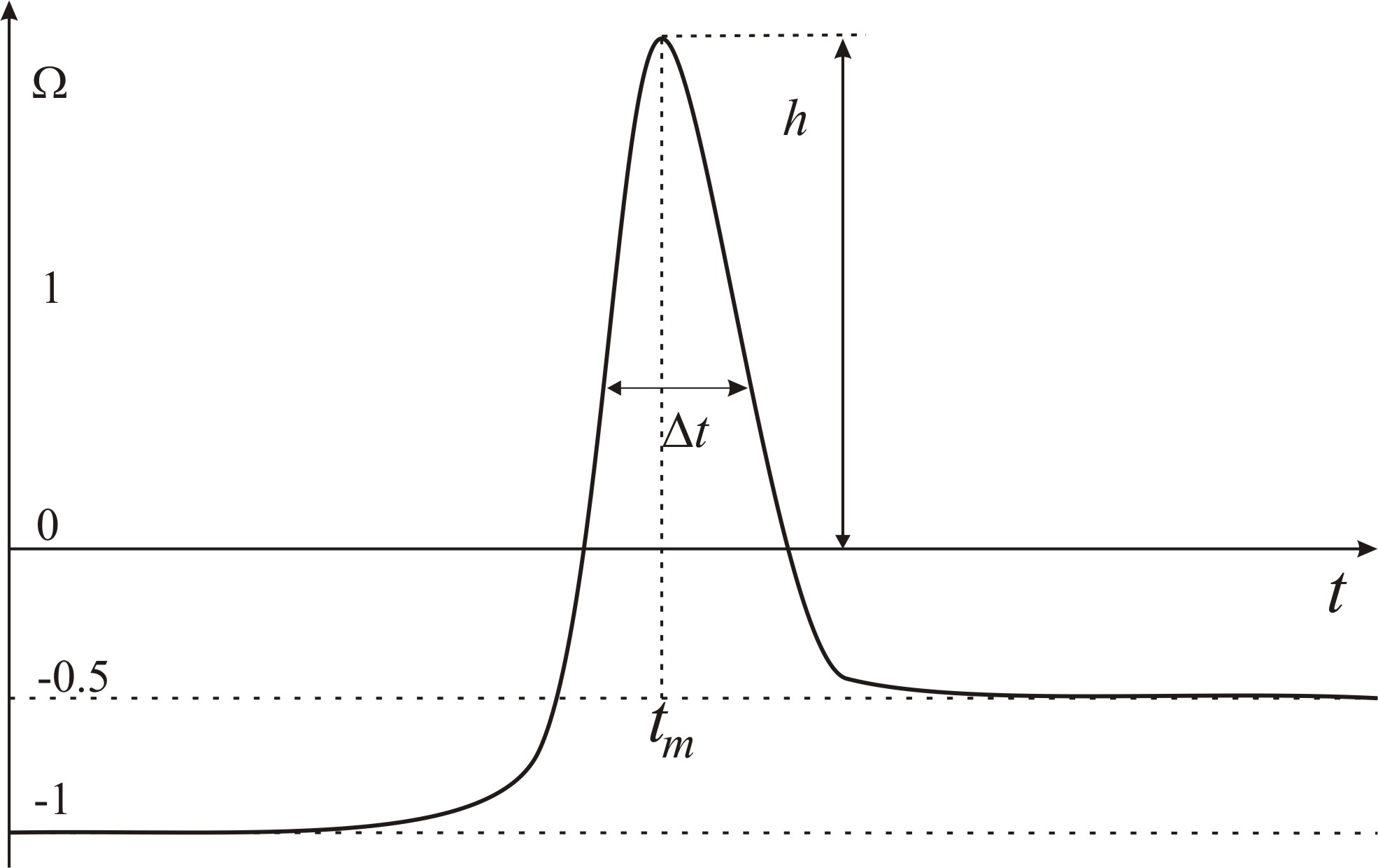}
\end{center}
{\small \textbf{Fig.  \thefigure}.\label{Fig69} The second type of the cosmological acceleration.\vspace{12pt}}
The following parameters are typical for this scenario: $t_m\sim 10^6\div 10^9$; $\Delta t \sim 10^6$; $h\sim 10$, i.e. sufficiently long intermediate stage of hyper-acceleration with final transition to non-relativistic stage.
\item \emph{3rd type}.\; Ultrarelativistic start  $\rightarrow$ smooth transition to non-relativistic stage $\rightarrow$ slight acceleration burst
$\rightarrow$ inflation stage (Fig. 70). This scenario is realized for a case of non-minimal interaction ($\sigma\not=0\rightarrow q\not=0$) and massive scalar field ($m_s\not=0$).
\begin{center}\refstepcounter{figure}
\includegraphics[width=0.8120mm]{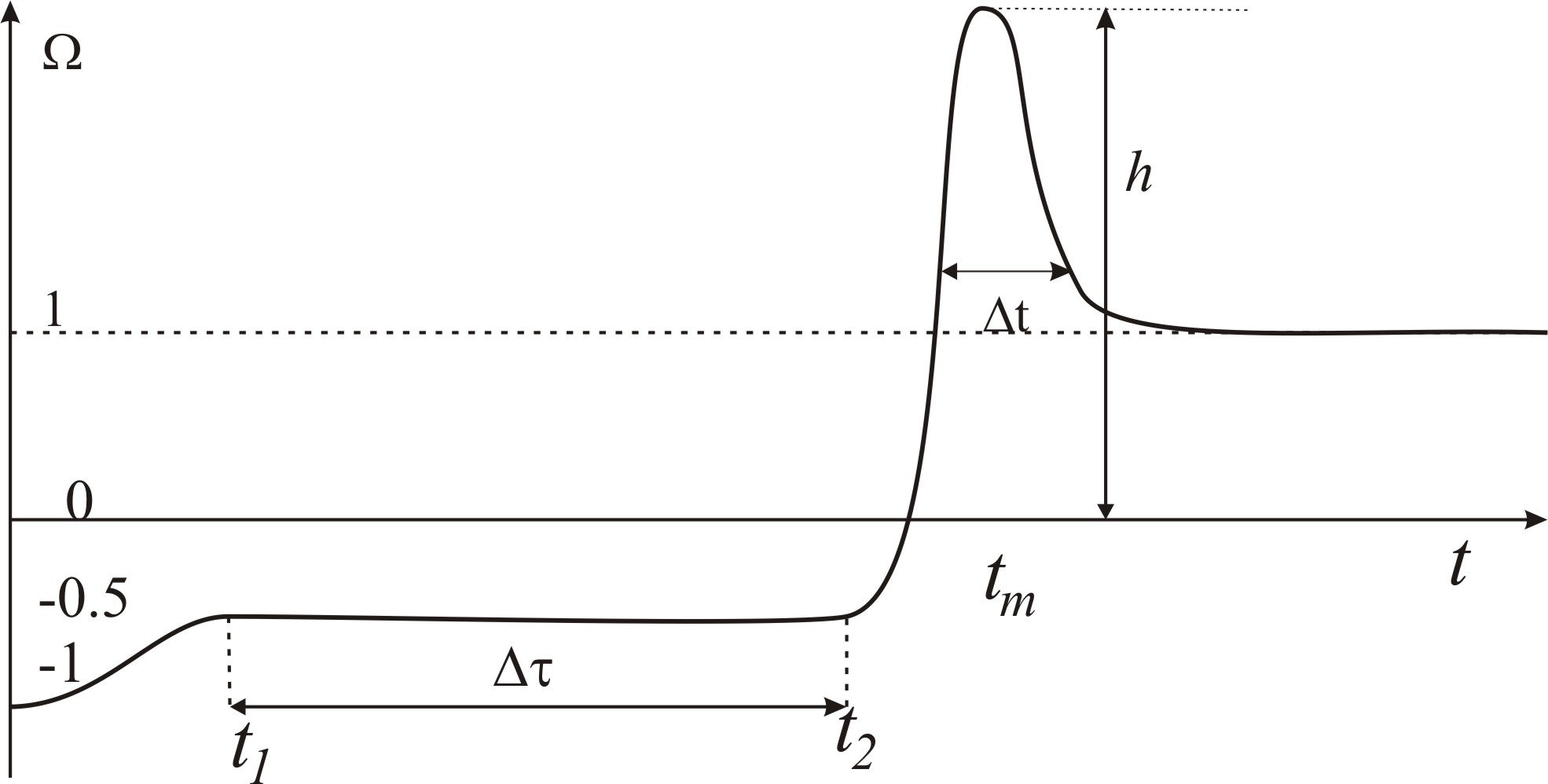}
\end{center}
{\small \textbf{Fig.  \thefigure}.\label{Fig70} The third type of the cosmological scenario.\vspace{12pt}}
The following parameters are typical for this scenario: $t_1\sim 10^2$ (time of change of the ultrarelativistic stage to non-relativistic); $\Delta\tau=10^3\div 10^8$ (non-relativistic stage's duration), $t_m\sim 10^2\div 10^9\div 10^9$, $h\sim 2\div3$, $\Delta t\sim 10^6\div 10^8$.
\item \emph{4th type}.\; Ultrarelativistic start $\rightarrow$ slight acceleration burst $\rightarrow$ non-relativistic stage
$\rightarrow$ transition to inflation stage (Fig. 71). This scenario is realized also for a case of non-minimal interaction ($\sigma\not=0\rightarrow q\not=0$) and massive scalar field ($m_s\not=0$).

\begin{center}\refstepcounter{figure}
\includegraphics[width=0.9120mm]{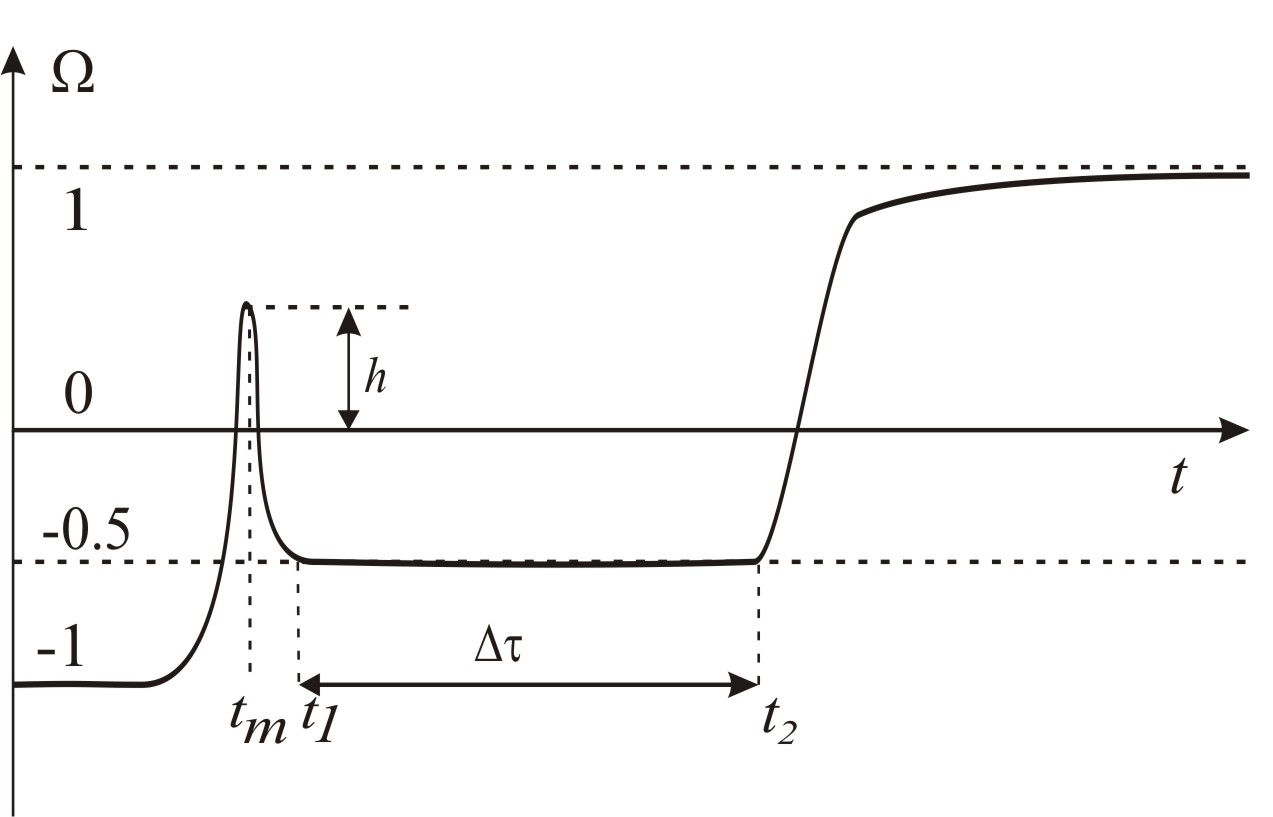}
\end{center}
{\small \textbf{Fig.  \thefigure}.\label{Fig71} The fourth type of the cosmological acceleration.\vspace{12pt}}

The following properties are typical for this scenario: $t_m\sim 10^2\div 10^6\div 10^9$, $t_1\sim 10^2\div 10^6$ (non-relativistic stage's beginning); $\Delta\tau\sim10^7$ (non-relativistic stage's duration), $h\sim 0.5$.
\end{itemize}
\item The largest and latest bursts of the cosmological acceleration (order of $\Omega\sim10^2$ and even more)\footnote{The examples of the cosmological acceleration bursts before $\Omega\sim 10^8\div10^{10}$ are described in
\cite{Ignatev_AAS}} at times $t_m\sim 10^5\div 10^9$ are appropriate to the 1st type scenario.
\end{enumerate}

Let us notice that actually all considered above phenomena take place At post-Planck times when the necessity of gravitation's quantization disappears.
Let is also notice that as is known (see e.g.\\ \cite{Gorb_Rubak}), in accordance with contemporary observations to solve problems of horizon and the Universe's flatness it is enough the duration of inflation $10^{-42}$ $\div 10^{-9}$ s (see e.g.\\ \cite{Gorb_Rubak}), i.e. post-Planck inflation $t\sim 10\div 10^{24}$ â in Planck time scales is enough. Even the cosmological acceleration bursts' durations fit these scales. Let us further notice the very important fact revealed in this research. During the cosmological evolution there naturally appear non-relativistic stages, both early ones (3rd type of the scenario),  intermediate ones (fourth type of the scenario) and final ones (second type of the scenario). The existence of these stages provides possibility of gravitation instability's progress and therefore generation of the cosmological structure. Herewith the 2nd and the 4th types of the scenario limit the scale of instability modes from above and below $kt>1$. Thus, on a basis of this model with antiparticle fantom scalar interaction it is apparently possible to create a more complete cosmological model able to describe basic observation data.

\section{Acknowledgments}
This work was funded by the subsidy allocated to Kazan Federal University for the state assignment in the sphere of scientific activities.

In conclusion, the Authors express their gratitude to the members of MW seminar for relativistic kinetics and cosmology of Kazan Federal University for helpful discussion of the work.


\begin{thebibliography}{}
%
\bibitem[Weinberg(2008)]{Weinberg}
    Steven Weinberg, Cosmology. Oxford University Press (2008).
%
\bibitem[Gor\-bunov, Rubakov(2011)]{Gorb_Rubak}
     D.S. Gorbunov and V.A. Rubakov, Introduction to the Theory of the Early Universe: Cosmological
     Perturbations and Inflationary Theory. Singapore: World Scientific (2011).
%
\bibitem[Einstein(1917)]{Einstein}
     A. Einstein, Sitzungsher. preus, Akad. Wiss., 1 (1917) 142.
%
\bibitem[Utiyama,Fukuyama(1971)]{Utiyama}
     R. Utiyama and T. Fukuyama, Progr. Theor. Phys., 45 (1971) 612.
%
\bibitem[Minkevich(2006)]{Minkevich}
     A.V. Minkevich, Gravitation \& Cosmology, 12 (2006) 11.
%
\bibitem[Minkevich,Garkun,Kudin(2007)]{Minkevich2}
     A.V. Minkevich, A.S. Garkun and V.I. Kudin, Class. Quantum Grav., 24 (2007) 5835.
%
\bibitem[Starobinsky(1980)]{Starobinsky}
    A.A. Starobinsky, Phys. Lett. B 91(1) (1980), 99.
%
\bibitem[Ignat'ev(1982)]{Ignatev1}
    Yu.G. Ignat'ev,  Russ. Phys. J., 25(4) (1982) 92 .
%
\bibitem[Ignat'ev(1983)]{Ignatev2}
    Yu.G. Ignat'ev, Russ. Phys. J., 26(8) (1983) 15.
%
\bibitem[Ignat'ev(1983a)]{Ignatev3}
    Yu.G. Ignat'ev, Russ. Phys. J., 26(8) (1983), 19.
%
\bibitem[Ignat'ev(1983b)]{Ignatev4}
    Yu.G. Ignat'ev, Russ. Phys. J., 26(12) (1983) 9.
%
\bibitem[Ivanov(1983)]{Ivanov}
    G.G. Ivanov, Russ. Phys. J.  26(1) (1983) 32.
%
\bibitem[Ignat'ev et al.(1984)]{kuza}
    Yu.G. Ignat'ev and R.R. Kuzeev, Ukr. Fiz. J., 29 (1984) 1021.
%
\bibitem[Ignat'ev and Miftakhov(2006)]{YuMif}
     Yu. Ignat'ev and R. Miftakhov,  Grav. and Cosmol., 12(2-3) (2006) 179; %
     arXiv:1011.5774[gr-qc].
     %
\bibitem[Ignat'ev and Miftakhov(2011)]{YuMif11}
     Yu.G. Ignatyev and R.F. Miftakhov, Grav. and Cosmol., 17(2)  (2011) 190.
%
\bibitem[Ignat'ev(2012)]{YuNewScalar1}
    Yu. G. Ignat'ev, Russ. Phys. J., 55(2) (2012) 166.
%
\bibitem[Ignat'ev(2012a)]{YuNewScalar2}
    Yu. G. Ignat'ev,  Russ. Phys. J., 55(5) (2012) 550.
%
\bibitem[Ignat'ev(2013)]{YuNewScalar3}
    Yu. G. Ignat'ev, Russ. Phys. J., 55(11) (2013) 1345; arXiv:1307.2509 [gr-qc]. 
%
\bibitem[Ignat'ev(2010)]{Yubook1}
    Yurii G. Ignatyev. Relativistic %
    Kinetic Theory of Nonequi\-lib\-rium Processes in Gravitational %
    Fields. Kazan, Foliant-Press, -- 2010;  http://rgs.vniims.ru/books/const.pdf.
%
\bibitem[Ignat'ev(2013a)]{Yubook2}
    Yurii G. Ignatyev. The Nonequilibrium Universe: The Kinetics Models of the Cosmological Evolution, Kazan: Kazan University Press, 2013;
 \\ http://www.stfi.ru/archive\_rus/2013\_2\_Ignatiev.pdf.
%
\bibitem[Yu.Ignat'ev and D. Ignatyev(2014)]{Ignatev14_1} 
    Yu.G. Ignatyev and D.Yu. Ignatyev, Grav. and Cosmol., {\bf 20}, 299 (2014).
%
\bibitem[Ignat'ev(2014)]{Yu_stfi14}  
    Yu.G. Ignat'ev, Space, Time and Found. Interact., No 1(6) (2014) 47.
%
\bibitem[Yu. Ignat'ev, A. Agathonov and D. Ignatyev (2014)]{Ignatev14_2}  
    Yu. G. Ignatyev, A. A. Agathonov and D. Yu. Ignatyev, Grav. and Cosm., 20, No. 4 (2014) 304–308.
%
\bibitem[Ignat'ev(2015)]{Ignatev_stfi15}
Yu. Ignat'ev, Space, Time and Found. Interact., No 1(10) (2015) 5.
%
\bibitem[Ignat'ev(2015a)]{Ignatev_15}
    Yu.G. Ignatyev, Grav. and Cosm., 21(4) (2015) 296.
%
\bibitem[Ignat'ev(2016)]{Ignatev_16_1}
    Yu. G. Ignat'ev, Grav. and Cosm., 22(1) (2016) 20.
%
\bibitem[Ignat'ev(2016)]{Ignatev_16_2}
    Yu.G. Ignat'ev,	Russ. Phys. J., 59(1)  (2016) 20.
%
\bibitem[Ignat'ev and Agathonov(2016)]{Ignatev_stfi16}
Yu.G. Ignat'ev and A.A. Agathonov, Space, Time and Found. Interact., 1(14) (2016) 91.
%
\bibitem[Ignat'ev and Agathonov(2015)]{Ignatev_Agathonov_2015}
    Yu. G. Ignatyev and A. A. Agathonov, Grav. and Cosmol., 21(2) (2015) 105.
%
\bibitem[Ignat'ev and Mikhailov(2015)]{Ignatev_Mikhailov_2015}
    Yu.G. Ignat'ev and M. L. Mikhailov,	Russ. Phys. J., 57(12) (2015) 1743.
%
\bibitem[Ignat'ev et al. (2015)]{Ignatev_AAS}
    Yurii Ignat'ev, Alexander Agathonov, Mikhail Mikhailov and Dmitry Ignatyev,	Astroph. Space Sci (2015) 357:61.
%
\bibitem[Synge(1957)]{Sing}
    Synge J.L.. The relativistic gas. Amsterdam, North-Holland
    Publishing Company, (1957).
\bibitem[Synge(1960)]{Sing1}
    Synge J.L., Relativity: The General Theory, Amsterdam, 1960.
%
\bibitem[Ignat'ev(2007)]{Bogolyub}
    Yu.G. Ignat'ev, Grav. and Cosmol.,  13(1) (2007) 59.
\bibitem[E. Cartan(1934)]{Cartan}
    E. Cartan, Les espaces de Finsler, Paris, 1934.
%
\bibitem[A. Vlasov(1966)]{Vlasov}
    A.A. Vlasov. Statistical Distribution Functions. Moskow, Nauka, 1966.
%
\bibitem[Ignat'ev and Popov(1990)]{Ignat_Popov}
    Yu.G. Ignat'ev and A.A. Popov. // Actrophysics and Space Science. -- 1990. -- Vol. %
    163. -- p. 153-174; Yu.G. Ignatyev, A.A. Popov. // arXiv:1101.4303v1 [gr-qc].
%
\bibitem[Landau and Lifshitz(1971)]{Land_Field}
    L.D. Landau, E.M. Lifshitz. The Classical Theory of Fields. Pergamon Press.
    Oxford$\cdot$ New York$\cdot$ Toronto$\cdot$ Sydney$\cdot$ Paris$\cdot$ Frankfurt, 1971
%
\bibitem[Landau  and Lifshitz(1980)]{Landau_Stat}
    L.D. Landau, E.M. Lifshitz. Statistical Physics. Vol. 5 (3rd ed.). Pergamon Press. %
    Oxford$\cdot$ New York$\cdot$ Toronto$\cdot$ Sydney$\cdot$ Paris$\cdot$ Frankfurt, 1980.
%
\bibitem[Ignat'ev(2015ñ)]{Ignatev_stfi15a}
Yu.G. Ignat'ev, Space, Time and Found. Interact., 3(12) (2015) 5; Yurii Ignat'ev, arXiv:1508.05375v1 [gr-qc].
%
\bibitem[Ignat'ev(2013b)]{Ignatev_13b}
Yu. G. Ignatyev	(Ignat'ev), Grav. and Cosmol., 19(4) (2013) 232; arXiv:1306.3633v1 [gr-qc].
%
\bibitem[Lebedev(1963)]{Lebed}
    N.N. Lebedev. Spetial Functions and Its Applications.  Moskow-Leningrad, GIFML, (1963).
\end{thebibliography}
\end{document}